\documentclass[usenatbib,usegraphicx]{mn2e}

\newcommand{\oi}{{\sc{O\,i}}}
\newcommand{\ci}{{\sc{C\,i}}}
\newcommand{\oiii}{{\sc{O\,iii}}}
\newcommand{\nii}{{\sc{N\,ii}}}
\newcommand{\niii}{{\sc{N\,iii}}}
\newcommand{\cii}{{\sc{C\,ii}}}
\newcommand{\hi}{{\sc{H\,i}}}
\newcommand{\hii}{{\sc{H\,ii}}}
\newcommand{\iso}{{\it{ISO}}}

\title[The \iso~LWS spectral survey towards Sgr~B2]
{The \iso~LWS high resolution spectral survey towards Sagittarius~B2\thanks{Based on observations with \iso, an ESA project with instruments funded by ESA Member States (especially  the PI countries: France, Germany, the Netherlands and the United Kingdom) with the participation of ISAS and NASA.}}

\author[E. T. Polehampton et al.]
{Edward~T.~Polehampton,$^{1,2}$
Jean-Paul~Baluteau,$^3$ 
Bruce~M.~Swinyard,$^1$ 
\newauthor
Javier~R.~Goicoechea,$^4$
John~M.~Brown,$^5$ 
Glenn~J.~White,$^{1,6}$ 
Jos\'{e}~Cernicharo,$^7$ 
\newauthor
Timothy~W.~Grundy,$^1$ \\
$^1$Space Science \& Technology Department, Rutherford Appleton Laboratory, Chilton, Didcot, Oxfordshire, OX11 0QX, UK\\
$^2$Department of Physics, University of Lethbridge, 4401 University Drive, Lethbridge, Alberta, T1J 1B1, Canada \\
$^3$Laboratoire d'Astrophysique de Marseille, CNRS \& Universit\'e de Provence, BP 8, F-13376 Marseille Cedex 12, France\\
$^4$LERMA, UMR 811, CNRS, Observatoire de Paris et \'{E}cole Normale Sup\'{e}rieure, 24 rue Lhomond, 75231 Paris Cedex 05, France\\
$^5$Physical and Theoretical Chemistry Laboratory, South Parks Road, Oxford, OX1 3QZ, UK\\
$^6$Department of Physics and Astronomy, Open University, Venables Building, Milton Keynes, MK6 7AA, UK\\
$^7$Departmento de Astrof\'{i}sica Molecular e Infrarroja, IEM, CSIC, Serrano 121, 28006, Madrid, Spain}

\date{Accepted ... Received ... ; in original form .....}

\pagerange{\pageref{firstpage}--\pageref{lastpage}} \pubyear{2007}

\begin{document}

\label{firstpage}
   
\maketitle

\begin{abstract}
A full spectral survey was carried out towards the Giant Molecular Cloud complex, Sagittarius B2 (Sgr~B2), using the \iso~Long Wavelength Spectrometer Fabry-P\'{e}rot mode. This provided complete wavelength coverage in the range 47--196~$\mu$m (6.38--1.53~THz) with a spectral resolution of 30--40~km~s$^{-1}$. This is an unique dataset covering wavelengths inaccessible from the ground. It is an extremely important region of the spectrum as it contains both the peak of the thermal emission from dust, and crucial spectral lines of key atomic (\oi, \cii, \oiii, \nii~and \niii) and molecular species (NH$_3$, NH$_2$, NH, H$_2$O, OH, H$_3$O$^+$, CH, CH$_2$, C$_3$, HF and H$_2$D$^+$). In total, 95 spectral lines have been identified and 11 features with absorption depth greater than 3$\sigma$ remain unassigned. Most of the molecular lines are seen in absorption against the strong continuum, whereas the atomic and ionic lines appear in emission (except for absorption in the \oi~63~$\mu$m and \cii~158~$\mu$m lines). Sgr~B2 is located close to the Galactic Centre and so many of the features also show a broad absorption profile due to material located along the line of sight. A full description of the survey dataset is given with an overview of each detected species and final line lists for both assigned and unassigned features.
\end{abstract}

                
%

\section{Introduction}

Sagittarius~B2 (Sgr~B2) is a well studied giant molecular cloud complex, located $\sim$120~pc from the Galactic Centre \citep[e.g.][]{lisf}. It is unique in our galaxy, being one of the most massive star forming regions, and has an extremely rich chemistry. Many of the molecular species detected in the interstellar medium have only been observed towards Sgr~B2. Furthermore, the line of sight towards Sgr~B2 crosses the main galactic spiral arms lying between the Sun and Galactic Centre. Cold clouds associated with these spiral arms are seen in absorption against the Sgr~B2 continuum emission \citep[e.g.][]{greaves94}. These facts make it a perfect target for systematic spectral surveys. 

The Sgr~B2 complex consists of three main clusters of compact \hii~regions and dense molecular cores aligned in a north-south direction \citep[e.g.][]{goldsmith1990}. These are surrounded by a diffuse envelope \citep[e.g. see][]{huttemeister}. The far-infrared (FIR) emission of Sgr~B2 is most intense close to the source Sgr~B2 M \citep{goldsmithb} where the spectrum is dominated by thermal continuum from dust and has a peak near 80~$\mu$m. The continuum opacity at 100~$\mu$m is high \citep[3.8$\pm$0.4;][]{goicoechea_d}, which means that only the external layers of the cloud can be seen in the FIR. This is in contrast to longer wavelengths, where the continuum is optically thin and emission lines from complex molecules in the hot cores are observed \citep[e.g.][]{nummelin_a}. Thus, the FIR spectrum of Sgr~B2 contains lines due to simple molecules observed in absorption against the dust continuum, as well as lines of atoms and ions in emission in the envelope.

Spectral surveys have previously been carried out towards Sgr~B2 in the mm and sub-mm region \citep{friedel,nummelin_b,nummelin_a,sutton,turner,cummins} but comprehensive, unbiased coverage at FIR wavelengths did not become possible until the launch of the {\it Infrared Space Observatory} ({\it ISO}). A program of several wide spectral surveys, including Sgr~B2, was carried out with the \iso~Long Wavelength Spectrometer \citep[LWS;][]{clegg} using its Fabry-P\'{e}rot (FP) mode, L03. These were very time consuming and so were conducted towards bright sources in order to obtain a high signal-to-noise in a reasonable integration time. Even so, only two sources were covered across the entire LWS wavelength range with no gaps; Sgr~B2 and Jupiter. In addition, two other objects were observed with almost complete wavelength coverage; the Kleinmann-Low nebula in the Orion Molecular Cloud 1 \citep{lerate} and the Galactic Centre (White et al. in preparation).

Here, we present the Sgr~B2 survey in full - this includes complete unbiased coverage over the entire wavelength range 47--196~$\mu$m. The detected lines are mostly due to rotational transitions between the lower energy levels of molecules, as well as low energy molecular vibrations and cooling lines of atomic species. The main identified features have already been published using results from this spectral survey as well as a series of targeted observations using the LWS FP L04 mode \citep[see ][and references therein]{goicoechea_d}. However, this is the first presentation of the entire spectrum (note that the FP spectrum in fig.~2 of Goicoechea et al. contains only the spectral lines, with gaps between them). A preliminary report on this survey was given by \citet{cox}. The data presented here will be available in their fully reduced form from the \iso~Data Archive\footnote{see {\tt www.iso.esac.esa.int/ida/}} as Highly Processed Data Products.

In Sect.~\ref{sect_obsns} we present the observing strategy and describe the observations. In Sect.~\ref{sect_datared} we describe in detail the data reduction with particular emphasis on the improvements in calibration developed for the Sgr~B2 dataset. In Sect.~\ref{sect_results} we present all the data with a full list of assigned features and summarise the results for each species. In addition, we present a list of unidentified features. This is particularly important in the 158--196~$\mu$m range that will be re-observed at much higher spatial and spectral resolution by the HIFI instrument on board the {\it Herschel Space Observatory}. The spectral range $>$57~$\mu$m will also be re-observed by the {\it Herschel} PACS instrument, with spectral resolution a factor of a few below that of our survey, although with $\sim$8 times better spatial resolution and much improved sensitivity.

\begin{figure}
\includegraphics[width=84mm]{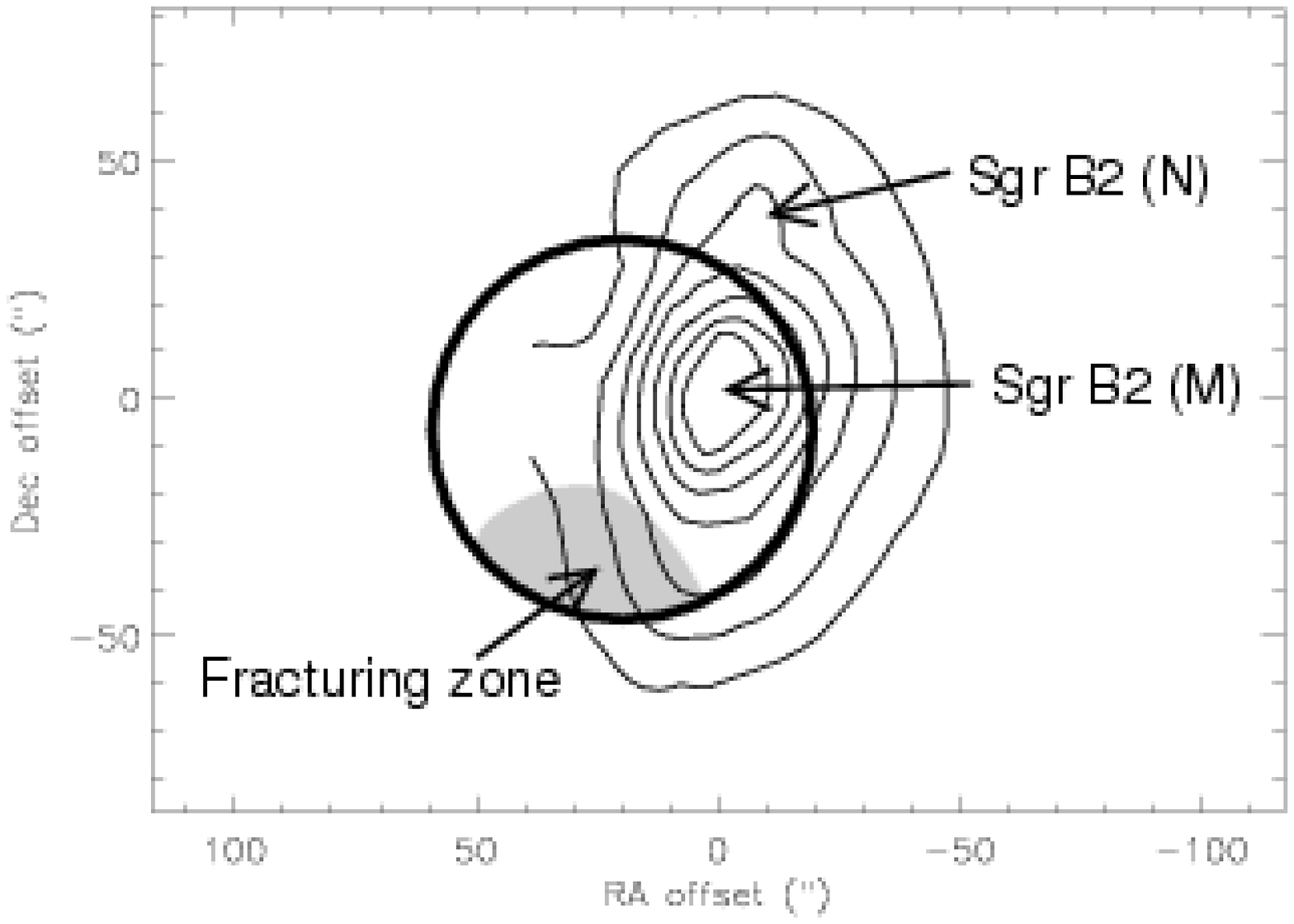}
\caption{The \iso~LWS beam footprint overlayed on the 100~$\mu$m contours from from \citet{goldsmithb}. Offsets are centred at the nominal position of Sgr~B2 (M) at $\alpha=17^{\mathrm{h}}47^{\mathrm{m}}20.2^{\mathrm{s}}$, $\delta=-28\degr 23\arcmin 7.3\arcsec$ (J2000). The contours are 0.2, 0.3, 0.4 ....0.8 of the peak intensity (18~kJy). The location of the \iso~beam quadrant where spectral fracturing occurs is indicated.}
\label{beam}
\end{figure}

\section{Observations}\label{sect_obsns}

Sgr~B2 was observed as part of the guaranteed time spectral surveys programme ISM\_V. Full wavelength coverage (47--196~$\mu$m) was achieved over the entire LWS spectral range using the FP mode, L03. In addition to this, extra observations were scheduled as a solicited proposal (SGRB2\_ZZ) in the range 167--194~$\mu$m. These extra observations aimed to increase the signal-to-noise in the long wavelength range. The first observation was carried out on 1997 March 6 (\iso~revolution 476) as a test of the L03 mode before it was fully commissioned. The remaining guaranteed time observations were carried out between 1997 April 3 and 1997 April 8 with the solicited proposal observations between 1998 February 26 and 1998 March 14. 

To operate the LWS at high resolution, a Fabry-P\'{e}rot (FP) interferometer was placed into the beam in front of the grating (which then acted as an order selector for the FP). To maintain high spectral resolving power across the full LWS wavelength range, one of two FPs was used, each with its etalons and spacing optimised for a particular waveband. The shorter wavelength FP, termed FPS, had a nominal range 47--70~$\mu$m and the longer wavelength FP, termed FPL, had a nominal range of 70--196~$\mu$m \citep{davis}. Radiation was detected over the whole spectral range using 10 detectors, each with its own band-pass filter. In each L03 observation both the LWS FP and grating were scanned to cover a wide range in wavelength. The final data consist of a series of many FP `mini-scans', each at a different grating angle. This is in contrast to the L04 mode, where the FP was scanned at only one or two grating angles, giving a narrow targeted observation.

The spectral resolution achieved was $30-40$~km~s$^{-1}$ across the range (see Sect.~\ref{throughput} for a detailed discussion). Each observation was carried out with a spectral sampling interval of a quarter of a resolution element with each mini-scan repeated 3 times for the guaranteed observations and 4--6 times for solicited observations. Each data point had a detector voltage ramp length of 0.5~s. The final dataset consists of 36 individual observations with a total of 53.6~hours of \iso~observing time. These are detailed in Table~\ref{alll03}. 

The LWS beam had an effective diameter of between 66$\arcsec$ and 86$\arcsec$ across its wavelength range \citep{gry} and was pointed at coordinates $\alpha=17^{\mathrm{h}}47^{\mathrm{m}}21.75^{\mathrm{s}}$, $\delta=-28\degr 23\arcmin 14.1\arcsec$ (J2000). This gave the beam centre an offset of 21.5$\arcsec$ from the main FIR peak of Sgr~B2. This pointing was used to exclude the source Sgr~B2~N from the beam. An in depth study of the LWS beam profile has shown that it was asymmetric with characteristics that vary for different positions of the source in the aperture \citep{lloyd_a}. In particular if the source was located in one quadrant of the aperture, there are problems of `spectral fracturing', where the spectral shape does not agree on different detectors - this is not a problem for on-axis sources \citep{lloyd_b}. As already mentioned, in order to exclude radiation from Sgr~B2 N, the brightest FIR peak was observed off-axis. For this reason, it was important to maintain the same orientation of the LWS aperture for all of the survey observations. A spacecraft roll angle of $\sim93\degr$ was used for all observations, ensuring that the source did not fall in the fractured part of the aperture. A representation of the LWS beam footprint on a 100~$\mu$m map of Sgr~B2 is shown in Fig.~\ref{beam}.

In addition to the L03 observations, the whole spectral range was observed at lower spectral resolution using the LWS grating mode L01 (resolving power, $\lambda/\Delta\lambda\sim$200-300). This was carried out with the same pointing and roll angle as the FP observations. Details of the L01 observation are given in Table~\ref{alll03}.

\subsection{Non-prime data}

During each observation, the instrument settings were optimised for one detector only (so that a single FP spectral order was scanned over the maximum in grating response at that detector position). This detector was designated as `prime'. However, at the same time, the other 9 non-prime detectors simultaneously recorded data. These data are also useful if the combination of grating angle and FP gap were such that an adjacent FP order was transmitted to the non-prime detector position. As the grating response covered a similar wavelength range to the separation between FP orders, this occurred frequently. 

In the course of the Sgr~B2 survey, all wavelengths in the LWS range were observed in prime data. There is then a huge additional dataset (9 times larger) containing non-prime data. Much of this was observed with high transmission through the instrument and so is extremely useful to increase the signal-to-noise ratio in the prime data and to confirm the detection of weak lines. This is particularly useful when data with the same wavelength can be recovered from different detectors.

The non-prime dataset is also useful for deriving an accurate flux calibration for the survey. The next section describes in detail how the standard FP calibration was extended by including non-prime data. In particular, dark current including stray light could be recovered where the combination of the FP and grating blocked all transmission to a non-prime detector. Although non-prime data had been used before both in L03 and L04 observations \citep[e.g.][]{ceccarelli}, they were always reduced in an ad hoc way. The comprehensive calibration developed for the Sgr~B2 survey allows both prime and non-prime data to be combined in a consistent and reliable way across the LWS wavelength range. This technique has since been used for the LWS FP survey towards Orion KL \citep{lerate}.

\section{Data reduction}\label{sect_datared}

Processing of the LWS FP data was carried out using the LWS offline pipeline (OLP) version 8 and routines developed for the LWS Interactive Analysis package \citep[LIA;][]{lim_d}. The basic OLP calibration, to the `Standard Processed Data' stage, is fully described in the LWS handbook \citep{gry}. This produced data still in engineering units (FP gap voltage and detector photocurrent). The basic calibration was extended specifically for the Sgr~B2 survey, with additional steps developed for the LIA FP processing routine, `{\sc fp\_proc}'. This routine applies the conversion from raw engineering units to flux and allows an interactive optimisation of each step. The extension to the standard calibration is described in Appendix~\ref{appendix_datared} (determination of accurate dark and stray light values, and characterisation of the instrumental response and throughput). The improvements made to the {\sc fp\_proc} routine were included as part of the version 10 release of the LIA software. Further optimisation (e.g. removal of glitches) was carried out using the \iso~Spectral Analysis Package \citep[ISAP;][]{sturm}.

\begin{figure*}
\includegraphics[width=168mm]{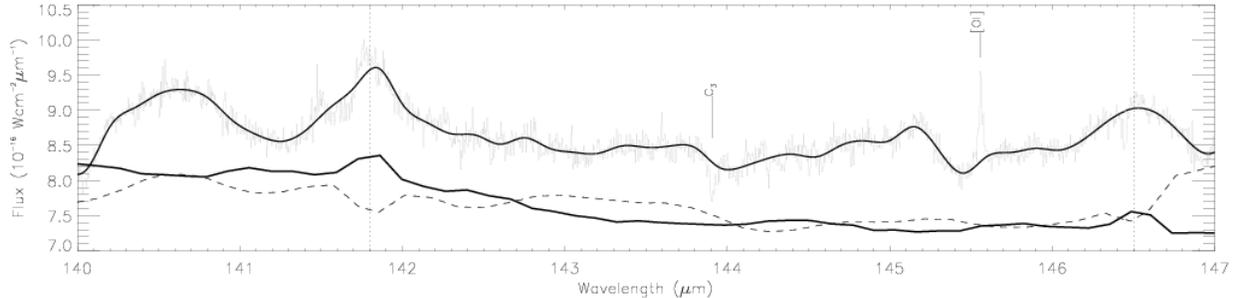}
\caption{Example of prime data from one observation (TDT 47600809; grey) with the fitted continuum. The L01 grating observation is shown below (thick line) and also the detector RSRF (dashed line). Two places where dips in the RSRF correspond with emission in both L01 and L03 observations are show by the dotted lines.}
\label{contdiv}
\end{figure*}

\subsection{Wavelength calibration and mini-scan shifting}

The wavelength calibration for the FP data was determined and monitored using observations of several standard sources as described in \citet{gry}. We have used the latest wavelength calibration coefficients from OLP 10. Adopting the most conservative value for the FPS wavelength uncertainty gives an accuracy of 6~km~s$^{-1}$. For FPL, a systematic test of the wavelength accuracy was performed using CO lines observed towards Orion - in this case, the observed residual velocity differences were never worse than 11~km~s$^{-1}$ \citep{gry}. 

However, note that the standard wavelength calibration was determined using spectral lines within the nominal operating range of each FP and the uncertainties quoted above only apply within these ranges. We have used non-prime data from FPL for the region $<$70~$\mu$m, and these data show a clear error in wavelength calibration that is larger than 11~km~s$^{-1}$. No correction has been made for this systematic shift - it is discussed further in Sect.~\ref{line_props}.

The wavelength calibration of the grating is also important in the reduction of FP data as the shape of the grating response function must be removed from each mini-scan. The precise location of the grating transmission maximum in wavelength is required so that the correct portion of the response shape can be used. However, the requirements for high resolution FP observations were not taken into account in the original design specification for the grating positional accuracy. In order to provide the necessary information for FP observations, the grating position should have been monitored roughly 50 times more accurately. The effect on data calibrated with the LWS OLP pipeline is that adjacent mini-scans do not necessarily join together. A small shift in the grating position can correct for this but there is no independent means of determining its value except for direct inference from the FP data. To provide the best estimate of these shifts, we have used the LIA FP reduction routine, {\sc fp\_proc}, to interactively process each mini-scan. This routine allows the shift to be adjusted until adjacent mini-scans show the best agreement. In the reduction of the Orion KL spectral survey data, an additional step to fit the shape of each mini-scan was applied \citep{lerate}. However, for Sgr~B2 the interactive shifting was enough to align adjacent mini-scans and the extra step was not applied.

The wavelength scale of each observation was corrected to the kinematical Local Standard of Rest (LSR) frame, to account for the motions of the Earth around the Sun and the Sun around the Galaxy. The corrections applied are detailed in Table~\ref{alll03}.

\subsection{Glitches}

Glitches in the data were caused by charged particle hits on the detectors. These caused a jump in the detector voltage ramp, changing its slope (and possibly the slope of subsequent ramps). These appear as data points with high (or low) photocurrent. Very large spikes in photocurrent were removed automatically in the pipeline processing \citep[see][]{gry} but smaller glitches and the decaying tails of large glitches could not be removed automatically due to the difficulty of finding an algorithm that could determine the difference between glitches and real spectral lines. This meant that the data had to be deglitched by hand.

In the L03 observations each mini-scan was repeated 3--6 times and this usually meant that there was at least one repeated scan with no glitch signature. In order to distinguish between glitches and real lines, the data were plotted with each repeated scan in a different colour using the histogram plot style within ISAP. In this way it was possible to make a decision on each glitch by comparing with the other repeated scans. The glitches that occurred in the LWS FP data can be described by four generic shapes and this provided a template against which to compare the data; a single spike (positive or negative), a sudden jump in photocurrent with decaying tail, or an upward spike with sudden fall and gradual recovery.

\subsection{L01 observation}

One L01 observation was included as part of the survey with the same pointing and spacecraft roll angle as the L03 data. However, due to the strength of the source, the voltage ramps for the long wavelength detectors showed non-linear behaviour (this was not a problem for the FP observations due to the low transmission of the FP etalons). These non-linear effects cause the spectrum to sag at long wavelengths \citep[see][]{leeks}.

In order to calibrate these data, we discarded the second half of each detector voltage ramp, effectively reducing the length from 0.5~s to 0.25~s. Further correction was then applied using the latest version of the `strong source correction' and the L01 post processing pipeline \citep{lloyd_d}. The final calibrated spectrum shows reasonable agreement in the overlap region between detectors, and the sagging has been removed.

Even though the grating observation required these extra measures to correctly calibrate it, the resulting spectrum has a lower uncertainty in the flux than the FP observations. This is because the intrinsic uncertainty in the calibration of FP absolute flux is much higher than that of the grating due to the additional throughput correction required \citep{gry}. The uncertainty in grating flux for the short wavelength detectors is $\sim$10\% \citep{gry}, and we estimate that it is $\sim$20\% for the strong-source-corrected long wavelength detectors.

The final L01 spectrum was only used to obtain the continuum level needed to determine the absolute flux of the atomic and ionic emission lines observed in the survey in Sect.~\ref{sect_atom}.

\begin{figure*}
\includegraphics[width=168mm]{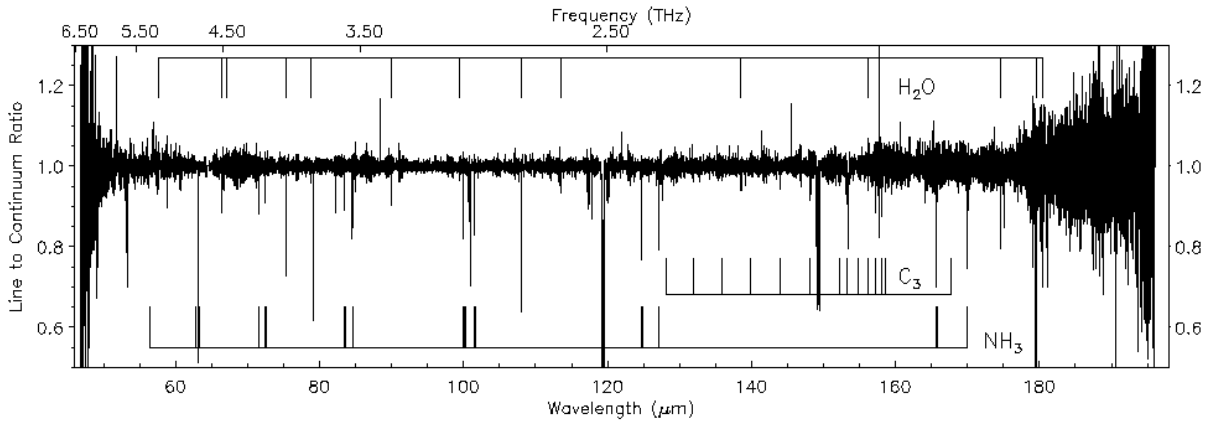}
\caption{The full L03 FP spectrum after co-adding prime and non-prime data using FPL for all LWS detectors (below 70~$\mu$m only non-prime data were used). The positions of detected features for several species are shown. The data were binned at 1/2 the instrumental resolution element.}
\label{fullsurvey}
\end{figure*}

\subsection{Continuum features in the FP data}\label{sect_contm}

In the final FP spectrum there are several jumps between data from different observations and between detectors when viewed on an absolute flux scale. These are due to uncertainties in the multiplicative calibration factors applied in the reduction \citep{swinyard_b}, especially the FP throughput, absolute responsivity and detector RSRFs. A good way to bypass these uncertainties is to divide the continuum level into the data to obtain relative line depth. The remaining error is dominated by statistical noise in the data with a small additional uncertainty due to the dark current determination (this has been well constrained: Sect.~\ref{sect_dark}) and continuum fit.

However, {\it within} each observation ($\sim$5--10~$\mu$m wide), there are some features on scales matching the grating resolution (0.3--0.6~$\mu$m) which complicate the continuum division process on large scales. These are probably due to structure in the RSRF of each detector. They contain some known features due to spectral lines present in the observations of Uranus (see Sect.~\ref{throughput}) but not included in the model (e.g. the HD line at 112~$\mu$m). These appear in the FP spectra as large scale features (see Fig.~\ref{contdiv}). In addition, sudden changes in the RSRF can cause large features due to the delay in responding to fast changes in flux caused by the transient response of the detectors. 

A further effect that could cause structure on large scales is the fact that the spectrum is built up from many narrow mini-scans, each of which is interactively adjusted in the data reduction. This could mean that spurious large scale features are inadvertently introduced by the shifts applied. However, in practise, for prime data where every mini-scan was centred near the maximum in grating transmission, the freedom in the shifting process is only in their slope and curvature and not in the absolute flux level. Therefore, drifts that were not present in the original data cannot be introduced purely by interactive shifts.

\begin{figure}
\includegraphics[width=84mm]{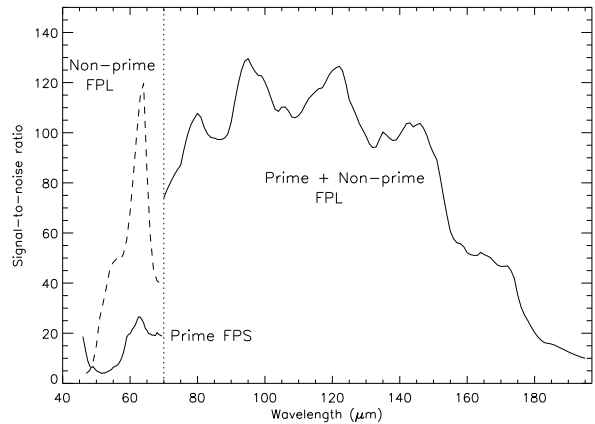}
\caption{The signal-to-noise ratio achieved across the survey range for data binned at 1/4 of the instrumental resolution element. The recovered non-prime data using FPL below 70~$\mu$m (dashed line) shows a clear improvement over the prime data using FPS.}
\label{signoise}
\end{figure}

In fitting the large scale continuum, we have assumed that all features in the FP data on scales of 0.5--1~$\mu$m are not real and were caused by multiplicative instrumental effects (the only possible additive effect is wavelength dependant stray-light and the optical design of the instrument should have ensured that this was negligible). The continuum fit was carried out by masking out strong lines matching the scale of the FP resolution (even wide absorption due to line of sight clouds can be clearly separated from the wide features on the scale of the grating as they have maximum width $\sim$10\% of the grating resolution). The masked data were then rebinned to a quarter of the resolution element width of the grating (the grating resolution element width is $\sim$0.3~$\mu$m for the SW detectors and $\sim$0.6~$\mu$m for the LW detectors). The rebinned spectrum was interpolated back to the original wavelength scale using a spline fit. An example for one observation is shown in Fig.~\ref{contdiv} with the RSRF and L01 grating observation. This figure shows that in this case two of the largest features have counterparts in the RSRF (in the opposite sense to the data), but also that these both show up as spurious features in the L01 observation.

The method of re-binning and spline interpolation was used in preference to fitting a polynomial because very high orders were necessary to cover all features and this meant that spurious features were introduced, particularly in the regions of masked lines.

\section{Results}\label{sect_results}

The full spectrum is shown in Fig.~\ref{fullsurvey}, and is broken up into narrower wavelength ranges with labelled features in Fig.~\ref{allsurvey}.

\begin{figure*}
\includegraphics[width=168mm]{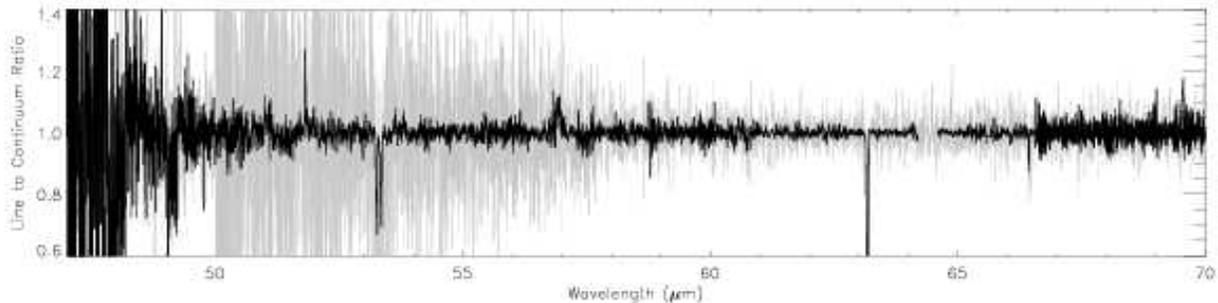}
\caption{Prime data (using FPS) in the region 47--70~$\mu$m are shown in grey with the recovered non-prime data using FPL shown in black. Both are binned at 1/4 of their respective resolution elements. Note the gap in the non-prime data (64.2--64.6~$\mu$m).}
\label{fplimprov}
\end{figure*}

\subsection{Signal-to-noise achieved}\label{sect_signois}

The final signal-to-noise (S/N) ratio achieved in line to continuum ratio data depends both on the transmission of the instrument and on the strength of the underlying continuum of Sgr~B2. Figure~\ref{signoise} shows the variation in S/N with wavelength based on the RMS noise in the final spectrum presented in Fig.~\ref{allsurvey} - ie. with a bin width of 1/4 of the instrumental resolution element. We have investigated the absorption depth/emission of the detected features relative to this RMS, rather than increasing the bandwidth of each bin to enable a detailed search for weak features. This is due to the other uncertainties in the data and the potential (partially resolved) velocity structure in the line shapes confusing any analysis of total flux using an equivalent width approach.

The maximum S/N ratio achieved is in the range 90--125. This falls off at longer wavelengths due to the decrease in continuum level. At short wavelengths in the range covered by FPS ($<$70~$\mu$m), the instrument throughput is low, as well as there being a sharp fall off in the continuum. The S/N in the prime data was improved significantly by including the non-prime observations. This is particularly useful in the wavelength region $<$70~$\mu$m where the throughput of FPL is higher than FPS. We have recovered and calibrated all useful non-prime data from FPL over the entire range of the survey. The resulting spectrum below 70~$\mu$m is compared with the prime data from FPS in Fig.~\ref{fplimprov} and the improvement in S/N shown in Fig.~\ref{signoise}. The non-prime spectrum was stitched together from many fragments observed using FPL with detectors SW2, SW3 and SW4. There is only one gap in the non-prime coverage between 64.2 and 64.6~$\mu$m. The only disadvantage in these data is a slight reduction in spectral resolution and some systematic error in wavelength calibration (by up to $\sim$25~km~s$^{-1}$ - see Sect.~\ref{line_props}). Table~\ref{table_signois} compares the S/N and spectral resolution for several lines in this region. For all the results presented in the following sections, only the higher S/N data using FPL are considered.

For wavelengths above 70~$\mu$m, most of the range was covered in at least one additional observation using a non-prime detector and these data have been co-added with the prime data in Fig.~\ref{allsurvey}. The non-prime data are particularly useful because they extend the overlap in spectral coverage by adjacent detectors, allowing features in the overlap region to be independently checked. All the data presented here were observed using the L03 mode and additional narrow L04 scans were not included. The L04 scans were generally observed with a higher number of repeated mini-scans per point and so have higher signal-to-noise \citep[although only around the targeted lines - see][for details of the L04 observations of Sgr~B2]{goicoechea_d}.

\begin{table}
\caption{The gain in signal-to-noise between prime and non-prime data at selected wavelengths below 70~$\mu$m. This is shown as the ratio of signal-to-noise using FPL to that using FPS. The resolution element for FPS and FPL is also shown.\label{table_signois}}
\begin{tabular}{cccc}
\hline
Wavelength & S/N Gain & $\Delta{v}$ FPS  & $\Delta{v}$ FPL\\
($\mu$m)   &          &  (km~s$^{-1}$)   &  (km~s$^{-1}$) \\
\hline
51.8 ([\oiii]) & 8.5 & 45 & 62 \\
53.3 (OH)      & 8.8 & 45 & 61 \\
57.3 ([\niii]) & 4.8 & 44 & 57 \\
63.2 ([\oi])   & 5.0 & 45 & 52 \\
\hline
\end{tabular}
\end{table}

\subsection{Assigned features}
\label{sect_assign}

\begin{table*}
\caption{Summary of assigned lines in the survey, giving the number of detected lines, an indication of whether absorption from the line of sight is observed in any of the lines, and the fitted column density at the velocity of Sgr~B2. The method used to calculate these values (fitting the ground state line, assuming local thermodynamic equilibrium (LTE), using a large velocity gradient model (LVG) or using a non-local radiative transfer model) and the reference is shown in the last column.}
\label{linsum}
\begin{tabular}{ccccc}
\hline
Species        & Features  & Line of sight & Column density           & Fitting method  \\
               & detected  & absorption    & in Sgr~B2 component      &  \& ref.  \\
               & in survey &               &        (cm$^{-2}$)       &    \\
\hline
NH$_{3}$       & 21   & no  & 3$\times$10$^{16}$        &  LVG [2]\\
NH$_{2}$       & 5    & no  & (1.5--3)$\times$10$^{15}$ &  LTE [1]\\
NH             & 3    & no  & 4$\times$10$^{14}$        &  LTE [1]\\
\hline
H$_{2}$O       & 13   & yes & (9$\pm$3)$\times$10$^{16}$&  Non-local [5]\\
OH             & 13   & yes & (1.5--2.5)$\times10^{16}$; 3.2$\times$10$^{16}$  &  Non-local [3]; Ground state [4]\\
H$_{3}$O$^{+}$ & 6    & yes & 1.6$\times$10$^{14}$      &  LTE [6]\\
\hline
CH             & 2    & yes & 9.3$\times$10$^{14}$      &  Ground state [7]\\
CH$_{2}$       & 2    & yes & 3.4$\times$10$^{14}$      &  Ground state [7]\\
C$_{3}$        & 16   & no  & (4--8)$\times$10$^{15}$   &  Non-local [8]\\
\hline
HF             & 1    & no  &  1.7$\times$10$^{14}$     & [9]\\
\hline
H$_2$D$^+$    & 1    & yes &  9$\times$10$^{13}$       & [10]\\
\hline
\end{tabular}

\medskip
[1] \citet{goicoechea_d}; [2] \citet{ceccarelli}; [3] \citet{goicoechea_b}; [4] \citet{polehampton_oh}; [5] \citet{cernicharo_e}; [6] \citet{goicoechea_c}; [7] \citet{polehampton_ch}; [8] \citet{cernicharo_d}; [9] assuming the abundance of 3$\times$10$^{-10}$ from \citet{neufeld_a} and their radial density profile between 0.6 and 22.5~pc. Note that Neufeld et al. used a slightly different beam position on the source to that adopted for the survey; [10] \citet{cernicharo_f}, calculated based on non-LTE excitation.
\end{table*}

The majority of lines that appear in the spectral range of the survey are due to rotational transitions between the lowest energy levels of simple hydride molecules (e.g. H$_{2}$O, OH). These are generally seen in absorption against the background continuum emission, due to the envelope of Sgr~B2 and in the case of the lowest excitation lines (normally the ground state transition), due to clouds along the line of sight. We also detect rotation-inversion transitions of the pyramidal molecules NH$_3$ and H$_3$O$^+$. In addition, ro-vibrational lines from the lowest energy bending modes of non-polar carbon chains occur in the survey range. In particular, several ro-vibrational transitions of C$_{3}$ in its $\nu_2$ bending mode were observed. Table~\ref{linsum} summarises the detected molecular lines and gives an estimate of the column density of each species. There are also emission lines from forbidden spin-transitions of atoms and ions (\cii, \nii, \niii, \oi, \oiii - see Sect.~\ref{sect_atom}). Finally, several isotopic lines of OH and H$_2$O are detected - see Sect.~\ref{isotopes}. These are important for modelling the physical conditions in the source as they have much lower opacity than the lines of the main isotopologues.

As already noted by \citet{ceccarelli}, no rotational transitions of CO were detected in the survey (the lowest energy transition in the range is $J$=14--13 at 186~$\mu$m, with upper energy of 581~K). In addition, \citet{goicoechea_d} did not detect CO anywhere in the extended region surrounding Sgr~B2 using the LWS grating spectrometer. However, \citet{cernicharo_e} have detected the $J$=7--6 line at 371.65~$\mu$m (806~GHz) from the ground. They performed non-local radiative transfer models to try to reproduce the lack of CO emission in the LWS range, showing that low H$_2$ density is required. If the $J$=7--6 emission occurs in gas with a kinetic temperature $\sim$100~K, the lack of FIR CO lines can be reproduced with $n({\rm{H_2}})=2\times10^4$~cm$^{-3}$, but if the temperature is higher, the density limit is $<10^4$~cm$^{-3}$.

The full list of detected lines is shown in Table~\ref{iden_lines}. This updates and extends the line lists already presented by \citet{polehampton_thesis} and \citet{goicoechea_d}. The remaining unidentified features are described in Sect.~\ref{sect_unassigned}. The transitions in Table~\ref{iden_lines} are described by the quantum numbers $N$, $J$ and $K$, where $N$ is the rigid body rotational quantum number, $J$ is the total angular momentum quantum number excluding nuclear spin, and $K$ is the projection of the total angular momentum. Further splitting due to nuclear spin is not included in Table~\ref{iden_lines}, and where this occurs, the wavelengths quoted are the average over the split levels weighted by the Einstein coefficients. The notation used for H$_2$O, CH$_2$ and NH$_2$ is $N_{KaKc}$, for NH is $N_{J}$, for H$_2$D$^+$ is $J_{KaKc}$, and for H$_3$O$^+$ and NH$_3$ is $J_K^{+/-}$, where $+/-$ specifies the parity of the inversion state. For C$_3$, the transitions are quoted as lines in the $P$-, $Q$- and $R$-branches of the $\nu_2$ bending mode. For OH and CH, the lines are quoted for the two rotational ladders due to spin-orbit coupling. For the low-$J$ levels involved, OH is close to Hund's coupling case (a) and the ladders can be labelled $^2\Pi_{1/2}$ and $^2\Pi_{3/2}$, where the subscript is the total (orbital + spin) angular momentum quantum number of the electrons, $\Omega$. CH is very close to Hund's case (b) in which there is much stronger coupling of electron spin with rotational motion. In this case the electron spin along the internuclear axis is not well defined, and the ladders are labelled $F_1$ and $F_2$ for the upper and lower spin components with a given $J$-value respectively. The transitions for both species are described by the total angular momentum quantum number, $J$.

Absorption due to the envelope of Sgr~B2 occurs at a LSR velocity of $\sim$65~km~s$^{-1}$, although other features have been observed centred at velocities between 50~km~s$^{-1}$ and 70~km~s$^{-1}$ \citep[e.g.][]{martin-pintado_a}. Additional absorption occurs in the ground state lines (e.g. for OH, CH and CH$_2$) at velocities between $-110$~km~s$^{-1}$ and $+30$~km~s$^{-1}$. This is due to the galactic spiral arms that cross the line of sight \citep[e.g. the Galactic Bar, the expanding molecular ring at 3--4~kpc from the Galactic Centre and the local spiral arms; see][]{greaves94}. Eight individual components have been detected in \hi~absorption observations towards Sgr~B2 M \citep{garwood}, although at very high spectral resolution, each of these components can be separated into many narrow features with velocity widths $\sim$1~km~s$^{-1}$ \citep[e.g. in the CS absorption measurements of][]{greaves94}.

\begin{figure*}
\includegraphics[width=168mm]{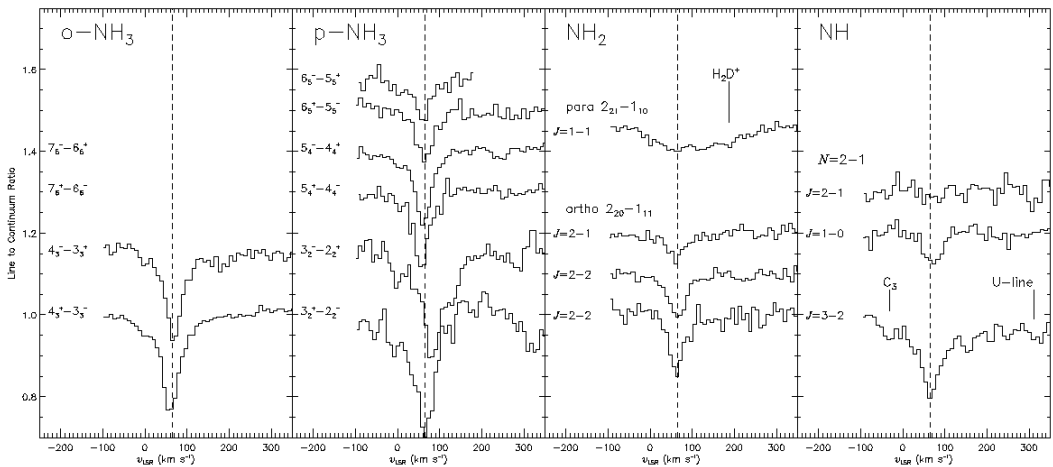}
\caption{Observed rotational transitions of NH$_3$, NH$_2$ and NH. Only the strongest NH$_3$ transitions to metastable levels are shown. Note that the central wavelength of the NH$_3$ 3$_2^-$--2$_2^+$ transition is shifted with respect to the other lines because it is a blend incorporating the 3$_1^-$--2$_1^+$ and 3$_0^-$--2$_0^+$ transitions. The dashed line shows the velocity of Sgr~B2 itself (65~km~s$^{-1}$).}
\label{nitrogen}
\end{figure*}

In order to determine the characteristics of each detected feature at the velocity of Sgr~B2, we performed a $\chi^2$ fit using the Lorentzian profile of the FP with three free parameters describing the line centre, resolving power and relative depth below the continuum. The results of these fits are shown in Table~\ref{iden_lines}. To improve the accuracy of the fit, fine adjustments were made to large scale baseline already achieved with the re-binning and spline technique described in Sect.~\ref{sect_contm}. This was carried out with low order polynomials. Note that no correction was made in Table~\ref{iden_lines} for systematic errors in wavelength calibration in the non-prime data below 70~$\mu$m. This can affect the quoted line centres by up to 25~km~s$^{-1}$.

This semi-automatic fitting technique was successful for the majority of lines. However, some weak features either required a more detailed analysis to fit \citep[e.g. HD;][]{polehampton_b}, or L04 data to increase the signal-to-noise \citep[e.g. OH $^2\Pi_{1/2}$ $J=5/2$--3/2 at 98.7~$\mu$m;][]{goicoechea_b}. These features are indicated in Table~\ref{iden_lines}. For lines in which line of sight absorption was observed, only an estimate for the peak absorption and where possible an estimate of the corresponding velocity, is reported in the table. More detailed analysis of the line of sight features is summarised in the following sections.

\begin{table*}
\caption{Identified Lines in the L03 prime data. See text for a description of the notation used to describe each transition. For transitions with line of sight absorption, only an approximate line to continuum value is given at the velocity of Sgr~B2 (in parenthesis). The line to continuum ratio, central velocity and line width were determined by fitting a Lorentzian profile to the L03 survey data (see text). No L04 data were used for these fits - features which were only detected in L04 data are noted. References in the literature are shown in the last column (both for {\it ISO} and KAO detections). References for the line wavelengths are given at the end of the table. Features not reported elsewhere are shown in bold font.}
\label{iden_lines}
\begin{tabular}{ccccccc}
\hline
Rest       & Species        & Transition                                & Line           & $v$           & $\Delta{v}$   & Reference \\
Wavelength &                &                                           & to             & (km~s$^{-1}$) & (km~s$^{-1}$) &           \\
($\mu$m)   &                &                                           & continuum      &               &               &           \\
\hline
51.8145    & [\oiii]         & $^{3}$P$_{2}$--$^{3}$P$_{1}$              & $1.28\pm0.02$  & $39\pm3^a$  & $82\pm7$  & [1]\\
53.2615    & OH             & $^2\Pi_{1/2}$--$^2\Pi_{3/2}~J=3/2^+$--$3/2^-$ & (0.69)     & $\sim47^a$  & broad     & [2]\\
53.3512    & OH             & $^2\Pi_{1/2}$--$^2\Pi_{3/2}~J=3/2^-$--$3/2^+$ & (0.69)     & $\sim47^a$  & broad     & [2]\\
56.3387    & p-NH$_{3}$     & $9_8^-$--$8_8^+$                          & $^b$           &           &           & [3]\\
{\bf{57.3295}} & \bf{[\niii]}& ${\bf{^2P_{3/2}-^2P_{1/2}}}$ &  ${\bf{1.12\pm0.01}}$  & ${\bf{42\pm3}}^a$  & ${\bf{50\pm6}}$  & \\
{\bf{57.6365}} & ${\bf{p-H_2O}}$ & ${\bf{4_{22}-3_{13}}}$               & $^c$           &           &   &\\
62.7274    & p-NH$_{3}$     & $8_7^+$--$7_7^-$                          & $0.960\pm0.006$& $49\pm4^a$  & $41\pm9$  & [3]\\
63.1837    & [\oi]           & $^3$P$_1$--$^3$P$_2$                      & (0.50)         &           & broad     & [4,5,6]\\
63.3765    & p-NH$_{3}$     & $8_7^-$--$7_7^+$                          & $0.976\pm0.003$& $39\pm5^a$  & $64\pm15$ & [3]\\
{\bf{65.1316}} & {\bf{OH}}    & ${\bf{^2\Pi_{3/2}~J=9/2^--7/2^+}}$      & ${\bf\sim0.98}$ & ${\bf{46\pm6}}^a$  & ${\bf \sim60}$       & \\
{\bf{65.2788}} & {\bf{OH}}    & ${\bf{^2\Pi_{3/2}~J=9/2^+-7/2^-}}$      & ${\bf{0.971\pm0.006}}$& ${\bf{46\pm6}}^a$  & ${\bf{60\pm20}}$ & \\
66.4377    & o-H$_2$O       & $3_{30}$--$2_{21}$                        & $0.866\pm0.01$ & $49\pm1^a$  & $45\pm3$  & [7]\\
67.0891    & p-H$_2$O       & $3_{31}$--$2_{20}$                        & $^c$   &             &           & [7]\\
{\bf{69.5377}}   & ${\bf{o-H_3O^+}}$   & ${\bf{4_3^+-3_3^-}}$           & ${\bf{0.950\pm0.03}}$ & ${\bf{51\pm14}}^a$ & ${\bf{36\pm28}}$ & \\
\hline
71.6084    & o-NH$_{3}$     & $7_6^+$--$6_6^-$                          & $0.865\pm0.02$ & $58\pm2$  & $27\pm5$  & [3]\\
72.4386    & o-NH$_{3}$     & $7_6^-$--$6_6^+$                          & $0.901\pm0.02$ & $58\pm2$  & $34\pm5$  & [3]\\
72.5238    & p-NH$_{3}$     & $7_5^-$--$6_5^+$                          & $^b$           &           &           & [3]\\
75.3807    & o-H$_{2}$O     & $3_{21}$--$2_{12}$                        & $0.695\pm0.009$& $59\pm1$  & $48\pm2$  & [7]\\
{\bf{78.7423}}    & ${\bf{o-H_2O}}$     & ${\bf{4_{23}-3_{21}}}$           & ${\bf{0.945\pm0.01}}$ & ${\bf{59\pm5}}$  & ${\bf{49\pm11}}$ & \\
79.1176    & OH             & $^2\Pi_{1/2}$--$^2\Pi_{3/2}~J=1/2^+$--$3/2^-$ & (0.6)      & $\sim55$  & broad     & [2]\\
79.1812    & OH             & $^2\Pi_{1/2}$--$^2\Pi_{3/2}~J=1/2^-$--$3/2^+$ & (0.6)      & $\sim55$  & broad     & [2]\\
{\bf{82.2742}}    & ${\bf{p-H_3O^+}}$   & ${\bf{3_2^+-2_2^-}}$             & {\bf{(0.875)}}   & ${\bf{\sim57}}$  & {\bf{broad}}     & \\
83.4320    & p-NH$_{3}$     & $6_5^+$--$5_5^-$                          & $0.860\pm0.01$ & $59\pm6$  & $42\pm9$  & [3]\\
83.5898    & p-NH$_{3}$     & $6_4^+$--$5_4^-$                          & $^b$           &           &           & [3]\\
84.4201    & OH             & $^2\Pi_{3/2}~J=7/2^+$--$5/2^-$                & $0.801\pm0.01$ & $58\pm1$  & $37\pm4$  & [2,8]\\
84.5441    & p-NH$_{3}$     & $6_5^-$--$5_5^+$                          & $0.885\pm0.04$ & $58\pm5$  & $26\pm8$  & [3]\\
84.5966    & OH             & $^2\Pi_{3/2}~J=7/2^-$--$5/2^+$                & $0.802\pm0.03$ & $53\pm4$  & $37\pm7$  & [2]\\
88.3564    & [\oiii]         & $^3$P$_1$--$^3$P$_0$                      & $1.193\pm0.01$ & $69\pm1$  & $52\pm2$  & [1]\\
89.9884    & p-H$_{2}$O     & 3$_{22}$--$2_{11}$                        & $0.897\pm0.03$ & $52\pm3$  & $36\pm5$  & [7]\\
98.7310    & OH             & $^2\Pi_{1/2}~J=5/2$--$3/2$                & $^d$           &           &           & [2]\\
{\bf{99.4931}} & ${\bf{o-H_2O}}$ & ${\bf{5_{05}-4_{14}}}$               & ${\bf{0.987\pm0.005}}$ & ${\bf{46\pm4}}$&${\bf{46\pm10}}$ &\\ 
99.9498    & p-NH$_{3}$     & $5_4^+$--$4_4^-$                          & $0.811\pm0.007$& $57\pm1$  & $37\pm2$  & [3]\\
100.1048   & o-NH$_{3}$     & $5_3^+$--$4_3^-$                          & $0.962\pm0.03$ & $54\pm2$  & $49\pm9$  & [3]\\
100.2128   & p-NH$_{3}$     & $5_2^+$--$4_2^-$                          & $^b$           &           &           & [3]\\
100.5767   & p-H$_3$O$^+$   & $2_1^+$--$1_1^-$                          & (0.905)        & $\sim60$  & broad     & [9]\\
100.8686   & o-H$_3$O$^+$   & $2_0^+$--$1_0^-$                          & (0.860)        & $\sim60$  & broad     & [9]\\
100.9828   & p-H$_{2}$O     & $2_{20}$--$1_{11}$                        & $0.695\pm0.006$& $64\pm1$  & $51\pm1$  & [7,9]\\
101.5337   & p-NH$_{3}$     & $5_4^-$--$4_4^+$                          & $0.819\pm0.005$& $60\pm1$  & $42\pm2$  & [3]\\
101.5965   & o-NH$_{3}$     & $5_3^-$--$4_3^+$                          & $0.976\pm0.004$& $57\pm3$  & $31\pm7$  & [3]\\
102.0050   & H$_2^{18}$O    & 2$_{20}$--1$_{11}$                        & $0.963\pm0.01$ & $49\pm3$  & $32\pm9$  & [7] \\
107.7203   & p-CH$_{2}$     & $2_{12}$--$1_{10}~J=3$--2                 & (0.97)         &           & broad     & [10]\\
108.0732   & o-H$_{2}$O     & $2_{21}$--$1_{10}$                        & $0.622\pm0.04$ & $63\pm1$  & $49\pm1$  & [7]\\
{\bf{109.3466}}   & ${\bf{o-H_2^{18}O}}$  & ${\bf{2_{21}-1_{10}}}$      & ${\bf 0.956\pm0.008}$& ${\bf 61\pm5}$  & ${\bf 40\pm12}$ & \\               
112.0725   & HD             & $J=1$--$0$                                & $^e$           &           &           & [11]\\
{\bf{113.5374}}   & ${\bf{o-H_2O}}$     & ${\bf{4_{14}-3_{03}}}$                        & ${\bf 0.944\pm0.007}$& ${\bf 58\pm2}$  & ${\bf 31\pm6}$  & \\
117.0648   & o-NH$_{2}$     & $2_{20}$--$1_{11}~J=2$--1                 & $0.936\pm0.006$& $63\pm2$  & $45\pm7$  & [1]\\
117.3831   & o-NH$_{2}$     & $2_{20}$--$1_{11}~J=2$--2                 & $0.889\pm0.006$& $63\pm1$  & $49\pm4$  & [1]\\
117.7918   & o-NH$_{2}$     & $2_{20}$--$1_{11}~J=1$--1                 & $0.864\pm0.01$ & $64\pm2$  & $38\pm4$  & [1]\\
119.2334   & OH             & $^2\Pi_{3/2}~J=5/2$--$3/2$                & (0.14)         &           & broad     & [2,8,12]\\
119.4409   & OH             & $^2\Pi_{3/2}~J=5/2$--$3/2$                & (0.14)         &           & broad     & [2,8,12]\\
119.621    & $^{17}$OH      & $^2\Pi_{3/2}~J=5/2$--$3/2$                & (0.97)         &           & broad     & [13]\\
119.828    & $^{17}$OH      & $^2\Pi_{3/2}~J=5/2$--$3/2$                & (0.97)         &           & broad     & [13]\\
\hline
\multicolumn{7}{l}{$^{a}$ note that no correction for systematic errors in wavelength calibration for non-prime data below 70~$\mu$m has been applied}\\
\multicolumn{7}{l}{$^{b}$ too weak to fit successfully, but observed by \citet{ceccarelli}; $^c$ fit unsuccessful}\\
\multicolumn{7}{l}{$^{d}$ too weak to fit successfully, but observed in L04 data by \citet{goicoechea_d}; $^{e}$ see \citet{polehampton_b}}\\
\end{tabular} 
\end{table*}

\begin{table*}
\caption{Identified Lines in the L03 prime data continued.}
\label{iden_lines2}
\begin{tabular}{ccccccc}
\hline
Rest       & Species        & Transition                      & Line            & $v$         & $\Delta{v}$   & Reference\\
Wavelength &                &                                 & to              &(km~s$^{-1}$)& (km~s$^{-1}$) &         \\
($\mu$m)   &                &                                 & continuum       &             &               &         \\
\hline
119.9664   & $^{18}$OH      & $^2\Pi_{3/2}~J=5/2$--$3/2$      & (0.9)           & $\sim65$   & broad     & [2,13,14]\\
120.1730   & $^{18}$OH      & $^2\Pi_{3/2}~J=5/2$--$3/2$      & (0.9)           & $\sim61$   & broad     & [2,13,14]\\
121.6973   & HF             & $J=2$--$1$                      & $0.971\pm0.006$ & $66\pm5$   & $28\pm8$  & [15]\\
{\bf{121.8976}}   & {\bf[\nii]}  & ${\bf{^3P_2-^3P_1}}$        &   $^f$         &            &           &  \\
124.6475   & o-NH$_{3}$     & $4_3^+$--$3_3^-$                & $0.755\pm0.004$ & $61\pm1$   & $47\pm1$  & [3]\\
124.7958   & p-NH$_{3}$     & $4_2^+$--$3_2^-$                & $0.973\pm0.004$ & $57\pm4$   & $33\pm8$  & [3]\\
124.8834   & p-NH$_{3}$     & $4_1^+$--$3_1^-$                & $0.951\pm0.01$  & $57\pm3$   & $21\pm11$ & [3]\\
126.8014   & p-NH$_{2}$     & $2_{21}$--$1_{10}~J=1$--1       & $0.965\pm0.01$  & $64\pm3$   & $68\pm19$ & [1]\\
126.8530   & o-H$_2$D$^+$   & $2_{12}$--$1_{11}$              & blend           &            & blend     & [19]\\
127.1081   & o-NH$_{3}$     & $4_3^-$--$3_3^+$                & $0.793\pm0.01$  & $68\pm1$   & $34\pm4$  & [3]\\
127.6461   & o-CH$_{2}$     & $1_{11}$--$0_{00}~J=2$--1       & (0.95)          &            & broad     & [10]\\
127.8582   & o-CH$_{2}$     & $1_{11}$--$0_{00}~J=1$--1       & (0.98)          &            & broad     & [10]\\        
{\bf{128.1949}}   & ${\bf{C_{3}}}$  & ${\bf{R(14)}}$          & ${\bf{0.967\pm0.01}}$  & ${\bf{64\pm6}}$   & ${\bf{42\pm14}}$ & \\
{\bf{131.9629}}   & ${\bf{C_{3}}}$  & ${\bf{R(12)}}$          & ${\bf{0.961\pm0.008}}$ & ${\bf{65\pm4}}$   & ${\bf{56\pm14}}$ & \\
{\bf{135.8356}}   & ${\bf{C_{3}}}$         & ${\bf{R(10)}}$   & ${\bf{0.955\pm0.01}}$  & ${\bf{67\pm4}}$   & ${\bf{31\pm12}}$ & \\
138.5278   & p-H$_{2}$O     & $3_{13}$--$2_{02}$              & $0.931\pm0.008$ & $64\pm2$   & $25\pm5$  & [7]\\
{\bf{139.8095}}   & ${\bf{C_{3}}}$         & ${\bf{R(8)}}$    & $^g$            &            &           & \\
143.8801   & C$_{3}$        & $R(6)$                          & $0.941\pm0.003$ & $69\pm1$   & $43\pm4$  & [16]\\
145.5254   & [\oi]           & $^3$P$_0$--$^3$P$_1$            & $1.174\pm0.006$ & $75\pm1$   & $54\pm3$  & [1,6]\\
148.0419   & C$_{3}$        & $R(4)$                          & $0.916\pm0.007$ & $70\pm2$   & $39\pm5$  & [16]\\
149.0912   & CH             & $F_2~3/2$--$1/2$        & (0.64)          & $\sim65$   & broad     & [1,10,17]\\
149.3899   & CH             & $F_2~3/2$--$1/2$        & (0.64)          & $\sim65$   & broad     & [1,10,17]\\
{\bf{151.5279}}   & {\bf{NH}} & ${\bf{2_1-1_0}}$                 & ${\bf{0.962\pm0.01}}$  & ${\bf{58\pm10}}$  & ${\bf{43\pm24}}$ & \\
152.2875   & C$_{3}$        & $R(2)$                            & $0.943\pm0.01$  & $84\pm3$   & $25\pm7$  & [16,18]\\
153.0961   & NH             & $2_2$--$1_1$                    & $0.925\pm0.02$  & $69\pm4$   & $31\pm11$ & [1]\\
153.2956   & C$_{3}$        & $Q(12)$                           & $^g$            &            &           & [16]\\
153.3444   & NH             & $2_{3}$--$1_{2}$                & $^g$            &            &           & [1,16]\\
154.8619   & C$_3$          & $Q(10)$                           & $0.973\pm0.01$  & $55\pm7$   & $45\pm15$ & [16]\\
156.1898   & C$_3$~-~blend  & $Q(8)$                            & $0.939\pm0.01$  & $69\pm3$   & $28\pm7$  & [16]\\
156.1940   & p-H$_2$O~-~blend& $3_{22}$--$3_{13}$             & blend           & ($61\pm3$) & blend     &  [16]\\
157.2609   & C$_{3}$        & $Q(6)$                            & $0.878\pm0.01$  & $67\pm3$   & $52\pm67$ & [16]\\
157.7409   & [\cii]          & $^2$P$_{3/2}$--$^2$P$_{1/2}$    & $1.655\pm0.01$  & $74\pm1$   & $49\pm1$  & [1,6]\\
158.0595   & C$_{3}$        & $Q(4)$                            & $0.894\pm0.01$  & $72\pm3$   & $33\pm6$  & [16]\\
{\bf{158.5735}} & ${\bf{C_{3}}}$   & ${\bf{Q(2)}}$             & ${\bf{0.937\pm0.02}}$  & ${\bf{68\pm3}}$   & ${\bf{\sim{18}}}$ & \\
163.1247   & OH             & $^{2}\Pi_{1/2}~3/2$--$1/2$      & $1.065\pm0.01$  & $65\pm4$   & $41\pm11$ & [2]\\
163.3973   & OH             & $^{2}\Pi_{1/2}~3/2$--$1/2$      & $1.060\pm0.01$  & $75\pm4$   & $35\pm10$ & [2]\\
165.5966   & p-NH$_{3}$     & $3_2^+$--$2_2^-$                & $0.711\pm0.01$  & $64\pm1$   & $54\pm4$  & [3]\\
165.7287   & p-NH$_{3}$     & $3_1^+$--$2_1^-$                & $0.911\pm0.03$  & $82\pm4$   & $44\pm13$ & [3]\\
167.6794   & C$_{3}$        & $P(4)$                            & $0.948\pm0.01$  & $70\pm7$   & $54\pm15$ & [16]\\
169.9676   & p-NH$_3$~-~blend & $3_2^-$--$2_2^+$              & blend           & $85\pm1$   & blend     & [3]\\
169.9887   & p-NH$_3$~-~blend & $3_1^-$--$2_1^+$              & $0.755\pm0.01$  & $48\pm1$   & $62\pm4$  & [3]\\
169.9961   & o-NH$_3$~-~blend & $3_0^-$--$2_0^+$              & blend           & $35\pm1$   & blend     & [3]\\
174.6259   & o-H$_{2}$O     & $3_{03}$--$2_{12}$              & $0.770\pm0.01$  & $65\pm1$   & $32\pm2$  & [7,8]\\
179.5267   & o-H$_{2}$O     & $2_{12}$--$1_{01}$              & (0.08)          &            & broad     & [7,8]\\
{\bf{180.2088}}   & {\bf{o-H}}${\bf{_3O^+}}$   & ${\bf{3_3^+-3_3^-}}$   & {\bf{(0.84)}}     & ${\bf{\sim60}}$   & {\bf{broad}}     & \\
180.4883   & o-H$_2$O       & 2$_{21}$--$2_{12}$              & $0.715\pm0.03$  & $63\pm3$   & $48\pm9$  & [7]\\
181.0487   & o-H$_2^{18}$O~-~blend & 2$_{12}$--$1_{01}$       & $0.714\pm0.03$  & $64\pm2$   & $36\pm7$  & [7,8,9]\\
181.0545   & p-H$_3$O$^+$~-~blend  & $1_1^+$--$1_1^-$         & blend           & ($55\pm2$) & blend     & [7,8,9]\\
\hline
\multicolumn{7}{l}{$^f$ broadened due to hyperfine structure; $^g$ contained within broader absorption profile}\\
\multicolumn{7}{l}{{\bf Line wavelengths:} 
[\oiii]: \citet[][]{pettersson}; 
OH: \citet[][]{brown,varberg}; NH$_{3}$: JPL catalogue;}\\
\multicolumn{7}{l}{[\niii]: \citet[][]{gry}; [\oi]: \citet[][]{zink}; H$_{2}$O: \citet[][]{matsushima,johns}; H$_{2}^{18}$O: \citet[][]{matsushima_b}; } \\
\multicolumn{7}{l}{H$_{3}$O$^{+}$: JPL catalogue; CH$_{2}$: \citet[][]{polehampton_ch}; HD: \citet[][]{evenson}; NH$_{2}$: \citet[][]{morino_b};}\\
\multicolumn{7}{l}{$^{17}$OH: \citet[][]{polehampton_c}; $^{18}$OH: \citet[][]{morino}; HF: \citet[][]{nolt}; [\nii]: \citet[][]{brown_f}; C$_{3}$: \citet[][]{giesen}; }\\
\multicolumn{7}{l}{CH: \citet[][]{davidson}; NH: JPL catalogue; [\cii]: \citet[][]{cooksy}; H$_2$D$^+$: \citet{cernicharo_f}.}\\ 
\multicolumn{7}{l}{{\bf References:} [1] \citet{goicoechea_d}; [2] \citet{goicoechea_b}; [3]
\citet{ceccarelli}; [4] \citet{baluteau_a}; }\\
\multicolumn{7}{l}{[5] \citet{lisc}; [6] \citet{vastel_b}; [7] \citet{cernicharo_e}; [8] \citet{cernicharo_a}; [9]
\citet{goicoechea_c};}\\
 \multicolumn{7}{l}{[10] \citet{polehampton_ch}; [11] \citet{polehampton_b};
[12] \citet{storey}; [13] \citet{polehampton_c}; }\\
\multicolumn{7}{l}{[14] \citet{lugten}; [15] \citet{neufeld_a}; [16] \citet{cernicharo_d}; [17] \citet{stacey}; [18] \citet{giesen};}\\
\multicolumn{7}{l}{[19] \citet{cernicharo_f}.}\\
\end{tabular} 
\end{table*}

Only a few of the lines have been previously observed towards Sgr~B2 with lower spectral resolution using the KAO \citep[OH at 119~$\mu$m; CH at 149~$\mu$m][]{storey,stacey} and so most were observed for the first time with \iso. The \iso~spectra of the main species have already been investigated in detail in previous papers (see Tables~\ref{linsum}~and~\ref{iden_lines} for references), however, there are several new features whose detection is only reported here. These are indicated in bold face in Table~\ref{iden_lines}. The following sections give an overview of the conclusions for each group of species.

\subsection{Molecular species}
\subsubsection{Nitrogen bearing molecules: NH$_3$, NH$_2$ and NH}\label{sect_nitrogen}

The greatest number of lines from a single molecule in the survey are due to rotation-inversion transitions of NH$_3$. In total there are 21 detected absorption transitions with upper energy levels in the range 65--688~K, including both metastable (see Fig.~\ref{nitrogen}) and non-metastable lower states \citep[see detailed results presented by][]{ceccarelli}. Pure inversion transitions have been detected in the radio region from many of these levels \citep[e.g.][]{huttemeister}, recently extending much higher in energy to $J_K$=18$_{18}$ at $\sim$3130~K above ground \citep{wilsonc}. A simplified model of the FIR lines, based on these radio lines showed that the higher energy transitions are optically thin but the lower energy lines have optical depths $\sim$1.6 \citep{ceccarelli}. They calculated rotational temperatures of 130$\pm$10~K for the metastable levels and 310$\pm$100~K for the non-metastable levels using a standard rotational diagram approach. 

Detailed large velocity gradient (LVG) modelling showed that the NH$_3$ absorption must originate in a high temperature layer in front of the FIR continuum emitted by Sgr~B2. The best fitting model parameters gave a temperature of 700$\pm$100~K and density $\le10^5$~cm$^{-3}$ for this layer. Further constraints on the density can be derived from the lack of CO rotational emission in the survey. \citet{ceccarelli} derive an upper limit on the density of the hot region of 10$^{4}$~cm$^{-3}$ based on the lack of FIR CO emission, in good agreement with that calculated by \citet{cernicharo_e}. The results of the modelling gave a total column density of NH$_3$ $\sim3\times10^{16}$~cm$^{-2}$, equally shared between the ortho and para forms (i.e. a non-LTE ortho/para ratio).

\begin{figure*}
\includegraphics[width=168mm]{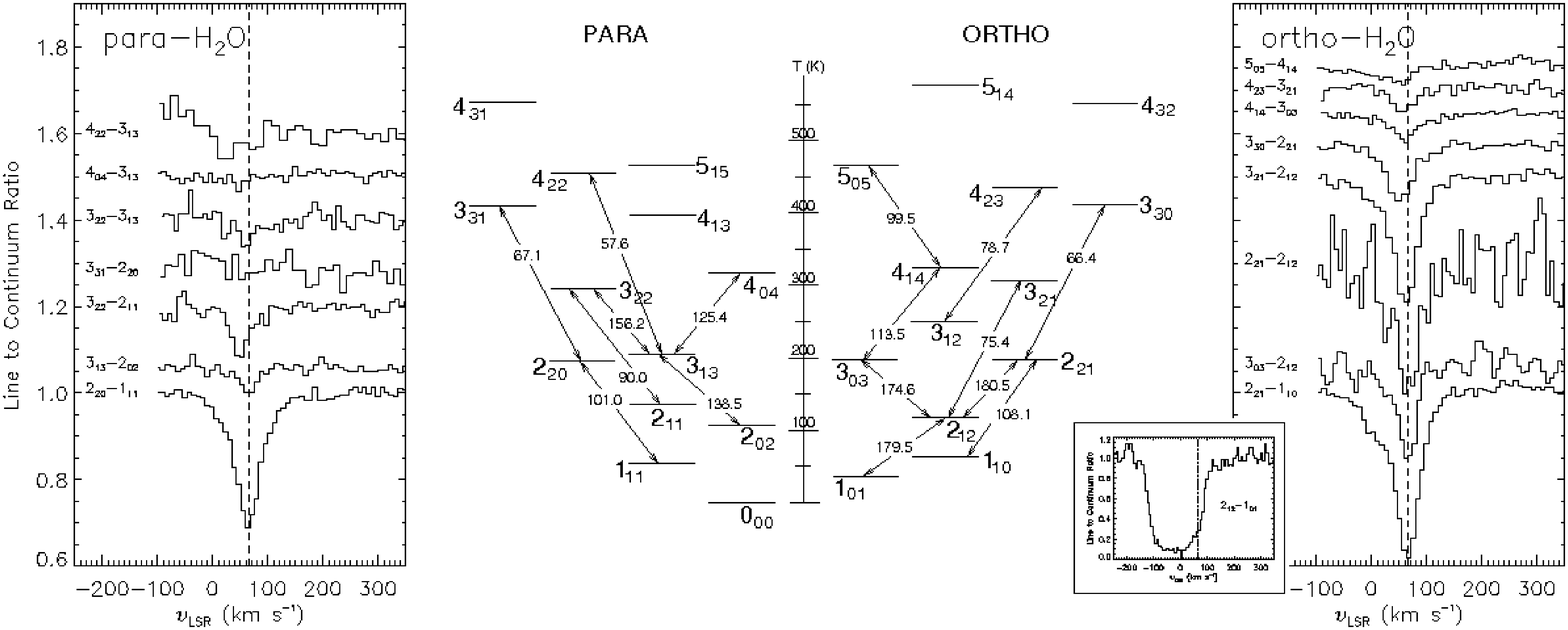}
\caption{Observed rotational transitions of para and ortho water. The energy level diagram for water is shown in the centre, with transition wavelengths given in microns. Not all of the para lines shown were detected. The dashed line shows the velocity of Sgr~B2 itself (65~km~s$^{-1}$). The insert on the right is the ground state ortho-H$_2$O line, which shows complete absorption at all line of sight velocities.}
\label{water}
\end{figure*}

There is also a clear detection of the closely related nitrogen hydrides NH$_{2}$ and NH in the survey (see Fig.~\ref{nitrogen}). The results for these lines have been presented by \citet{cernicharo_d} and \citet{goicoechea_d}. The survey shows a detection of both ortho- and para-NH$_2$, the first such observation of the ortho form. One para-NH$_2$ line ($2_{21}$--$1_{10}$ $J=1$--1) is detected at 126.8~$\mu$m. The profile of this line is shown in Fig.~\ref{nitrogen}, and is much broader than the ortho lines. This is due to a blend with the ground state line of ortho-H$_2$D$^+$ (see Sect.~\ref{sect_deuterium}).

Para-NH$_2$ has been previously observed in Sgr~B2 from the ground via its mm wave transitions \citep{vandishoeck_a}. These mm observations indicated a total column density of ortho+para NH$_2$ of ($5.2\pm1.2$)$\times$10$^{15}$~cm$^{-2}$, assuming an ortho/para ratio of 3. It is difficult to use the \iso~observations to provide measured confirmation of this ortho/para ratio due to the uncertainties in the one detected para line. However, the fact that the other components of the NH$_2$ $2_{21}$--$1_{10}$ transition were not detected indicates that a ratio of 3 is a reasonable estimate. \citet{goicoechea_d} estimated the total column density of NH$_2$, taking account of both the \iso~and mm data, to be $N({\rm NH_2})=(1.5$--$3)\times10^{15}$~cm$^{-2}$. The column density they estimated for NH was $\sim4\times10^{14}$~cm$^{-2}$, giving a final ratio NH$_3$/NH$_2$/NH approximately equal to 100/10/1. Goicoechea et al. show that these ratios cannot be explained by either dark cloud models \citep[e.g.][]{millar}, or by UV illuminated PDRs \citep[e.g.][]{sternberg}. However, low-velocity shocks can heat the gas up to 500~K (consistent with the temperatures derived from \iso~NH$_3$ observations) as well as satisfactorily reproducing the observed NH$_3$ and NH$_2$ abundances, providing that ammonia is efficiently formed on grain surfaces and sputtered back to the gas phase by the shock passage \citep[see detailed models of][]{flower}. This implies that shocks contribute to the energy balance of the Sgr~B2 envelope as well as other mechanisms such as those associated with the presence of far-UV radiation fields.

All three nitrogen species only show absorption at the velocity of Sgr~B2, and no line of sight absorption is observed. However, the ground state ammonia line has been observed by the ODIN satellite towards Sgr~B2 \citep{hjalmarson}. This does show absorption along the whole line of sight.

\subsubsection{Oxygen bearing molecules: H$_2$O, OH and H$_3$O$^+$}\label{sect_oxygen}

Water is an extremely important species in the survey as it has the largest associated column density of all the detected molecules. It is also an important research field in its own right, particularly in the warm neutral gas driven by star formation, where it plays a dominant role in the thermal balance. As water is an abundant species in the Earth's atmosphere, only space telescopes have access to its thermal rotational lines and only very limited studies were possible before \iso.

Many rotational lines of water have been observed in the survey. These are mostly due to Sgr~B2 itself, although in the lowest energy lines there is some absorption at the velocities of line of sight clouds. The water transitions detected in the survey are shown in Fig.~\ref{water}. These lines have been analysed in detail by \citet{cernicharo_a,cernicharo_e} using data from the LWS L04 mode. They presented data for 14 water lines and 2 lines of H$_2^{18}$O observed with \iso, and combined these with a map of emission by the $3_{13}$--$2_{20}$ 183.31~GHz maser line observed with the IRAM 30-m telescope. The FIR lines are all in absorption and optically thick (particularly the $2_{12}$--$1_{01}$ at 179.5~$\mu$m which has an optical depth of $\sim10^3$--$10^4$). The H$_2$O absorption traces the outer surface layers of the Sgr~B2 envelope (i.e. the same hot, low density layer observed in NH$_3$), whereas the 183~GHz line traces denser gas closer to the hot cores in the inside of the cloud. 

\citet{cernicharo_e} carried out LVG and non-local radiative transfer modelling and found that IR photons from the dust are the dominant source of excitation of the water rotational levels. This showed that the observed lines were not very sensitive to the gas temperature, and detection of weaker lines that would be more sensitive are needed to better constrain the physical conditions of the water absorbing layers. However, assuming that the FIR water lines arise in the same warm gas traced by OH \citep{goicoechea_b}, the best fitting column density for the \iso~observations was $(9\pm3)\times10^{16}$~cm$^{-2}$, with an abundance of (1--2)$\times10^{-5}$. 

Previous observations of the ground state water line with {\it SWAS}, and HDO lines with ground based telescopes \citep{neufeld_b, neufeld_f, comito} indicate that there is also a component of the absorption that is due to the warm envelope surrounding the Sgr~B2 cores as well as the hot layer observed in NH$_{3}$ lines. If the majority of the water absorption is located in the hot foreground layer, the estimated column density of H$_2$O associated with Sgr~B2 calculated from the {\it SWAS} observations is (2.5--4)$\times$10$^{16}$~cm$^{-3}$ \citep{neufeld_f}, of the same order of magnitude as that calculated from \iso~by \citet{cernicharo_e}. 
 
\citet{cernicharo_a}, \citet{goicoechea_b} and \citet{goicoechea_c} calculate an abundance for H$_2$O of a few 10$^{-5}$. This abundance gives a ratio with OH of H$_2$O/OH=2--4. This relatively low ratio can be reproduced if the Sgr~B2 envelope is illuminated by a strong far-UV radiation field, and the gas is warm enough to activate neutral-neutral reactions involving H$_2$O and OH. The presence of such a radiation field is inferred from the extended atomic and ionic fine structure line emission (see Sect.~\ref{sect_atom}). Detailed photochemical models adapted to the conditions of the Sgr~B2 envelope reproduce the above ratios if most of the observed FIR H$_2$O and OH absorption arises from the photoactive surface of the cloud \citep{cernicharo_e}. Note that the same shock models that apparently reproduce the observed NH$_3$ and NH$_2$ abundance ratios (Sect.~\ref{sect_nitrogen}) fail to reproduce the results for water and related species (at least in the averaged picture contained within the large \iso~beam). In particular, the predicted H$_2$O column density is almost 2 orders of magnitude larger than observed by \iso, while the predicted OH column density is an order of magnitude lower than observed. In the line of sight clouds, the ratio of H$_2$O to OH is even lower, 0.6--1.2 \citep{polehampton_oh}, almost as low as the ratio calculated for line of sight features towards W51 and W49 of 0.3--0.4 \citep{neufeld_c, plume}.

\begin{figure}
\includegraphics[width=84mm]{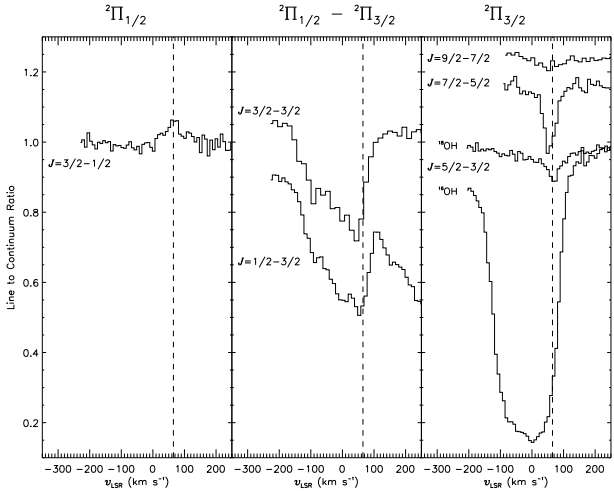}
\caption{Observed rotational transitions of OH within and between the $^{2}\Pi_{1/2}$ and $^{2}\Pi_{3/2}$ ladders. The ground state line in the $^{2}\Pi_{3/2}$ ladder is shown for both $^{16}$OH and $^{18}$OH. The dashed line shows the velocity of Sgr~B2 itself (65~km~s$^{-1}$).}
\label{oh}
\end{figure}

The survey also contains lines due to the pure rotational transitions of the OH molecule up to 400~K above ground (see Fig.~\ref{oh}). Only the strongest ground state transition had previously been observed towards Sgr~B2, using the KAO \citep{storey}. The ground state lines show absorption due to the entire line of sight, but at higher energies only Sgr~B2 itself is detected. The detected OH transitions have 2 components due to the $\Lambda$-doublet splitting of each rotational level, and in most cases these were resolved by the LWS FP. No asymmetry in the absorption was observed between any of these doublets.

\citet{goicoechea_b} have carried out a detailed radiative transfer analysis of these lines, and conclude that they also arise from the low density ($<$10$^4$~cm$^{-3}$) envelope of Sgr~B2 where the temperature should range from T$_k$=40~K in the innermost regions to T$_k$=600~K in the cloud edge. The high abundance of OH, $\sim$(2--5)$\times$10$^{-6}$, and large OH/H$_2$O abundance ratio is attributed to the presence of clumpy PDRs on the edge of the Sgr~B2 envelope.

Here, we present a weak detection of an additional absorption doublet of OH at 65~$\mu$m ($^{2}\Pi_{3/2}$ $J$=9/2--7/2). According to the models of \citet{goicoechea_b} this line is only predicted to be in absorption if it arises from the colder regions of the envelope, or if the FIR continuum level at $\sim$65~$\mu$m was underestimated for the warm gas. The wide range of conditions traced by the observed OH lines mean that the additional 65~$\mu$m detection does not provide significant new constraints on the previous models.

Overall, the OH excitation is dominated by FIR pumping, with transitions in the $^2\Pi_{1/2}$ ladder seen in emission. This is due to the upper levels being populated by the cross ladder transitions at 53~$\mu$m ($^{2}\Pi_{1/2}$--$^{2}\Pi_{3/2}$ $J$=3/2--3/2) and 79~$\mu$m ($^{2}\Pi_{1/2}$--$^{2}\Pi_{3/2}$ $J$=1/2--3/2) - see Fig.~\ref{oh}. 

\begin{figure*}
\includegraphics[width=168mm]{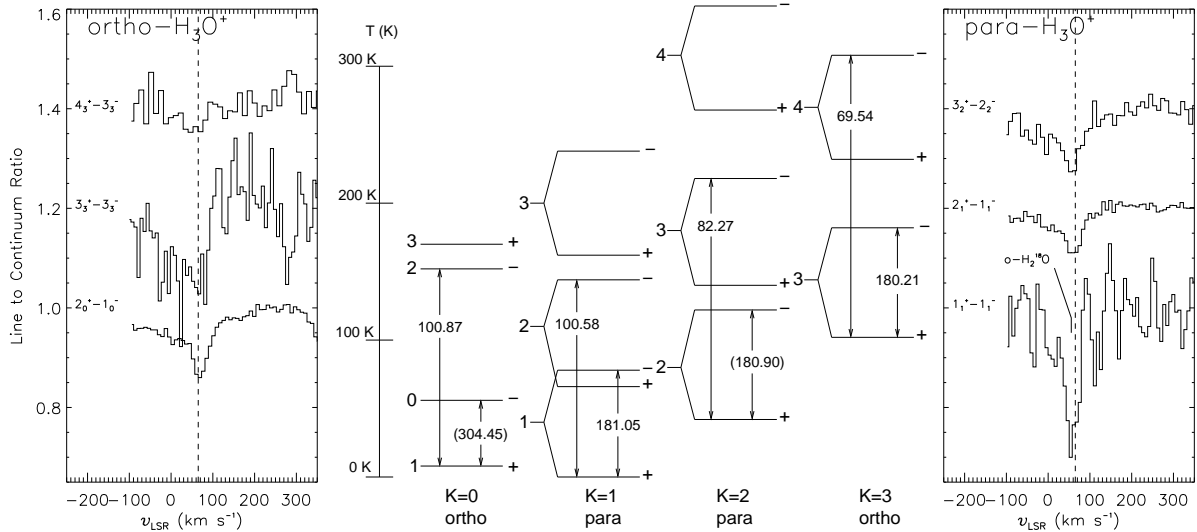}
\caption{Observed rotation-inversion transitions of H$_3$O$^+$. The energy level diagram in the centre shows transition wavelengths in microns. Note that the 1$^+_1$--1$^-_1$ para line is blended with H$_2^{18}$O. The dashed line shows the velocity of Sgr~B2 itself (65~km~s$^{-1}$).}
\label{h3o+}
\end{figure*}

In addition to absorption from Sgr~B2, the transitions from the ground rotational state show absorption at velocities associated with clouds along the line of sight. This includes the two cross ladder transitions at 53~$\mu$m and 79~$\mu$m and the fundamental transition in the $^{2}\Pi_{3/2}$ ladder at 119~$\mu$m, which is highly optically thick at all line of sight velocities. The higher energy transitions are not seen in these clouds, indicating that they are colder with lower excitation. The line of sight components were fitted in the 53~$\mu$m and 79~$\mu$m lines by constructing a high resolution model of 10 velocity components, based on the line widths and velocities from the \hi~measurements of \citet{garwood}. This model was then convolved to the resolution of the LWS FP \citep[see][]{polehampton_oh}. As only the ground state lines were observed in the line of sight clouds, the total column density of OH can be estimated simply from the best fitting high resolution model. The optical depth for each component, $\tau$, was adjusted to find the best fit with the line to continuum ratio set by,
\begin{equation}
I = I_{\rm{c}}\exp{(-\tau)}
\end{equation}
where $I_{\rm{c}}$ is the intensity of the continuum. Ground state column densities for each component (in cm$^{-2}$) were calculated assuming a Doppler line profile with Maxwellian velocity distribution \citep{spitzer},
\begin{equation}
N_j = \frac{8\pi\sqrt{\pi}}{2\sqrt{\ln{2}}}10^{17}\frac{\tau\Delta{v}}{A_{ij}\lambda_{ij}g_i/g_j}
\end{equation}
where $\Delta{v}$ is the line width in km~s$^{-1}$, $A_{ij}$ is the Einstein coefficient for spontaneous emission in s$^{-1}$, $\lambda_{ij}$ is the wavelength in $\mu$m and $g_i$ is the statistical weight of state $i$. The fitted components were all optically thin, except for the one at the velocity of Sgr~B2. The column density for each line of sight feature are shown in Table~\ref{lineofsight}. The OH abundance in the line of sight clouds was estimated using molecular hydrogen column densities from \citet{greaves96} to be in the range 10$^{-7}$--10$^{-6}$, whereas at the velocity of Sgr~B2, it is (2--5)$\times10^{-6}$ \citep{goicoechea_b}. OH absorption also occurs across the extended region surrounding Sgr~B2 as seen by the LWS grating spectrometer \citep{goicoechea_d}. The final column density determined by \citet{polehampton_oh} at the velocity of Sgr~B2 was 3.2$^{+0.6}_{-0.4}\times10^{16}$~cm$^{-2}$, in reasonable agreement with \citet{goicoechea_b} who derived a value of (1.5--2.5)$\times10^{16}$~cm$^{-2}$. 

Both the isotopic species of OH, $^{17}$OH and $^{18}$OH, were also detected in the survey and are described in Sect.~\ref{isotopes}.

Another species closely linked to water and hydroxyl is H$_3$O$^{+}$. This molecule is similar in structure to ammonia but has a very large inversion barrier, meaning that it shows pure inversion transitions in the FIR. The results for several of the lower energy lines have already been presented by \citet{goicoechea_c}, and show absorption due to the entire line of sight. Here, we present several additional detections of H$_3$O$^+$ lines at shorter wavelengths and higher energy levels. The detection of higher excitation H$_3$O$^+$ lines in absorption indicates that H$_3$O$^+$ is also likely to arise in the warm gas traced by H$_2$O and OH in the Sgr~B2 envelope. Further detailed analysis and modelling is required.

Both water and OH are a product of H$_3$O$^+$ dissociative recombination. However, as noted by \citet{cernicharo_e}, the large H$_2$O and OH column densities found towards Sgr~B2 can only be reproduced if a significant fraction of the gas is warm enough ($T_k\sim$300--500~K) to activate additional neutral-neutral reactions that can efficiently form larger amounts of water and OH than H$_3$O$^+$ dissociative recombination alone. Detailed photochemical models assuming that the Sgr~B2 envelope is illuminated by 10$^3$--10$^4$ times the mean interstellar radiation field \citep{goicoechea_d,cernicharo_e} satisfactorily reproduce the H$_2$O/OH abundance ratios and absolute column densities, as well as the \oi~line intensities (see Sect.~\ref{sect_atom}). Therefore, the oxygen chemistry seems to be dominated by the presence of warm PDRs in the external layers of Sgr~B2 (at least in the averaged picture provided by the large beam of \iso~observations).

The region around Sgr~B2 has recently been mapped in the $3_2^+$--$2_2^-$ (364~GHz) and $1_1^-$--$2_1^+$ (307~GHz) lines of H$_3$O$^+$ using the APEX telescope \citep{vandertak}. However, these lines appear in emission and trace denser gas than the absorption lines presented here in the core of Sgr~B2. The line ratio indicates a high excitation temperature, and at the conditions of the core of Sgr~B2 M give a total H$_3$O$^+$ column density of $1.4\times10^{16}$~cm$^{-2}$, much higher than the value estimated for the envelope of Sgr~B2 from the \iso~lines: 1.6$\times10^{14}$~cm$^{-2}$ \citep{goicoechea_c}.

\begin{figure*}
\includegraphics[width=168mm]{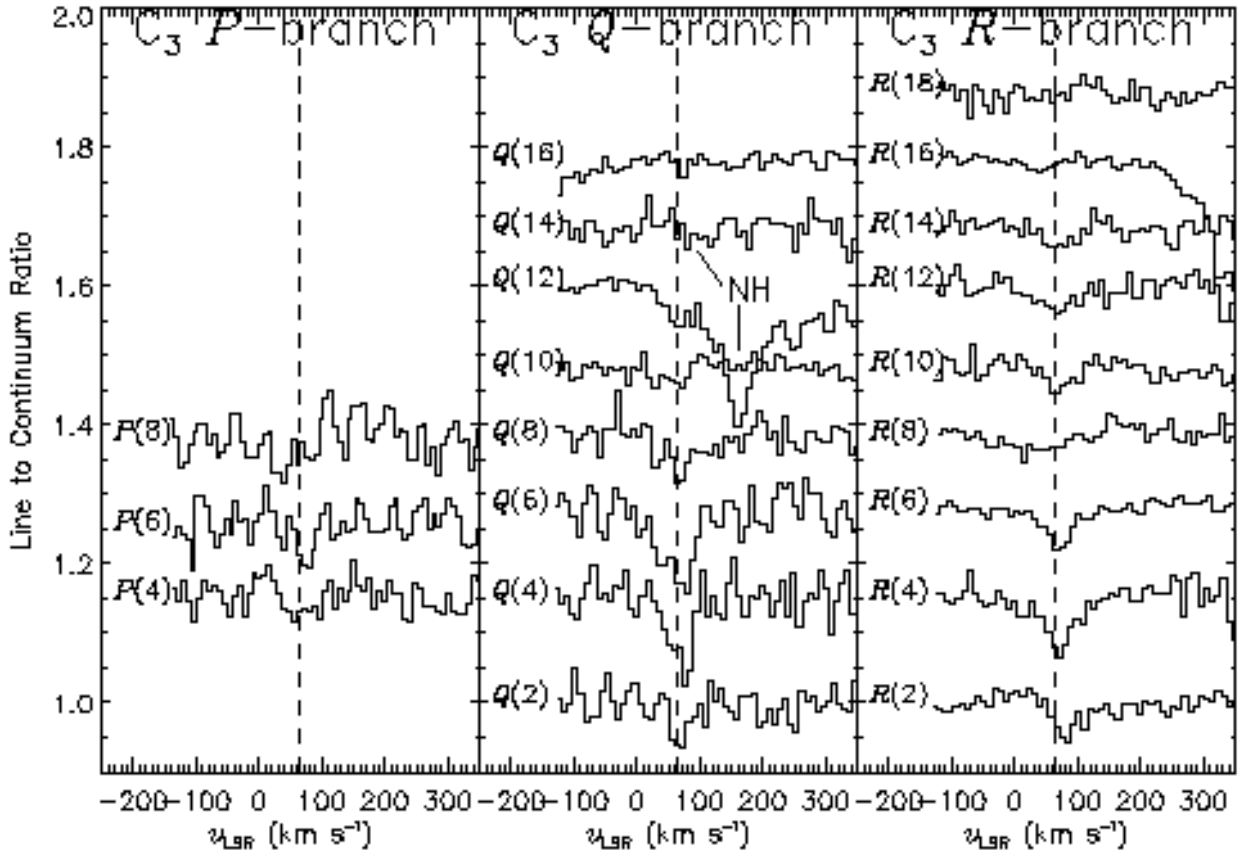}
\caption{The observed $P$, $Q$ and $R$ branch lines of C$_3$ in the survey. Data are shown around a wide range of lines, not all of which are claimed as detections. The nominal velocity of Sgr~B2 (65~km~s$^{-1}$) is shown by the dashed line.}
\label{c3}
\end{figure*}

\subsubsection{Carbon bearing molecules: CH, CH$_2$ and C$_3$}\label{carbon}

There is a clear detection in the survey of the two $\Lambda$-doublet components of the $J=3/2$--1/2 transition of CH from the ground to first rotational state of the $F_2$ ladder, at 149~$\mu$m. These lines were first observed using the KAO by \citet{stacey}. The results from the \iso~survey have been presented by \citet{polehampton_ch} and in L04 observations by \citet{goicoechea_d}. Goicoechea et al. also observed CH in the extended Sgr~B2 region using the LWS grating mode. 

In contrast to OH, no higher energy rotational transitions of CH have been detected. The next highest energy transition in the $F_2$ ladder, $\sim$96~K above ground, is $J=5/2$--3/2 at 115~$\mu$m. Taking an upper limit for the absorption in this line shows that $T_{\rm{rot}}<20$~K \citep{polehampton_ch}. The lowest energy line in the $F_1$ ladder of CH at 180~$\mu$m ($J$=5/2--3/2) is also not observed. The lower energy level of this transition would be populated by absorption in the cross ladder transition at 560~$\mu$m ($F_1$--$F_2$ $J$=3/2--1/2). However, as noted by \citet{gonzalezalfonso}, the continuum in Sgr~B2 at 560~$\mu$m is not strong enough to produce a significant population in the $F_1$ ladder (in contrast to the ultraluminous galaxy Arp220).

The lack of absorption from higher energy levels indicates that the ground state transition should give a good measure of the total column density. A similar method to that applied to OH was used by \citet{polehampton_ch} to determine the column density of CH in the line of sight features in the range (0.3--3.1)$\times10^{14}$~cm$^{-2}$, and (9.3$\pm$0.9)$\times10^{14}$~cm$^{-2}$ in the Sgr~B2 component. The results are shown in Table~\ref{lineofsight}. \citet{goicoechea_d} found integrated column densities of (0.8--1.8)$\times10^{15}$~cm$^{-2}$ across the extended region, peaking at the central Sgr~B2 M and Sgr~B2 N positions.

The survey also incudes the lowest energy rotational transitions of CH$_2$, the detection of which was described by \citet{polehampton_ch}. This represents the first detection of these low energy transitions and is only the second definitive detection of CH$_2$ in space: \citet{hollis} detected emission from CH$_2$ towards Orion KL and W51M via its 4$_{04}$--3$_{13}$ transition at 68--71~GHz. Other than this, only a tentative assignment of CH$_2$ absorption bands in the UV spectrum towards HD154368 and $\zeta$~Oph has been reported \citep{lyu}. Both CH and CH$_2$ show broad absorption due to the entire line of sight towards Sgr~B2. They are chemically related as both are formed by the dissociative recombination of CH$_{3}^{+}$ and destroyed by reaction with atomic oxygen. The shape of the CH$_2$ line could be well fitted by assuming a constant CH/CH$_2$ ratio along the whole line of sight and the best fit gave CH/CH$_2$ equal to 2.7$\pm$0.5, contrary to the expectation of the CH$_{3}^{+}$ branching ratios measured in the lab \citep[CH: 30\%, CH$_2$: 40\%, C: 30\%;][]{vejby}. In order to explain the low abundance of CH$_2$, other formation/destruction routes are required. One solution is to include a high UV radiation field (possibly applicable in Sgr~B2 itself), or to include the formation of hydrocarbons on dust grain surfaces \citep[e.g. the models of][give CH/CH$_2$ ratios of 7.7--13.6 for $A_{\rm{v}}$=4]{viti}.

Finally, 16 lines of triatomic carbon, C$_3$, have been detected in the survey. C$_3$ has its lowest energy bending mode ($\nu_2$=1--0) centred well within the \iso~survey range at 157.69~$\mu$m \citep[63.42~cm$^{-1}$;][]{schmuttenmaer} and this is the only molecule in the survey detected through vibrational transitions.

C$_3$ has previously been observed in the C-rich evolved star IRC+10216 through its high energy stretching mode at 4.9~$\mu$m \citep{hinkle}, as well as via emission in some of the same FIR transitions as seen in Sgr~B2 \citep{cernicharo_d}. C$_3$ has also been observed in optical absorption spectra of diffuse and translucent clouds in the sight lines of stars \citep[][and references therein]{adamkovics}. However, in molecular clouds where the expected mid-IR flux is too low to allow systematic searches for the 4.9~$\mu$m stretching mode, and the optical extinction is very high, the FIR bending mode is the best way to detect C$_3$ and other such nonpolar species (e.g. the C$_n$ carbon chains). These species do not have a permanent dipole moment and so no pure rotational lines. As noted by \citet{cernicharo_d}, the low-lying vibrational bending modes of polyatomic molecules could dominate the spectra to be observed by future FIR space missions, which will have higher spectral and spatial resolution than \iso.

C$_3$ has been analysed in detail by \citet{cernicharo_d}, who observed 9 of the FIR lines using the LWS L04 mode. They used a non-local radiative transfer model to calculate a fractional abundance of $X({\rm{C_3}})=3\times10^{-8}$ (or even higher if C$_3$ only arises from the warm surface of the cloud). In the L03 survey, we detect several more higher energy transitions of C$_3$, bringing the total number of lines to 16. Figure~\ref{c3} shows the survey data centred on the C$_3$ wavelengths from the Cologne Database for Molecular Spectroscopy \citep{mueller} up to $P(8)$, $Q(16)$ and $R(18)$. Only 3 lines in the $P$-branch are detected, but in the $Q$- and $R$-branches some absorption is observed in lines up to $Q(14)$ and $R(16)$ (Fig.~\ref{c3} also shows the data around several additional transitions which are not detected).

The $R(2)$ line of C$_{3}$ has also been observed using the KAO with high spectral resolution \citep{giesen}. This line appears narrow (FWHM of 8.3~km~s$^{-1}$) and centred at the velocity of Sgr~B2 at 63.7~km~s$^{-1}$. \citet{cernicharo_d} indicate that the $R(2)$ line at 152.3~$\mu$m may show absorption due to the entire line of sight, whereas the other lines are only detected at the velocity of Sgr~B2 itself. No broad absorption is detected in the L03 survey data presented here for any of the detected C$_3$ features. The discrepancy can be explained by the small wavelength coverage in the L04 data meaning that it is difficult to determine the true continuum level. The narrow absorption is also confirmed by the KAO observation of this line, which shows only the Sgr~B2 velocity component \citep{giesen}. However, the central velocity observed by the KAO was 63.7~km~s$^{-1}$, which is in approximate agreement with the other C$_3$ lines observed in our survey, but does not match our $R(2)$ line velocity. This may indicate that the assignment for the $R(2)$ line in the survey may be wrong. The line observed by \iso~occurs centred at a velocity of 84~km~s$^{-1}$, higher than all the other observed lines. This could either be due to blending with a line from another (unknown) species, or due to some contribution from a spurious instrumental effect very close to the expected C$_3$ wavelength.

\subsubsection{Hydrogen Fluoride}

The survey shows a clear detection of the $J$=2--1 rotational transition of HF at 121.6973~$\mu$m. This line was also observed with higher signal-to-noise using the LWS L04 mode \citep{neufeld_a}, with the telescope pointed half way between Sgr~B2 M and Sgr~B2 N (the L03 survey data were pointed so as to only include Sgr~B2 M in the beam). The absorption line in the survey agrees very well in velocity with the L04 data ($66\pm5$~km~s$^{-1}$ compared with 67~km~s$^{-1}$) but the equivalent width is lower ($0.5\pm0.2$~nm compared with 1.0~nm). This difference in equivalent width could indicate that the absorption varies with the position of the beam, related to difference between the M and N sources. There may also be a contribution to the line from the ortho-H$_2$O $4_{32}$--$4_{23}$ transition at 121.7191~$\mu$m (separation of $\sim54$~km~s$^{-1}$ from HF). \citet{neufeld_a} attribute a 5$\sigma$ emission feature to this line. However, \citet{cernicharo_e} have fitted it as an absorption component. In any case, the contribution must be weak because the peak absorption occurs centred at the expected HF position rather than shifted towards the H$_2$O wavelength and the signal-to-noise in the L03 survey is not enough to distinguish any H$_2$O contribution.

\citet{neufeld_a} used an excitation model to determine the HF level populations as a function of position in the source. HF has a high critical density for excitation of the $J$=1 level by collisions and so radiative pumping dominates. They assumed a radial density and temperature profile for the envelope of Sgr~B2 to calculate an HF abundance of $3\times10^{-10}$. They show that HF is expected to be the major reservoir of fluorine and so this result indicates a depletion of fluorine by a factor of $\sim$50 from the Solar System value. However, \citet{ceccarelli} suggest that there could be a contribution to the absorption from the hot layer seen in NH$_3$. This would lead to a higher abundance in that layer and so a lower depletion factor \citep[although this would increase the depletion in the core of the cloud;][]{neufeld_f}.

The related molecule HCl, has been observed using the KAO via its $J=1-0$ transition at 478.96~$\mu$m \citep{zmuidzinas}. However, the lowest energy transition of HCl in our survey range is $J=3-2$ at 159.785~$\mu$m \citep{nolt} and no feature was detected at this wavelength. 

\subsection{Atomic and ionic lines}\label{sect_atom}

\begin{figure*}
\includegraphics[width=168mm]{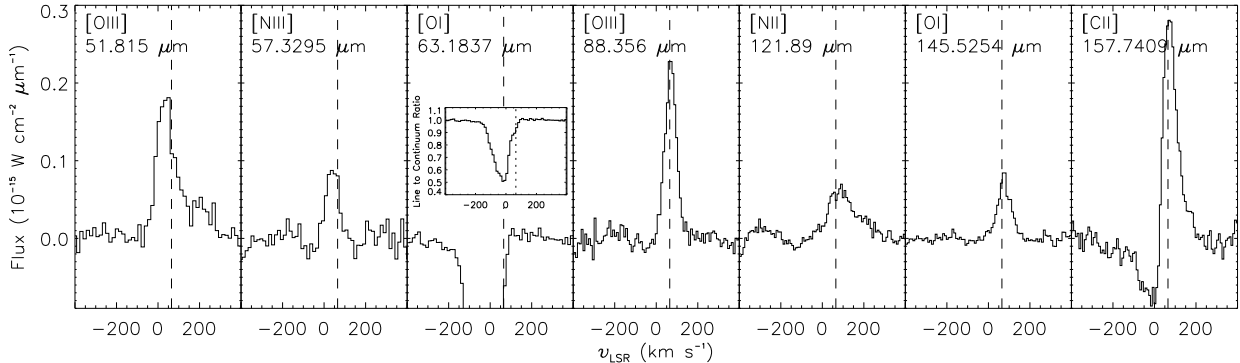}
\caption{Atomic and ionic lines observed in the survey range, shown with an absolute flux scale based on the LWS grating observation of Sgr~B2. The vertical dashed line is at the nominal velocity of Sgr~B2 at 65~km~s$^{-1}$.}
\label{fig_atom}
\end{figure*}

\begin{table}
\caption{Summary of atomic species observed in the survey. The survey fluxes were calculated using the line to continuum ratio from Table~\ref{iden_lines} with an estimate of the continuum flux from the LWS grating mode observation. The last column shows the line fluxes measured using the LWS grating mode by \citet{goicoechea_d}.}
\label{atomsum}
\begin{tabular}{cccc}
\hline
Species      & Estimated      & Survey Flux & Grating Flux$^a$    \\
             & Continuum      & 10$^{-18}$  & 10$^{-18}$          \\
             & 10$^{-15}$     & W~cm$^{-2}$ & W~cm$^{-2}$         \\
             & W~cm$^{-2}$    &             &                    \\
\hline
\oiii~51.815 &  0.66$\pm$0.07 & 4.1$\pm$0.6 & 2.2$\pm$0.6   \\
\niii~57.330 &  0.81$\pm$0.08 & 1.5$\pm$0.3 & $<$1.7        \\
\oi~63.184   &  1.0$\pm$0.1   & -absn-      & -absn-        \\
\oiii~88.356 &  1.3$\pm$0.1   & 6.0$\pm$0.6 & 4.2$\pm$1.7   \\
\nii~121.898 &  0.77$\pm$0.15 & 3.7$\pm$0.7 & $<$4.4        \\
\oi~145.525  &  0.54$\pm$0.11 & 3.9$\pm$0.8 & 7.3$\pm$0.8   \\
\cii~157.741 &  0.49$\pm$0.1  & 13.0$\pm$2.7 & $^b$          \\
\hline
\end{tabular}

\medskip
$^{a}$ from \citet{goicoechea_d}\\
$^{b}$ not detected in the grating due to combination of absorption and emission\\
\end{table}

The survey range includes the important atomic cooling lines from \oi~and \cii, as well as lines from ionised oxygen and nitrogen. Figure~\ref{fig_atom} shows the observed atomic lines on an absolute flux scale. The continuum flux was determined from the LWS grating observation of Sgr~B2. The atomic lines were also observed with the LWS grating in the extended region surrounding Sgr~B2 where they show widespread emission over $9\arcmin\times27\arcmin$ \citep{goicoechea_d}. However, the higher resolution FP observations are very important for the central Sgr~B2 M position where the weak lines such as \niii~(57.317~$\mu$m) and \nii~(121.898~$\mu$m) were not detected at the spectral resolution of the grating. Also, the lines of \oi~(63.184~$\mu$m) and \cii~(157.741~$\mu$m) show structure due to absorption along the line of sight \citep{vastel_b} which was not resolved with the LWS grating. The fitted emission line fluxes are shown in Table~\ref{atomsum}.

\subsubsection{\oi~and \cii}

The survey spectral range contains the 3 important cooling lines that trace photodissociation regions (PDRs) at the interfaces of molecular clouds, \oi~$^3$P$_1$--$^3$P$_2$ 63.2~$\mu$m, \oi~$^3$P$_0$--$^3$P$_1$ 145.3~$\mu$m and \cii~$^2$P$_{3/2}$--$^2$P$_{1/2}$ 157.7~$\mu$m. \citet{goicoechea_d} noted that the shocked gas in the Sgr~B2 envelope only makes a minor contribution to the observed lines fluxes, and therefore the extended \oi~and \cii~emission is dominated by the PDR scenario. In particular, comparison of the \cii~and \oi~lines with PDR models indicates a far-UV radiation field 10$^3$-10$^4$ times the mean interstellar field at the edge of Sgr~B2 \citep{goicoechea_d}. This is consistent with the origin of water and OH in these photoactive layers.

These lines are also particularly interesting towards Sgr~B2 because the 63~$\mu$m \oi~line and 158~$\mu$m \cii~line are seen in absorption in the line of sight clouds, and \oi~is self-absorbed at the velocity of Sgr~B2 itself. This \oi~absorption was also observed in the grating observations \citep{baluteau_a}, indicating a large fraction of the oxygen was in atomic form. The FP data confirm this, and show that not only is there absorption in the line of sight but the emission from Sgr~B2 itself is completely cancelled by absorption in its outer envelope \citep{lisc,vastel_b}. In contrast, emission is observed throughout the rest of the extended cloud \citep{baluteau_a,lisc,goicoechea_d}.

The higher excitation \oi~145~$\mu$m line is seen purely in emission at the velocity of Sgr~B2. This is because the lower energy level of the transition is not sufficiently populated in the cool line of sight gas to show absorption.

In the standard PDR view, there are 3 main layers going from atomic at the outside of the cloud, to where \oi~and \cii~coexist, to where all carbon is locked into CO \citep{hollenbach}. In this last layer, \oi~coexists with CO until the \oi/O$_2$ transition which only occurs at much higher extinction levels. The high resolution FP observations of \oi~and \cii~in absorption towards Sgr~B2 are a very useful probe of the PDRs along its line of sight. The absorption lines are particularly useful as they give a direct measure of the column density in each cloud. \citet{vastel_b} fitted the \oi~absorption using CO and \hi~observations as a template to disentangle the different PDR layers that contain atomic oxygen in each cloud. First, \hi~absorption observations were used to fit the component associated with the outer atomic layers and then the remaining O absorption was associated with CO observations tracing the cold molecular cores. The results for each of the velocity features in the line of sight are shown in Table~\ref{lineofsight}.

A similar method was used by \citet{lisc}, who compared the column density of \oi~ in three velocity ranges to that of CO (after subtracting the \oi~component associated with \hi) to give an apparent correlation of \oi/CO$\sim$9$\pm$1.3. No significant intercept was found for the relationship between \oi~and CO, indicating the lack of an intermediate PDR layer where \oi~is present between the CO region and completely atomic skin traced by \hi. This layer is predicted by PDR models to contain the transition CO/\ci/\cii.

However, the L03 observations of the \cii~line at 158~$\mu$m can be used to provide additional information, and using this additional line (as well as the \oi~line at 145~$\mu$m), \citet{vastel_b} could separate the predicted excess of \oi. They calculated a slightly lower ratio of \oi/CO in the cloud cores of 2.5$\pm$1.8, and an excess of \oi~between molecular and atomic regions of the clouds which indicated $N($\ci$)=(2.4\pm0.9)\times10^{17}$~cm$^{-2}$. This is in approximate agreement with ground based observations of \ci~towards Sgr~B2 \citep{vastel_b}. These results indicate $\sim$70\% of the oxygen is in atomic form and not locked into CO. Uncertainties in this analysis are large due to the spectral resolution of the LWS FP, and observations of \cii~and \ci~with {\it Herschel} HIFI will give a clearer picture of the different PDR layers.

It has been proposed that the absorption of \cii~by foreground clouds as observed here could also explain observations of bright galaxies which show a deficiency in \cii~flux compared with their total FIR luminosity \citep{vastel_b}. The spectrum of the bright ultraluminous galaxy Arp220 shows a very similar spectrum to Sgr~B2 with strong absorption by molecules (OH, H$_2$O, CH, NH and NH$_3$) as well as absorption by \oi~and weak emission by \cii~\citep{gonzalezalfonso}. Gonz\'{a}lez-Alfonso et al. found that they required some contribution from an absorbing `halo' to account for the observed \oi~absorption. However, the deficiency in \cii/FIR could mostly be explained by the non-PDR component of the FIR continuum, although some effect from extinction in the halo probably also contributes.

\begin{table*}
\caption{Summary of the column densities calculated for the line of sight features. The velocity ranges are based on the \hi~measurements of \citet{garwood}. Only features $<$0~km~s$^{-1}$ were considered for the atomic lines. Columns 2--6 are from \citep{vastel_b}, column 7 and 9 are from \citep{polehampton_oh} and column 8 is from \citet{polehampton_ch}. Estimates of the uncertainties are given in those references.}
\label{lineofsight}
\begin{tabular}{cccc|cc|ccc}
\hline
Velocity     & \hi        &    \oi             & \cii                & $^{13}$CO           & \oi          & OH                  & CH  \\
(km~s$^{-1}$)&(10$^{21}$~cm$^{-2}$)&Atomic part&(10$^{18}$~cm$^{-2}$)&(10$^{15}$~cm$^{-2}$)&Molecular part&(10$^{15}$~cm$^{-2}$)&(10$^{14}$~cm$^{-2}$) \\
             &            &10$^{18}$~cm$^{-2}$)&                     &             &(10$^{18}$~cm$^{-2}$) &                     &\\
\hline
$-$110 to $-$60 & 2.07 & 1.27 & 1.36 & 0.82 & 2.49  & 9.56 & 3.5  \\
$-$52 to $-$44  & 3.61 & 1.66 & 0.72 & 2.90 & 2.88  & 3.5  & 1.8  \\
$-$24           & 2.05 & 1.36 & 0.59 & 0.20 & 1.09  & 5.4  & 1.9  \\
$-$4 to $+$6    & 8.34 & 4.24 & 1.84 & 0.17 & 15.72 & 7.6  & 2.0 \\
$+$16           &      &      &      &      &       & 3.6  & 3.1  \\
$+$31           &      &      &      &      &       & 5.4  & 1.5  \\
$+$53 to $+$67  &      &      &      &      &       & 32.0 & 9.3  \\
\hline
\end{tabular}
\end{table*}

\subsubsection{Ionised lines of oxygen and nitrogen}

These lines are associated with the warm ionised gas beyond the edges of the PDRs, physically distinct from the region emitting/absorbing \oi. The \nii~line at 122~$\mu$m was first observed by \citet{rubin_a} using the KAO towards G333.6$-$0.2 (see also \citet{erickson_b} and \citet{colgan_b}). \nii~is an important coolant in the ISM: the {\it COBE} satellite found that the two \nii~fine structure lines at 122~$\mu$m and 205~$\mu$m are the brightest in the galaxy after the \cii~line at 158~$\mu$m \citep{wright_b}. The \nii~emission is preferentially produced in lower density gas than the pair of \oiii~lines observed at 52 and 88~$\mu$m. 

Low spectral resolution LWS grating rasters of the \nii~and \oiii~lines around Sgr~B2 revealed a very extended component of ionised gas, with average electron densities of $\sim$240~cm$^{-3}$ \citep{goicoechea_d}. Detailed photoionisation models of these lines show that the ionising radiation has an effective temperature of $\sim$36,000~K and a Lyman continuum flux of $\sim$10$^{50.4}$~s$^{-1}$ \citep{goicoechea_d}. However, the location and distribution of the ionising sources in the Sgr~B2 envelope remains unclear, and will require much higher angular resolution observations. The LWS L04 mode was also used to observe \nii~towards several other Galactic Centre sources \citep{rodriguez_b}. 

\begin{figure}
\includegraphics[width=84mm]{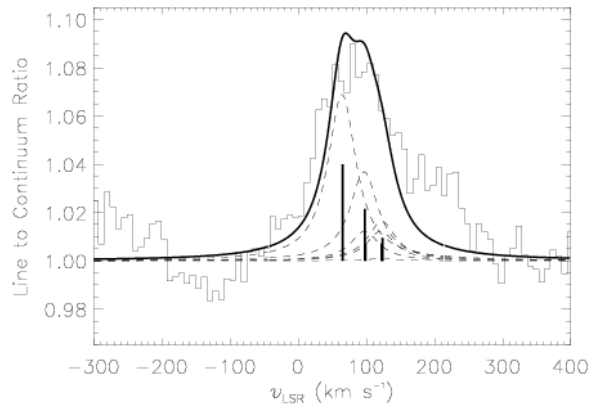}
\caption{The \nii~line at 121.9~$\mu$m. The solid line shows a fit including hyperfine splitting, with the relative strength of hyperfine components set to their expected line strengths (the thick bars show the location of the individual components). The line width at each component was fixed to 51~km~s$^{-1}$.}
\label{nii}
\end{figure}

The \nii~line is particularly interesting as it has several hyperfine structure components that produce a noticeable broadening in the observed profile. This is due to the nuclear spin of the $^{14}$N atom, resulting in 3 groups of components which are separated by $\sim30$~km~s$^{-1}$ \citep{brown_f}. In order to fit the \nii~line, we fixed the relative contribution of each component to its expected line strength taken from \citet{brown_f}, and fixed the line width to 51~km~s$^{-1}$ (which is the average of that observed for the other atomic lines above 70~$\mu$m) and the centre velocity to be 65~km~s$^{-1}$. The strongest line component occurs at a wavelength of 121.8884~$\mu$m. Figure~\ref{nii} shows the result of the fit, giving a total \nii~flux of $3.7\pm0.7\times10^{-18}$~W~cm$^{-2}$ (adopting a continuum level estimated from the L01 grating observation of $0.77\times10^{-15}$~W~cm$^{-2}$). Even accounting for the hyperfine components and a line width wider than the LWS FP spectral resolution at 122~$\mu$m ($\sim$34~km~s$^{-1}$), the fitted profile does not explain all the observed emission. There is some extra emission at a velocity of $+$200~km~s$^{-1}$ that is difficult to explain. A trace of this high velocity emission is also present in the \oiii~line at 51.8~$\mu$m (see Fig.~\ref{fig_atom}). 

\subsection{Global line properties and kinematics}\label{line_props}

\begin{figure}
\includegraphics[width=84mm]{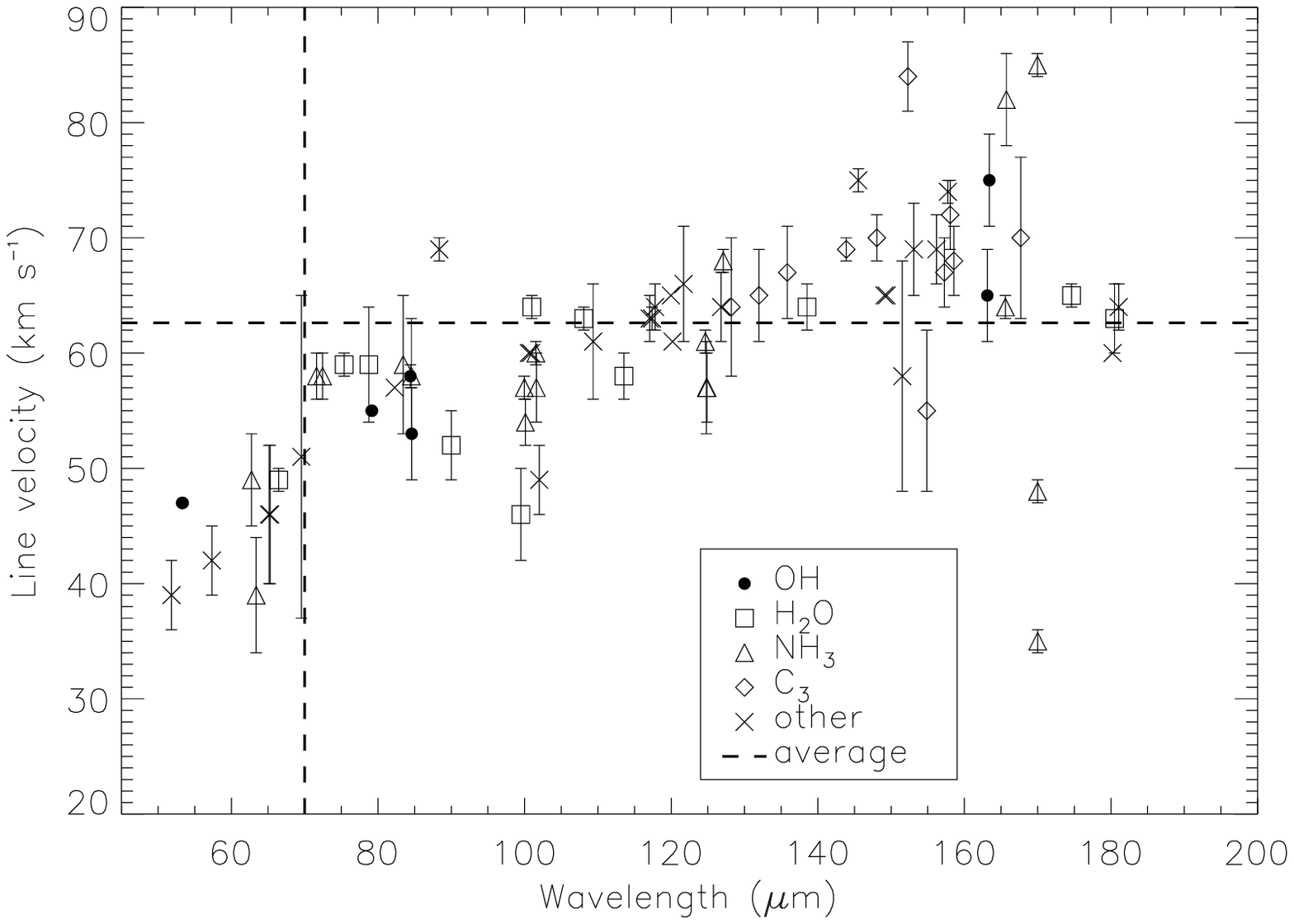}
\includegraphics[width=84mm]{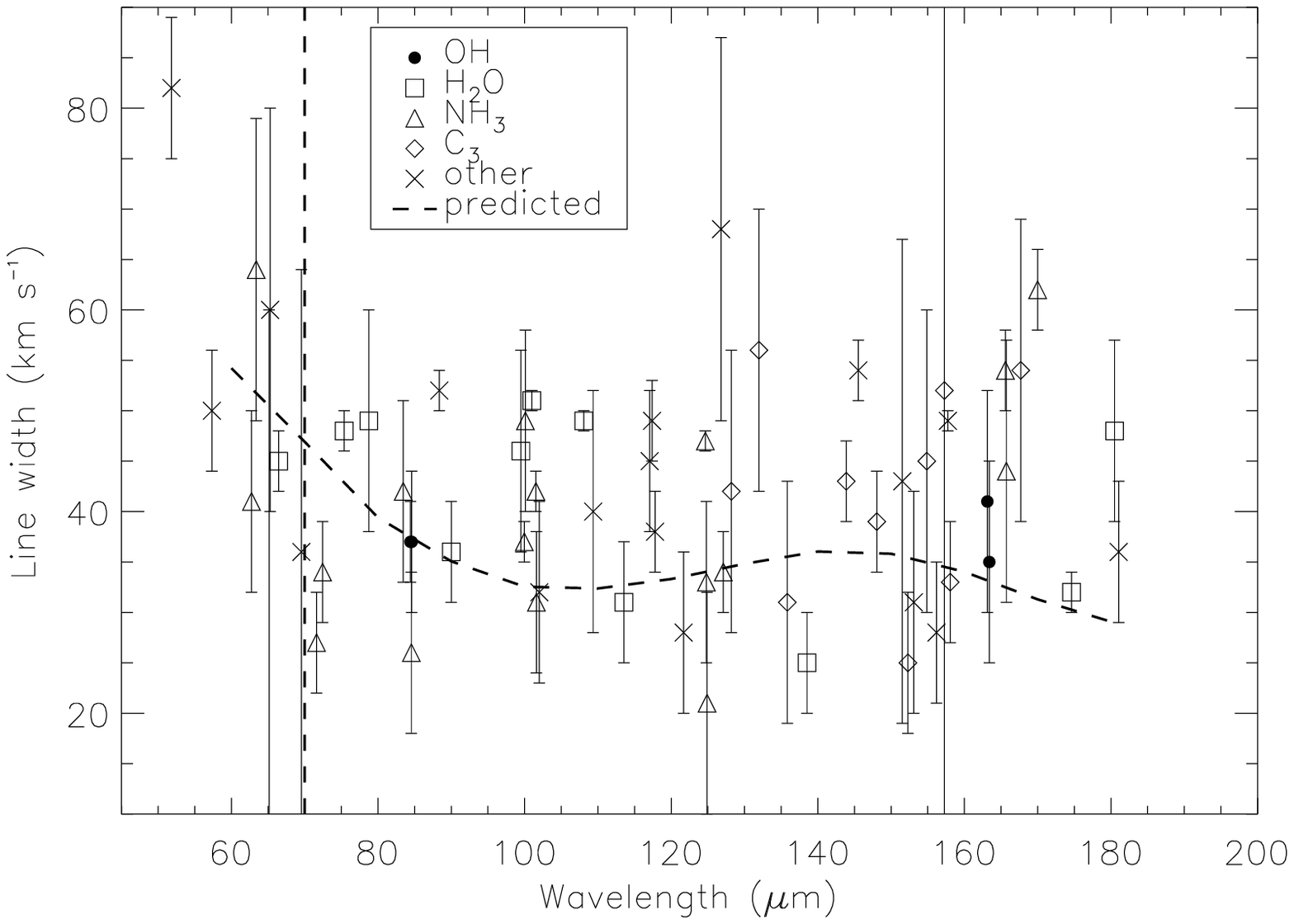}
\caption{Best fitting centre velocity (top) and line width (bottom) of identified features from Table~\ref{iden_lines}. In the top plot the average velocity in the range 70--196~$\mu$m is shown as the horizontal dashed line (62.7$\pm$5.6~km~s$^{-1}$). The vertical dashed line shows the edge of the nominal operating mode of FPL (70~$\mu$m), below which the wavelength calibration has significant systematic errors. In the bottom plot the dashed line shows the predicted width based on the instrument resolving power.}
\label{widvel}
\end{figure}

Figure~\ref{widvel} shows how the results of the line fits vary across the survey range. In the upper plot showing the central velocity, there is a clear systematic error in the wavelength calibration for the non-prime data below 70~$\mu$m. This can be explained because the polynomial used to fit the relationship between the instrumental FP encoder setting and gap between the FP plates for FPL was based on data above 70~$\mu$m. This effect is also clear when the non-prime lines are directly compared with the equivalent prime FPS observation \citep[e.g. the 53~$\mu$m OH lines - see ][]{polehampton_oh}. A correction to the wavelength calibration has not been applied in the data reduction process for the survey, or in the results presented in Table~\ref{iden_lines} \citep[however, it was taken into account where data below 70~$\mu$m were analysed in detail; e.g.][]{polehampton_oh}.

In the wavelength range 70--196~$\mu$m, the average velocity over all lines is 62.6~km~s$^{-1}$ with an RMS of 5.5~km~s$^{-1}$. This was calculated by including only the absorption from Sgr~B2 itself and ignoring the outlying lines above 80~km~s$^{-1}$ and below 50~km~s$^{-1}$. These outliers are: the H$_2$O line at 99.5~$\mu$m (46~km~s$^{-1}$); the H$_2^{18}$O line at 102.0~$\mu$m (49~km~s$^{-1}$); the C$_3$ $R(2)$ line at 152.3~$\mu$m (84~km~s$^{-1}$), already discussed in Sect.~\ref{carbon}; the NH$_3$ line at 165.7~$\mu$m (82~km~s$^{-1}$); and the 3 NH$_3$ lines at 170~$\mu$m which are blended, distorting the central velocity. Apart from these lines, the spread in velocity about the average is consistent with a study of the accuracy of the wavelength calibration using CO lines in the FP spectrum of Orion - in that case, the line centres were always less than 11~km~s$^{-1}$ from the expected position \citep{gry}. 

We have also calculated the average velocity separately for each individual species. However, it is difficult to determine if differences are real or due to systematic effects. In Fig.~\ref{widvel} there appears to be a trend with more lines above average $>$120~$\mu$m and more lines below average $<$120~$\mu$m. This is reflected in the average velocities because all the C$_3$ lines are at wavelengths longer than 120~$\mu$m and most of the NH$_3$ lines are in the shorter wavelength bracket. The average velocities for each species are, NH$_3$ 59.1$\pm$3.6~km~s$^{-1}$; NH$_2$ 63.5$\pm$0.6~km~s$^{-1}$; H$_2$O 60.8$\pm$4.2~km~s$^{-1}$; OH 60.2$\pm$8.4~km~s$^{-1}$; C$_3$ 66.7$\pm$4.8~km~s$^{-1}$. The atomic and ionic lines are generally higher in velocity than the molecular lines, \oiii~(88~$\mu$m) 72~km~s$^{-1}$; \oi~(145~$\mu$m) 75~km~s$^{-1}$; \cii~(158~$\mu$m) 74~km~s$^{-1}$.

Variation in the velocity of lines from different species was also found in the 330--355~GHz survey of \citet{sutton} and the 218--263~GHz survey of \citet{nummelin_b}. Sutton et al. found an average velocity of 60.6~km~s$^{-1}$ for Sgr~B2 M and Nummelin et al. found an average velocity of 61.6~km~s$^{-1}$, with smaller molecules in range 55--60~km~s$^{-1}$. However, at these wavelengths the lines are more likely to trace the material in the cores of the cloud rather than in the outer regions of the envelope as for our survey. 

The average line widths found in these mm surveys were 18.5~km~s$^{-1}$ \citep{sutton} and 17~km~s$^{-1}$ \citep{nummelin_b}, of order half the spectral resolution of our survey. The lower plot in Fig.~\ref{widvel} shows our measured line widths from Table~\ref{iden_lines}, with the predicted instrumental profile width based on the measured resolving power (from Fig.~\ref{resolv}). The measured line widths cluster at or above this line, consistent with broadening caused by the width associated with the source.

\subsection{Isotopic species}\label{isotopes}

Several isotopic variants of the observed species were detected in the survey - these are due to the two species showing the strongest absorption in their main lines, H$_2$O and OH. Observations of isotopologues can be used to determine intrinsic isotope ratios, if good estimates of the column density of the main species can be made. However, if these ratios are already known, the isotopologue lines are also valuable in modelling as their optical depth is much lower than the main lines. 

\subsubsection{Oxygen isotopes}

The ground state lines of both $^{17}$OH and $^{18}$OH are observed in the survey, showing absorption in both Sgr~B2 and the line of sight clouds. OH is particularly good for tracing the intrinsic oxygen isotopic ratio, because chemical fractionation reactions that might act to distort their values (e.g. as occurs for $^{13}$C/$^{12}$C) are not thought to be important \citep{langer}. In addition, optically thin lines from all three isotopologues were observed as part of the survey (for $^{16}$OH these were the cross ladder transitions). 

In order to calculate the $^{16}$O/$^{18}$O ratio, the ground state line profile of $^{16}$OH was compared with that of $^{18}$OH \citep{polehampton_oh}. The isotopic ratio in each velocity component of the lines was determined by carrying out a simultaneous fit of the $^{16}$OH 53~$\mu$m and 79~$\mu$m lines with the $^{18}$OH 120~$\mu$m line. The fit was carried out in a similar way to that described for $^{16}$OH in Sect.~\ref{sect_oxygen} using \hi~observations as a basis for the line shapes of each velocity component. The results show values of $^{16}$O/$^{18}$O in the Galactic Disc that are broadly consistent with previous measurements (in the range 360--540). However, in velocity components associated with the Galactic Centre, the ratio is higher than previous estimates \citep[320$^{+70}_{-30}$ compared with the standard value of 250;][]{wilson}. The standard ratio for the Galactic Centre was derived from measurements of the $\Lambda$-doublet transitions of OH \citep[e.g.][]{williams,whiteoak_b}. However, \citet{bujarrabal} have suggested that these could be an underestimate due to excitation anomalies in the hyperfine levels caused by FIR rotational pumping. The survey results would appear to support this (the rotational lines at the resolution of the LWS FP are not affected by anomalies in the hyperfine levels).

The only previous measurements of $^{17}$OH in the ISM were made towards Sgr~B2 via its microwave $\Lambda$-doublet transitions \citep{gardner76,bujarrabal} and the \iso~observations are the first time its pure rotational transitions have been seen in space. Previous measurements of the value of $^{18}$O/$^{17}$O in the Galactic Disk using CO indicated a constant isotopic ratio of 3.65$\pm$0.15 \citep{penzias} and this agrees with the microwave observations of $^{17}$OH towards Sgr~B2 \citep[3.6$\pm$0.5;][]{bujarrabal}. Analysis of the lines from the survey has been presented by \citet{polehampton_c}. Due to the weakness of the $^{17}$OH lines, they did not fit them directly, but used the $^{18}$OH observations with a ratio of 3.5 (the weighted average of all previous measurements in the Galactic Centre) to predict the line shape. This is complicated by the hyperfine splitting in $^{17}$OH (due to the spin of the $^{17}$O nucleus) which causes a series of line components with similar separation to the velocity features along the line of sight. Using a ratio of $^{18}$OH/$^{17}$OH=3.5 shows good agreement at the velocity of Sgr~B2 but some disagreement for the line of sight, appearing to underestimate the observed absorption (indicating a lower ratio is needed). However, this could also be due to uncertainties in the data reduction distorting the line shape \citep{polehampton_c}.

In comparison to OH, water is not a good tracer of the true oxygen isotopic ratio. This is because even though transitions of both H$_{2}^{16}$O and H$_{2}^{18}$O were detected, they are highly optically thick and the excitation of the observed water lines is highly complex \citep{cernicharo_e}. The model of \citet{cernicharo_e} predicts that even the H$_{2}^{18}$O line at 181.05~$\mu$m (2$_{12}$--1$_{01}$) is optically thick, with $\tau\sim5$ (in addition, this line is blended with H$_3$O$^+$). The equivalent transition for H$_{2}^{16}$O at 179.53~$\mu$m has $\tau\sim10^3$--$10^4$. The next strongest line is 2$_{21}$--2$_{10}$, and the model shows that $\tau\sim300$ for H$_{2}^{16}$O. Therefore, water is much more difficult to use as a probe of the isotopic ratio.

\subsubsection{Carbon isotopes}

The ratio $^{12}$C/$^{13}$C has been found to be 25$\pm$1 in Sgr~B2 derived from CO observations \citep{langer2} and so we might expect isotopologues of the carbon bearing species to be easily detected. However, $^{13}$CH is not detected, with a lower limit on $^{12}$CH/$^{13}$CH of 36 \citep{polehampton_ch}. This is higher than the value derived from CO but could be due to isotopic fractionation reactions which would increase $^{12}$CH/$^{13}$CH over the true $^{12}$C/$^{13}$C isotopic ratio \citep{langer}.

The other carbon bearing lines in the survey such as C$_{3}$ are too weak for their isotopologues to be detected.

\subsubsection{Nitrogen isotopes}

The isotopic ratio, $^{14}$N/$^{15}$N, is about 400--500 in Sgr~B2. This has been measured in radio observations towards Sgr~B2 N in both emission and absorption lines \citep{peng}. Therefore, we would not expect to detect any lines from $^{15}$NH$_{3}$ above the noise level in the survey. However, the ground state ($1_0$--$0_0$) transition of both $^{14}$NH$_3$ and $^{15}$NH$_3$ has been observed with the ODIN satellite towards Sgr~B2 \citep{hjalmarson}. They observed strong absorption from $^{14}$NH$_{3}$ at all line of sight velocities and weak absorption from $^{15}$NH$_3$ at the velocity of Sgr~B2.

\subsubsection{Deuterium}\label{sect_deuterium}

The key deuterated ion, H$_2$D$^+$, has been detected via its ortho ground state rotational transition 2$_{12}$--1$_{11}$ at 126.853~$\mu$m \citep{cernicharo_f}. This line appears blended with the NH$_2$ 2$_{21}$--1$_{10}$ line at 126.8014~$\mu$m, but is clearly detected because the broad blended feature cannot be due to NH$_2$ alone. This is because all other detected NH$_2$ lines arising from excited states are narrow and appear at the velocity of Sgr~B2 itself. Removal of the contribution of NH$_2$ shows that the H$_2$D$^+$ line is absorbed by all components along the line of sight \citep{cernicharo_f}.

H$_2$D$^+$ is an important molecule as it plays a crucial role in the deuteration of other species in cold gas \citep[e.g.][]{gerlich}. The calculated column density at the velocity of Sgr~B2 is 9$\times$10$^{13}$~cm$^{-2}$ and in the line of sight clouds is (2.7--6.7)$\times$10$^{13}$~cm$^{-2}$ \citep{cernicharo_f}.

The only other isotopic species containing deuterium that has possibly been detected in the survey is HD. A $\sim3\sigma$ emission feature is seen in co-added prime and non-prime data corresponding to the HD $J$=1--0 line at 112.0725~$\mu$m \citep{polehampton_b}. This leads to a D/H ratio in the Sgr~B2 envelope in the range (0.2--11)$\times10^{-6}$. 

For other molecules, the formation process favours deuteration and so the abundances can be dramatically enhanced over the D/H ratio \citep{roberts}. 

{\bf HDO:} The lowest transition in the range is 2$_{21}$--2$_{02}$ at 159.354~$\mu$m, which is not detected. The equivalent transition to the saturated line in H$_{2}$O at 179~$\mu$m 2$_{12}$--1$_{01}$ occurs outside the LWS range at 235~$\mu$m. However, other transitions of HDO have been observed from the ground \citep[][and references therein]{comito}, and some of the higher energy transitions in the FIR have been observed in the \iso~spectral survey towards Orion~KL \citep{lerate}. 

{\bf OD:} Chemical models predict that OD/OH should be enhanced over the D/H ratio. In diffuse clouds OD/OH may be between 3$\times$10$^{-4}$ and 4$\times$10$^{-3}$ \citep{croswell}. This would mean that the ground state rotational transition would be clearly detected towards Sgr~B2. However, only the cross ladder transitions from the ground state occur within the LWS range and these are too weak to be detected. The ground state transition occurs at 215.4~$\mu$m \citep{brown_od}, unfortunately falling in the unobservable gap between bands 5 and 6 of {\it Herschel} HIFI.

{\bf NH${_2}$D:} This has been detected towards the Sgr~B2 N hot core with a NH${_2}$D/NH$_3$ ratio of 1.1$\times$10$^{-3}$ \citep{peng}. This is enhanced compared with the local D/H ratio but is an order of magnitude smaller than in hot core sources outside of the Galactic Centre. This ratio is too low to expect a detection in the survey.

\subsection{Unassigned features}\label{sect_unassigned}

We have catalogued the unidentified features in the survey which have a strength greater than $\sim$3$\sigma$ above the RMS noise in data. All are seen in absorption and there are no unassigned emission lines ($>3\sigma$). Possible assignments for these absorption features were checked using the Cologne Database for Molecular Spectroscopy \citep{mueller}, the JPL spectral line database \citep{pickett} and the catalogue of molecular transitions maintained by J. Cernicharo. Table~\ref{ulines} gives a list of these features with their fitted line widths and depths, which were determined in a similar way to the assigned lines: the line centre, width and depth were free parameters in a Lorentzian fit. The wavelengths were calculated assuming that the lines were centred at 65~km~s$^{-1}$. All the wavelengths for unidentified features mentioned in the following text have been corrected by this velocity. In addition to the strong features listed in Table~\ref{ulines}, there are many more possible lines with a significance $<3\sigma$.

The features in Table~\ref{ulines} were fitted in the binned data, but have also been checked in the raw unaveraged data to make sure they were not due to glitches, thin patches in the data where glitches had been removed or bad matching where mini-scans overlap. Such spurious features are easy to spot in the unaveraged data if they are examined mini-scan by mini-scan. All the features in Table~\ref{ulines} are considered to be real, and in most cases appear in at least 2 independent observations. The level of each detection above the RMS noise in the adjacent continuum is also shown. The line profile of each unidentified line are shown in detail in Fig.~\ref{fig_ulines}. Some of the features are narrow, with a width close to the FP resolution, and some are broad indicating possible contribution from the line of sight clouds (or multiple blended line components). The 139.12~$\mu$m feature appears to have 2 distinct components.

The feature at 180.338~$\mu$m is close to the $2_{12}$--$1_{01}$ line of H$_2^{17}$O at 180.3302~$\mu$m, but if this is really the true identification, the line would peak at 79~km~s$^{-1}$ rather than the expected 65~km~s$^{-1}$.

\begin{figure}
\includegraphics[width=84mm]{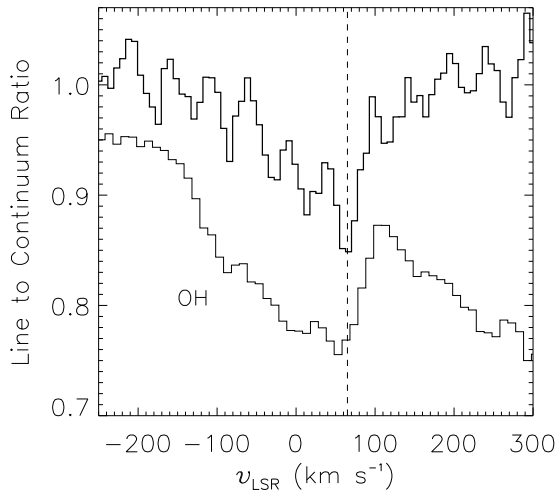}
\caption{The line shape of the unidentified feature at 175.07~$\mu$m (above)
  compared with the $^{2}\Pi_{1/2}$--$^{2}\Pi_{3/2}$ $J$=1/2--3/2 line of OH at
  79~$\mu$m (below). The OH line has been scaled to match the unidentified feature.}
\label{uline175}
\end{figure}

\begin{figure*}
\resizebox{16.8cm}{!}{\includegraphics{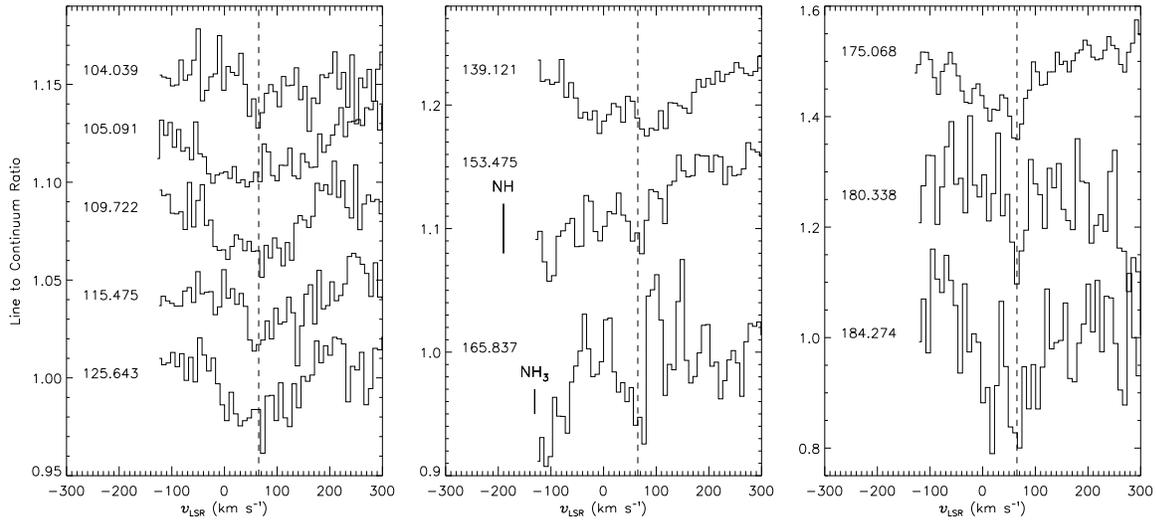}}
\caption{Unidentified lines in the survey data plotted versus velocity
  (assuming they are centred at 65~km~s$^{-1}$). Note that the y-axis has a different scale for each of the three plots.}\label{fig_ulines}
\end{figure*}

There are three features in the list that are particularly notable because they clearly appear in multiple observations, including ones made on different detectors: these are at 153.47~$\mu$m, 175.07~$\mu$m and 184.27~$\mu$m. The feature at 175.07~$\mu$m is particularly strong ($\sim$6$\sigma$), clearly showing a broad profile, similar in shape to the lines of CH and OH due to absorption along the whole line of sight to Sgr~B2. Figure~\ref{uline175} shows the line shape compared with that of OH. This indicates that the line must be due to a low energy transition, probably from the ground state. In all the other molecules observed there are no higher energy transitions that show line of sight absorption. However, we have not been able to identify any candidates that fit this description. The closest matches found in the line catalogues are all too high in energy: H$_3$O$^+$ $6_6^+$--$6_6^-$ at 175.0633~$\mu$m; H$_2$O $5_{05}$--$4_{32}$ at 174.9200~$\mu$m. 

One possible group of species that could be responsible for some of the unidentified features is the ions of the neutral hydride species that were observed in the survey. These species may occur in the external layers of the Sgr~B2 envelope where the volume density is low (in the inner parts they would be destroyed by reaction with H$_2$). In the following paragraphs we present examples of some of the hydride cations that could be thought to be detected. However, none of these molecules match with any of our unassigned lines, showing that further study and lab work will be needed to identify the true carriers.

Rotational transitions of the CH$^+$ ion have been detected in emission in the planetary nebula NGC 7027 \citep{cernicharo}. However, its lowest energy transition in our survey range, $J=2$--1, at 179.611~$\mu$m overlaps with the much broader saturated absorption line due to water and so cannot be separated.  

The OH$^+$ ion is isoelectronic to the NH radical and could be produced in the outer layers of interstellar clouds or in shocks \citep[][and references therein]{almeida_b, almeida_c}. The fundamental transition, $N=1$--0, occurs at 329.8~$\mu$m ($J=0$--1), 308.5~$\mu$m ($J=2$--1) and 290.2~$\mu$m ($J=1$--1) \citep{bekooy} outside the range of our survey. However, there is also a cluster of features that could match with the broad U-line at 153.47~$\mu$m, due to the $N=2$--1, $J=1$--2 transition (Cologne Database for Molecular Spectroscopy). Unfortunately, this transition has a very low line strength, and it would be much more likely to detect the stronger feature due to the $N=2$--1, $J=3$--2 transition at 152.9897~$\mu$m. This is not detected, ruling out any identification of OH$^+$ at 153.47~$\mu$m. A much better chance of detecting OH$^+$ will be provided by the HIFI instrument on board the {\it Herschel} satellite, as it will be able to observe the ground state lines. 

The NH$^+$ ion is isoelectronic to CH and \citet{almeida} have proposed that photoionisation, charge-transfer, radiative association and ion-molecule reactions in the gas phase may lead to a significant concentration of NH$^+$ in the ISM. Its ground state lines ($N=1$--1, $J=3/2$--1/2) occur outside our survey range at 294.14~$\mu$m and 296.08~$\mu$m \citep{verhoeve}, but the second rotational transitions, $N=2$--1, $J=5/2$--3/2, do occur within the range at 160.01~$\mu$m and 159.70~$\mu$m \citep{kawaguchi}. However, these do not correspond to any of the unassigned lines.

Calculations for CH$_2^+$ show that its low energy rotational transitions all occur within the survey range. However, accurate measurements of its rotational levels have not been made in the laboratory. The best measurements so far have been made by photoelectron spectroscopy \citep{willitsch} and by near-IR spectroscopy \citep{gottfried}. Using the parameters from these measurements, we calculate that the ground state $1_{11}$--$0_{00}$ transitions should be at $\sim$128.9~$\mu$m ($J=3/2$--1/2) and $\sim$134.6~$\mu$m ($J=1/2$--1/2). The next higher energy transitions are $2_{11}$--$1_{01}$ at $\sim$109.6~$\mu$m and $1_{10}$--$1_{01}$ at $\sim$157.2~$\mu$m and 164.5~$\mu$m. There is considerable uncertainty in these predictions (up to 2~$\mu$m), in particular because the $A$-rotational constant and the spin-rotation splittings are very poorly determined or not even known, and more lab experiments are needed before a firm identification could be made. The molecule is predicted to have a reasonably sized dipole moment, calculated to be 0.629~D by \citep{osmann} and 0.701~D by \citep{brinkmann}. 

The H$_2$O$^+$ ion has been observed in optical emission spectra of comets \citep[e.g.][]{cremonese}. However, its ground state rotational transitions, $1_{11}$--$0_{00}$, occur at 268.84~$\mu$m and 263.08~$\mu$m \citep{murtz}, outside our survey range. The next higher energy transitions are $2_{21}$--$1_{10}$ at 105.74~$\mu$m and 104.72~$\mu$m, and $2_{20}$--$1_{11}$ at 100.91~$\mu$m and 99.88~$\mu$m \citep{murtz}. None of these are detected.

In addition to the lines from simple molecules, there could be features from the bending modes of carbon clusters and PAHs \citep{mulas}. In emission, these lines show a band structure due to the $Q$-branch lines (which would be resolved by the LWS FP) and a broader 'grass'. No emission features matching the predicted band structure of PAH molecules has been observed in the survey, however, it is possible that the broader component could exist and was removed in the reduction process. This is because very broad features in the continuum in the FP data are impossible to distinguish from artifacts of the reduction process (see Sect.~\ref{sect_contm}) and so were removed. However, in absorption by cold gas, only the ground state $Q$-branch line would be observed and the profile could be narrow. In order for this absorption to be detected, a molecule with high transition intensity would be required. Concrete identification of a U-line with a single PAH molecule would be very difficult in this case, and require further laboratory measurements of the candidate species.

\begin{table}
\caption{Unidentified Lines in the survey data. The wavelengths given were
  calculated assuming the line centres are at 65~km~s$^{-1}$. Approximate
  widths and line depths were derived by fitting the lines with a Lorentzian
  profile in the same way as the assigned lines in Table~\ref{iden_lines}. The 4th
  column gives the detection above the RMS noise in the surrounding
  continuum. The lines in bold are the most secure (i.e. detected in
  multiple observations).}
\label{ulines}
\begin{tabular}{cccc}
\hline
Rest     & $\Delta{v}$   & Line to     & Level of    \\
Wavelength ($\mu$m)& (km~s$^{-1}$) & Continuum    & detection ($\sigma$)   \\
\hline
104.039  &  25     &  0.976  & 3  \\
105.091  &  160    &  0.975  & 4  \\
109.722  &  120    &  0.967  & 3.5  \\
115.475  &  40     &  0.975  & 3  \\
125.643  &  110    &  0.962  & 4  \\
139.121  &  160    &  0.957  & 6  \\
{\bf 153.475} & \multicolumn{3}{l}{blend with NH}   \\                
165.837  &  30     & 0.926   & 3  \\
{\bf 175.068} & 120 &  0.872  & 6  \\
180.338  &  20     &  0.836   & 3  \\
{\bf 184.274} & 60 &  0.816  & 3  \\
\hline
\end{tabular}
\end{table}

\section{Summary}

We have presented the data from a complete spectral survey of Sgr~B2 between 47 and 196~$\mu$m, observed with the \iso~LWS FP spectrometer. The spectral resolution was 30--40~km~s$^{-1}$ across the range and the signal-to-noise achieved in the line to continuum spectrum was $\sim$100. A total of 95 lines have been identified in the spectrum, with a further 11 unassigned features. The main results can be summarised as follows:

\begin{itemize}
\item The survey is dominated by molecular lines in absorption due to NH$_3$, NH$_2$, NH, H$_2$O, OH, H$_3$O$^+$, CH, CH$_2$, C$_3$, HF and H$_2$D$^+$ (and in emission for OH at 163~$\mu$m). The ground state lines of OH, CH, CH$_2$, ortho-H$_2$O, H$_3$O$^+$ and H$_2$D$^+$ show absorption due to the entire line of sight as well as Sgr~B2 itself. Isotopic lines are observed from $^{18}$OH, $^{17}$OH and H$_2^{18}$O.
\item Atomic and ionic lines of \oi, \oiii, \cii, \nii~and \niii~are seen in emission at the velocity of Sgr~B2 itself, with the exception of the \oi~63~$\mu$m line which is completely self-absorbed. This line and the \cii~158~$\mu$m line show absorption due to the line of sight clouds.
\item The species with greatest column density in the survey is H$_2$O, with $N({\rm{Sgr~B2}})\sim9\times10^{16}$~cm$^{-2}$.
\item The average velocity for all the assigned lines which only show absorption due to Sgr~B2 is 63$\pm$6~km~s$^{-1}$. There appears to be some variation in the velocity of different species but it is difficult to separate this from systematic effects.
\item Several unassigned features remain, all of which are in absorption. These could be due to simple molecules or bending modes of carbon chains/clusters. 
\end{itemize}

This survey provides a basis for future follow up at higher spectral and spatial resolution (e.g. with HIFI, or instruments on SOFIA). The planned {\it Herschel} guaranteed time spectral survey of Sgr~B2 with HIFI and PACS will overlap with this survey above 57~$\mu$m, and extend it into the sub-mm. It will be interesting to check the unassigned features in the overlap region, particularly the line at 175~$\mu$m which shows a very similar profile to the ground state lines of OH and CH.

\section*{Acknowledgements}

We would like to thank our referee for useful suggestions, and Christine Joblin, Holger M\"{u}ller and Mercedes Lerate for helpful discussions.
JRG was supported by an individual Marie Curie fellowship, contract MEIF-CT-2005-515340.
The \iso~Spectral Analysis Package (ISAP) is a joint development by the LWS and SWS Instrument Teams and Data Centres. Contributing institutes are CESR, IAS, IPAC, MPE, RAL and SRON. LIA is a joint development of the \iso-LWS Instrument Team at the Rutherford Appleton Laboratory (RAL, UK - the PI Institute) and the Infrared Processing and Analysis Center (IPAC/Caltech, USA).

\bibliographystyle{aa}
\bibliography{thesis_references}

\appendix

\section{Detailed data reduction}\label{appendix_datared}

\begin{table}
\caption{Dark current and stray light values for FPS and FPL. The nominal dark
  currents are given for comparison - these are the combination of results
  made throughout the mission using three independent means of measurement
  given by \citet{gry}.}
\label{tab_darks}
\begin{tabular}{cccc}
\hline
Detector & Nominal dark   & FPS              & FPL \\
         & (10$^{-16}$~A) & dark+stray light & dark+stray light \\
         &                & (10$^{-16}$~A)   & (10$^{-16}$~A) \\
\hline
SW1 &  4.96$\pm$0.54  &   $^{(a)}$    &   -  \\
SW2 &  2.08$\pm$0.43  &   $^{(a)}$    & 3.0$^{(b)}$   \\
SW3 &  2.20$\pm$0.21  & 2.48$\pm$0.09 & 3.3$\pm$0.25$^{(b)}$  \\
SW4 &  1.18$\pm$0.34  & 1.33$\pm$0.04 & 1.6$\pm$0.1  \\
SW5 &  1.56$\pm$0.24  & 1.80$\pm$0.03 & 1.9$\pm$0.2 \\
LW1 &  2.50$\pm$0.29  & 2.8$\pm$0.1   & 2.7$\pm$0.1$^{(c)}$   \\
LW2 &  0.07$\pm$0.27  & 0.60$\pm$0.07 & 0.46$\pm$0.03 \\
LW3 &  0.53$\pm$0.39  & 2.2$\pm$0.4   & 1.41$\pm$0.2  \\
LW4 &  1.76$\pm$0.42  & 4.1$\pm$0.6   & 2.52$\pm$0.02 \\
LW5 &  1.21$\pm$0.25  & 1.8$\pm$0.1   & 1.43$\pm$0.02 \\
\hline
\end{tabular}

\medskip
$^{(a)}$ Nominal value assumed\\
$^{(b)}$ Estimated from comparison of lines measured with both FPs\\
$^{(c)}$ Special modelling performed for LW1 - see \citet{polehampton_thesis}.
\end{table}

\subsection{Dark current and stray light} \label{sect_dark}

Careful determination of the detector dark signals was very important for the accurate calibration of relative line depths in the survey. It is a particular problem for Sgr~B2 due to the strength of the source and the fact that it was observed slightly off-axis. Scattering of off-axis rays \citep[due to the partially reflecting substrate of one of the LWS mirrors, M2; see ][]{gry} caused a series of reflections within the body of the instrument. These reached the detectors outside of the main beam without passing through the grating or FP and hence produced a background signal independent of wavelength. This was particularly important during FP observations because the radiation not transmitted by the FP ($>99$\% of the incident flux) was reflected back towards M2. Stray light within the beam (wavelength dependant) was not important because on-axis scattering was minimised by the design of the instrument.

In order to characterise the dark + stray light towards Sgr~B2, we used data recorded on non-prime detectors where the separation between FP orders was enough that no order occurred within the grating response function. At wavelengths where the FP order separation was sufficiently large that any contribution from the FP profile wings was negligible, the source radiation in the main beam was completely blocked. In this way, the dark + stray light signal could be accurately determined for most detectors. For the remaining detectors, it was estimated where possible by carefully comparing lines that had been measured using both FPs. For the detector LW1 using FPL, special modelling was used to estimate the dark signal and this is described in full by \citet{polehampton_thesis}. The final values adopted are shown in Table~\ref{tab_darks}.

A similar method has been adopted to calculate the dark + stray light signal for the \iso~LWS FP survey towards Orion \citep{lerate}. A comparison of the results shows that the contribution from stray light is much higher for Orion. This probably reflects both the differences in source morphology in the beam, and the different spectral shapes - Orion is very bright at shorter wavelengths with a peak in thermal emission $<$47$\mu$m.

\subsection{Characterising the instrumental response and throughput}\label{throughput}

\begin{figure}
\includegraphics[width=84mm]{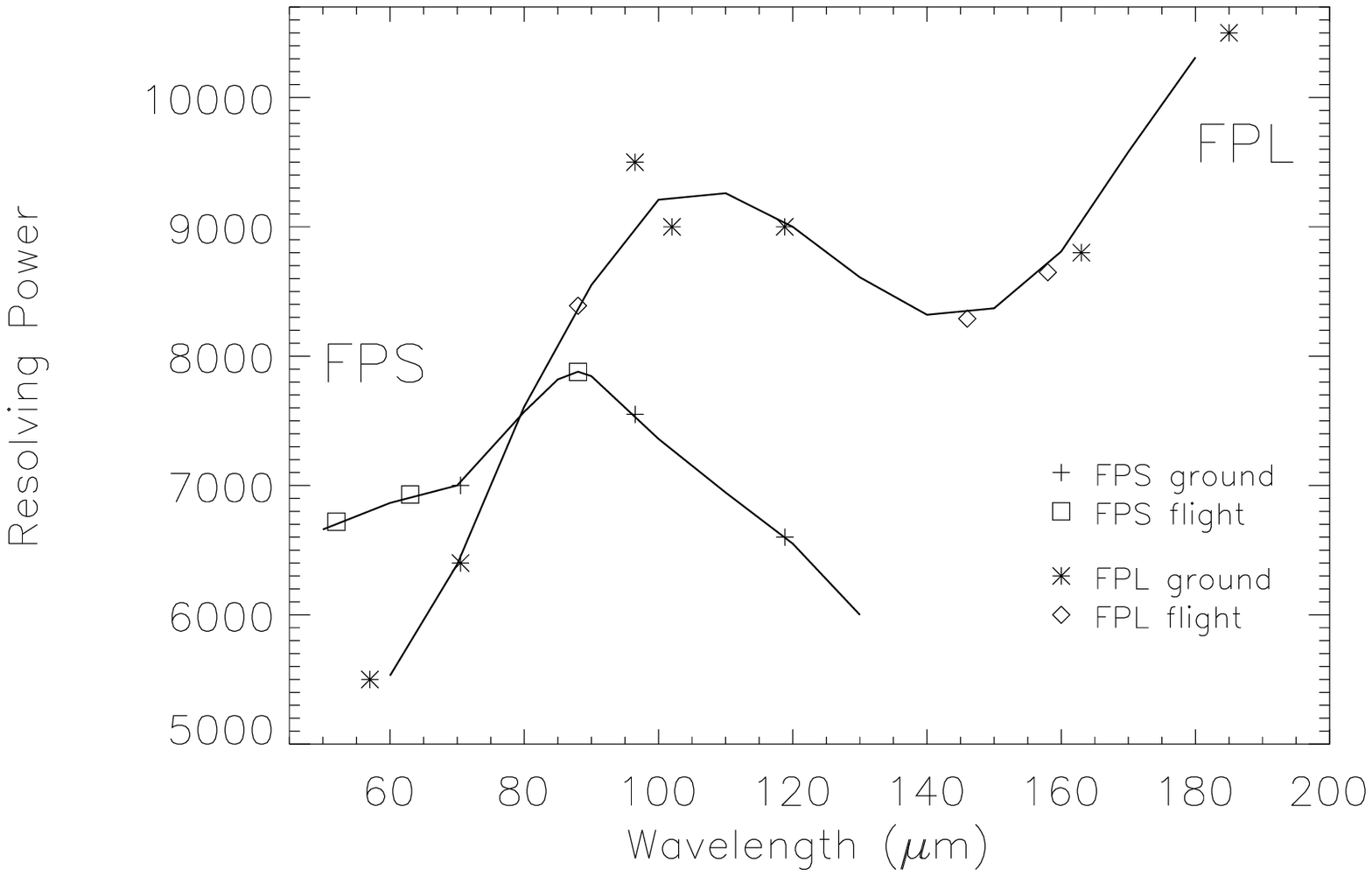}
\includegraphics[width=84mm]{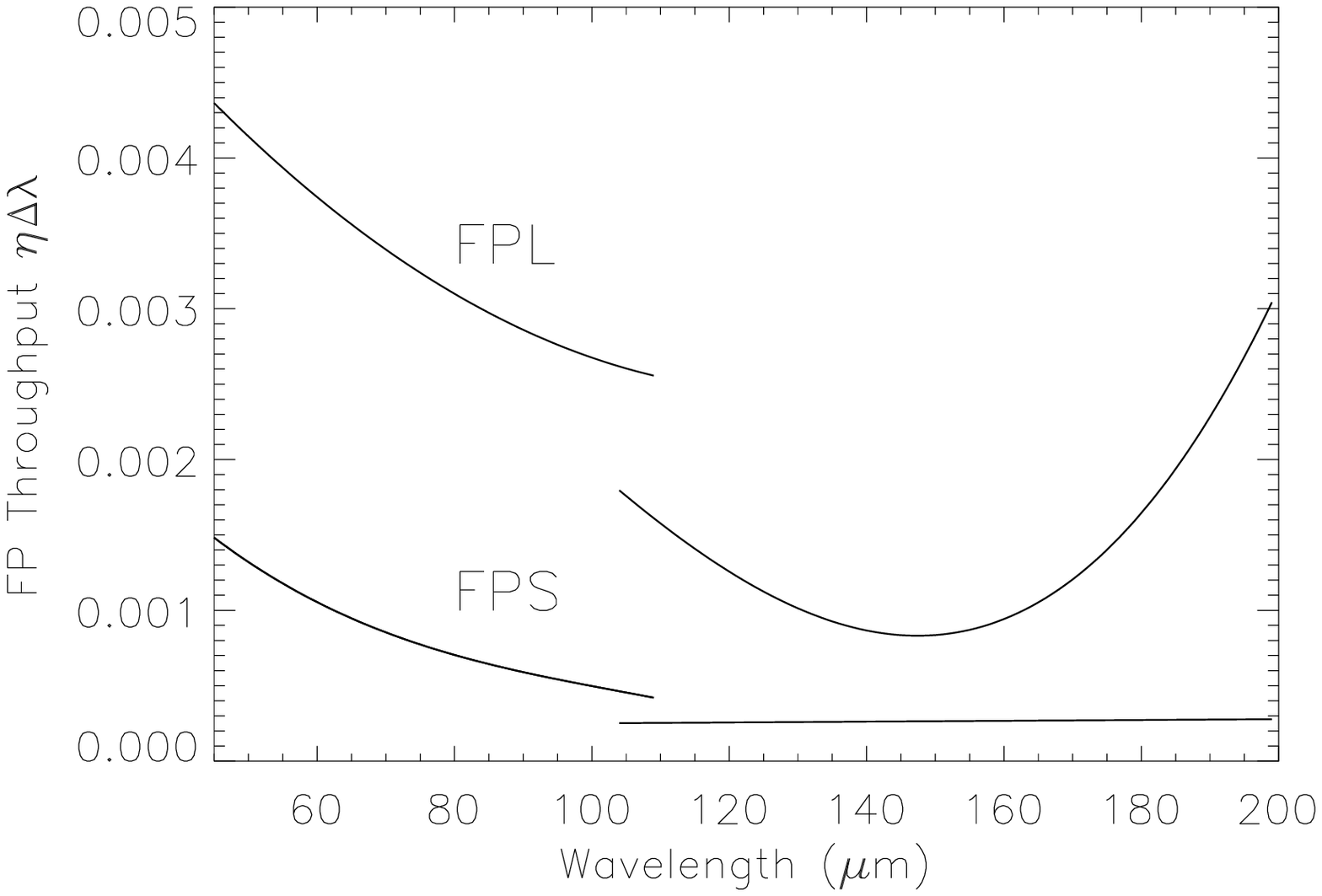}
\caption{Top: Resolving power of FPS and FPL (defined as $\lambda/\Delta\lambda_{\rm{FWHM}}$) from ground-based and in-flight measurements. Bottom: Third order polynomial fit to the total measured FP throughput for FPS and FPL (defined as the product of FP transmission efficiency and resolution element equivalent width).}
\label{resolv}
\end{figure}

The relative spectral response function (RSRF) of each LWS detector was determined in the grating mode using observations of Uranus, and monitored throughout the mission using this and several other sources. Uranus was used as the primary calibrator because it is bright, point-like in the LWS beam and almost spectrally featureless and has a well established atmospheric emission model \citep[e.g][]{griffin}. The RSRFs were obtained by removing a model of the Uranus spectrum from the observations, giving the wavelength dependant response for each detector band \citep{gry,lim_c}. These are dominated by the transmission of the band-pass filters placed in front of each detector. An additional correction was then applied to reference the absolute responsivity of each detector to its value when the RSRF calibration observations were made. This was determined using standard illuminator flashes before and after each observation.

The spectral response function of the LWS grating is also important for the calibration of FP data as it was used as an order selector for each FP and its shape is imprinted on all FP mini-scans. We used this fact to recover the grating response from non-prime FP data measured during the survey. This is possible because the FP scanned different regions of the profile in different non-prime observations. These were then recombined and the profile reconstructed. Data containing strong lines were excluded from the analysis and the Sgr~B2 continuum spectrum was removed using the L01 grating observation. The resulting data were fitted using a theoretical profile (the convolution of a Gaussian profile for the grating with the FP Airy function) with its width as a free parameter. This gave a good fit and sets of grating profiles were built up for different grating angle ranges. This accurate knowledge of the grating response in FP mode allowed a further `side order correction' to be made in the calibration, accounting for neighbouring FP orders that may have been transmitted through the wings of the grating response profile \citep[see][]{gry, polehampton_a}.

The remaining calibration needed for FP data is due to the properties of the FPs themselves. During the ground testing of the LWS, the transmission efficiency and spectral response of each FP were measured at a number of spot wavelengths using a FIR laser \citep{emery,davis}. However, in flight, it was difficult to confirm these two quantities independently of each other due to the lack of sufficiently well characterised bright narrow lines in astronomical sources. Only a few such sources (G36.3+0.7, IC3568, NGC7023, NGC7027, G0.6$-$0.6) were observed through the mission and the resulting values for the FP spectral resolving power versus wavelength are shown in the top panel of Fig.~\ref{resolv} \citep[note that the LWS handbook only contains measurements of resolving power made during ground testing of the instrument; ][]{gry}. Detailed fitting of the lines showed that the FP response function could be accurately described by the theoretical Airy profile with coefficient of finesse set by the resolving power. In the case of NGC7023, the lines were sufficiently narrow to allow a direct comparison with the grating leading to a determination of the FP transmission at two wavelengths (8.3\% at 63.2~$\mu$m for FPS and 3.5\% at 157.7~$\mu$m for FPL).

Due to the limited wavelength coverage of these flight measurements, only the product of transmission efficiency and spectral resolution element was used in the standard calibration procedure. This quantity, the FP `throughput', was easier to measure in practice by observing a bright source with known continuum spectrum by scanning the grating whilst keeping the FP gap fixed. This produces a convolution of the FP Airy function with the grating response, the peaks of which trace the FP throughput. These observations were carried out towards Mars and the source continuum removed using the Martian model developed by \citet{rudy} adapted for the LWS spectral range \citep{sidher_c}. The procedure has been described previously in the LWS handbook \citep{gry} but was improved and extended to wavelengths outside the nominal range of each FP to allow non-prime data from the spectral survey to be calibrated. In addition, an equivalent dataset was observed towards Sgr~B2 which shows reasonable agreement with the Sgr~B2 L01 spectrum when processed with the extended throughput correction derived from Mars \citep{polehampton_a}. The final throughput results are shown in the bottom panel of Fig.~\ref{resolv}. The final transmission efficiency of each FP can be determined by combining both plots in Fig.~\ref{resolv} and this gives excellent agreement with the two points measured towards NGC7023.

Figure~\ref{resolv} shows that there is an apparent break in the throughput where the coverage by detector LW1 ends and LW2 begins. One possible explanation for this discontinuity is that detectors SW1--LW1 were mounted on a physically separate structure to LW2--LW5 and may have had different optical alignment with respect to the pupil. Another possibility is due to the detectors themselves - LW2--LW5 all used stressed (Ge:Ga) detectors whereas the shorter wavelength channels were all unstressed \citep{clegg}. However, neither explanation has any corroborating evidence from other observations or from the literature.

Comparing the final throughput for the two FPs clearly shows that at short wavelengths it is advantageous in terms of signal-to-noise to use FPL non-prime data rather than FPS. The only cost is a slight reduction in spectral resolution but this is outweighed by the signal-to-noise gain (see Sect.~\ref{sect_signois}).

\section{Data}

\begin{figure*}
\resizebox{14.8cm}{!}{\includegraphics{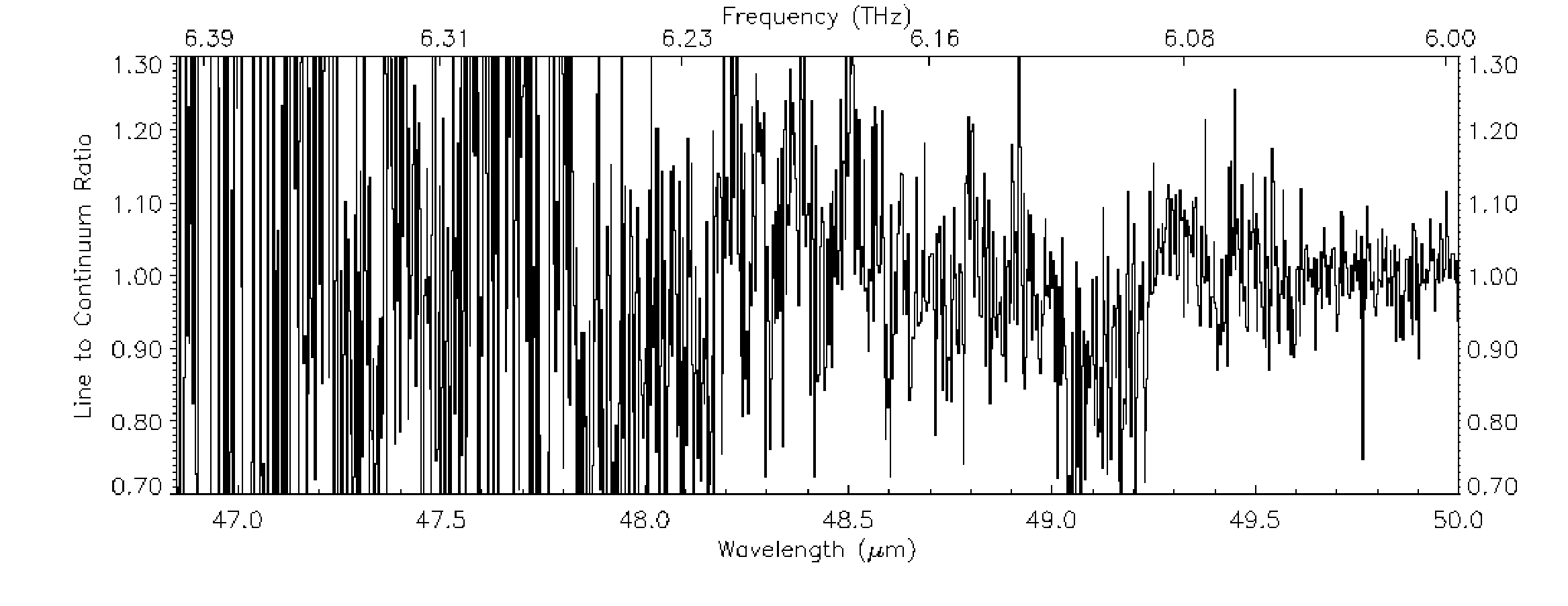}}
\resizebox{14.8cm}{!}{\includegraphics{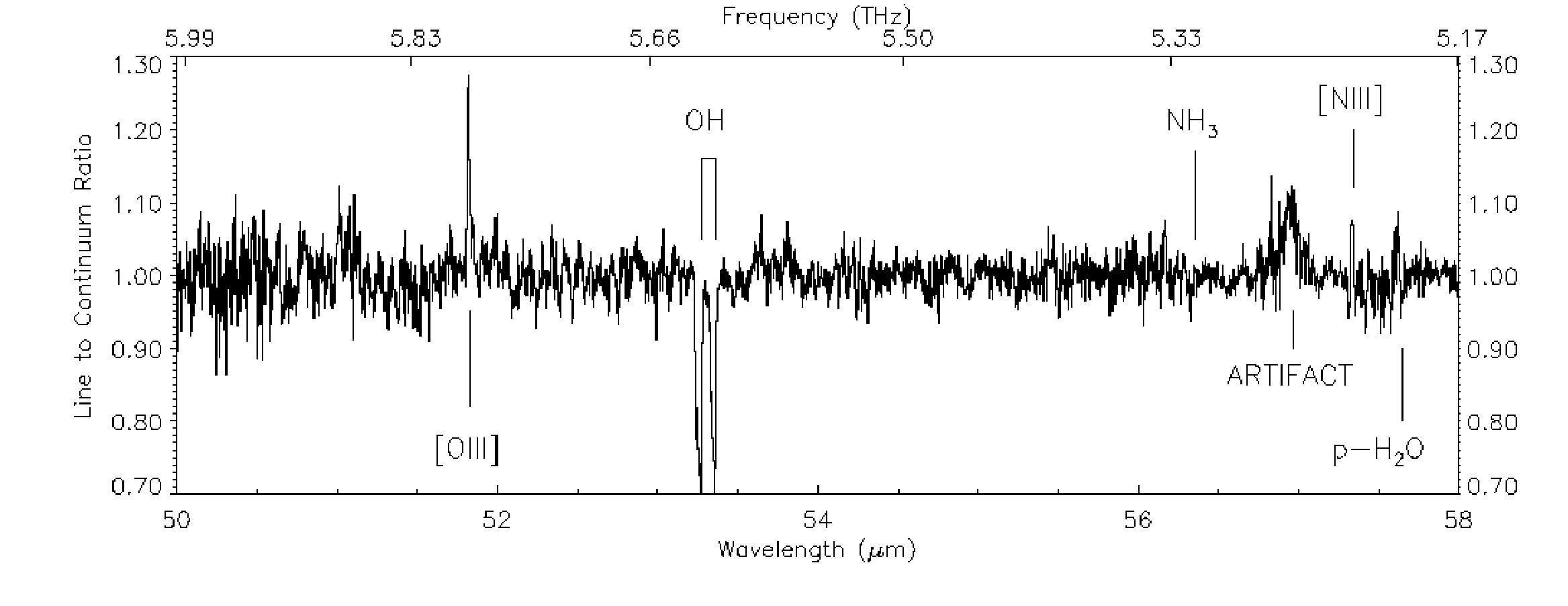}}
\resizebox{14.8cm}{!}{\includegraphics{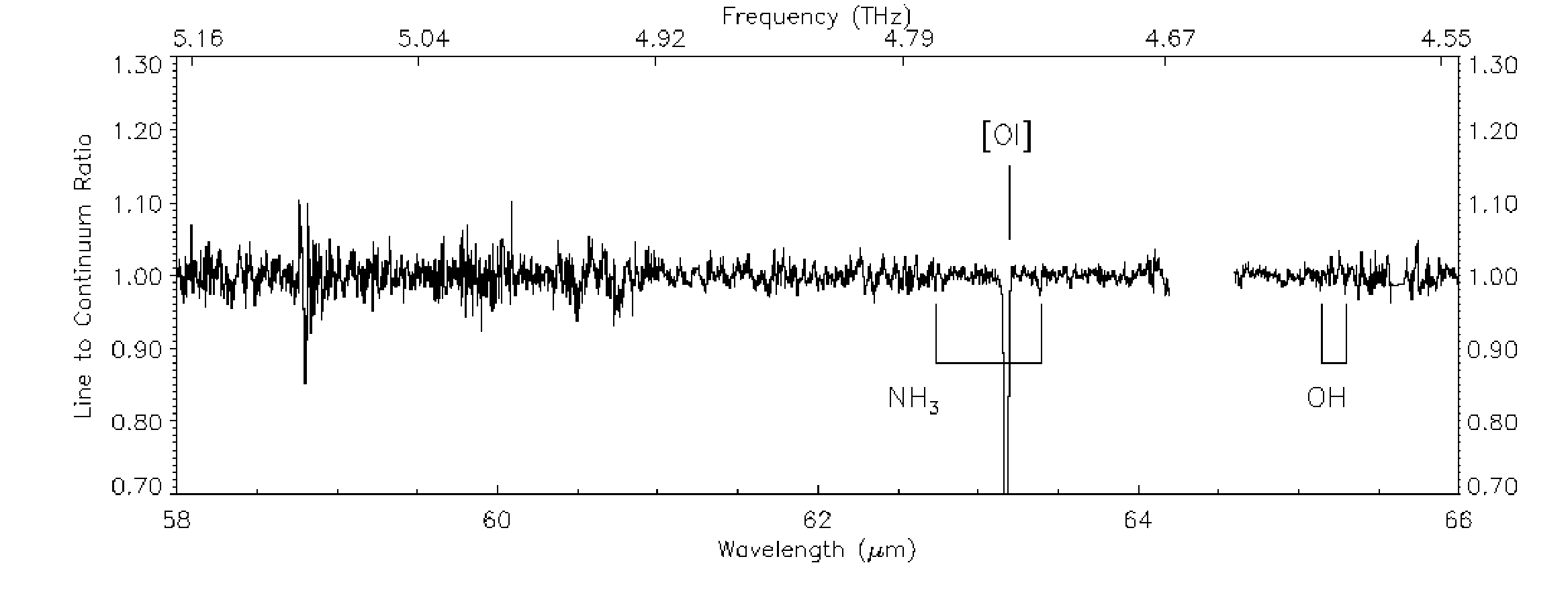}}
\resizebox{14.8cm}{!}{\includegraphics{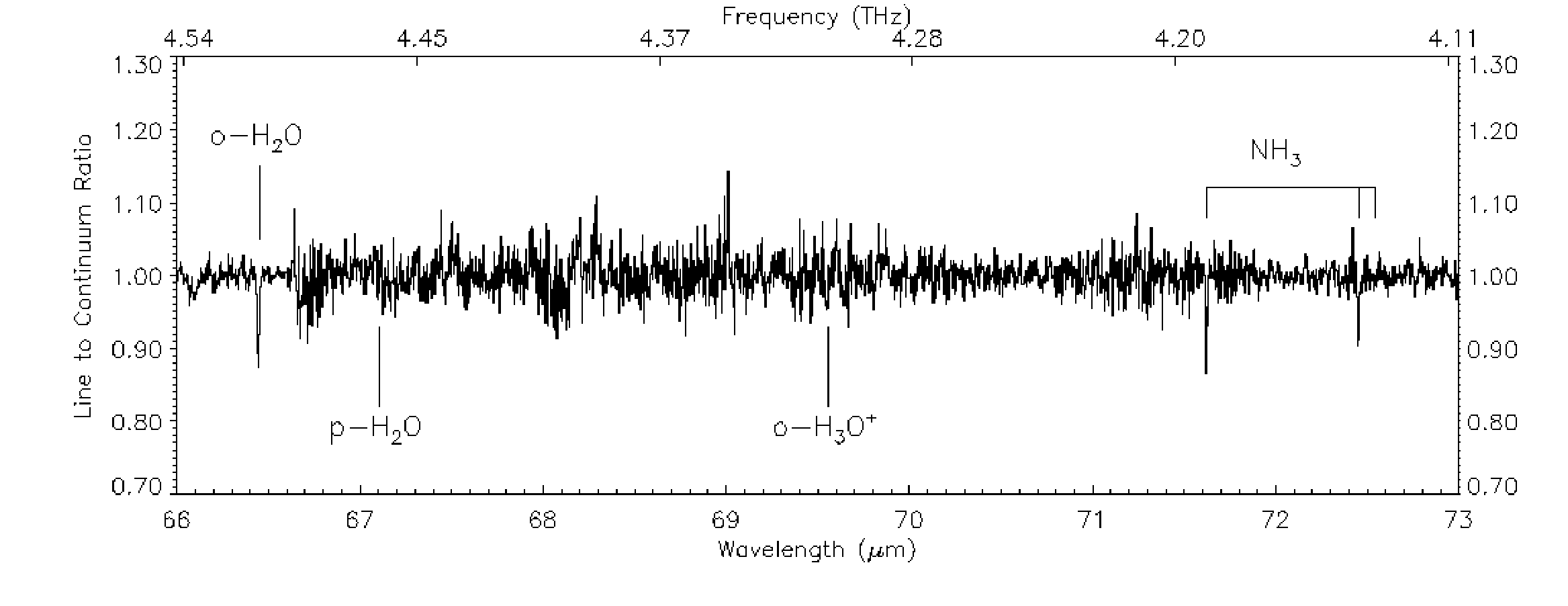}}
\caption{Full L03 FP spectrum after co-adding prime and non-prime data using FPL for all LWS detectors (below 70~$\mu$m only non-prime data were used). The data are binned at 1/4 of the instrumental resolution element. Unassigned features detected at $>$3$\sigma$ are labelled `U-line'. Two features are labelled `ARTIFACT' (at 57~$\mu$m and 141.7~$\mu$m). These are residual features due to the data reduction and are not real.\label{allsurvey}}
\end{figure*}

\begin{figure*}
\resizebox{14.8cm}{!}{\includegraphics{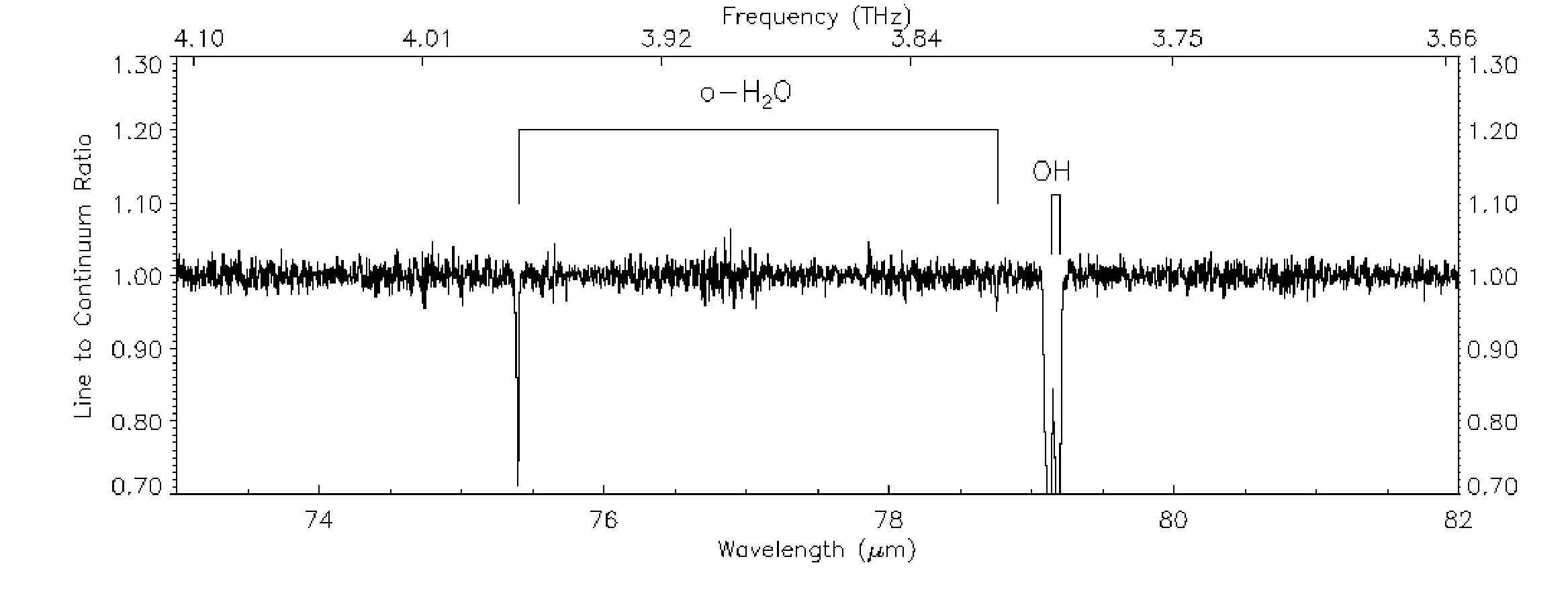}}
\resizebox{14.8cm}{!}{\includegraphics{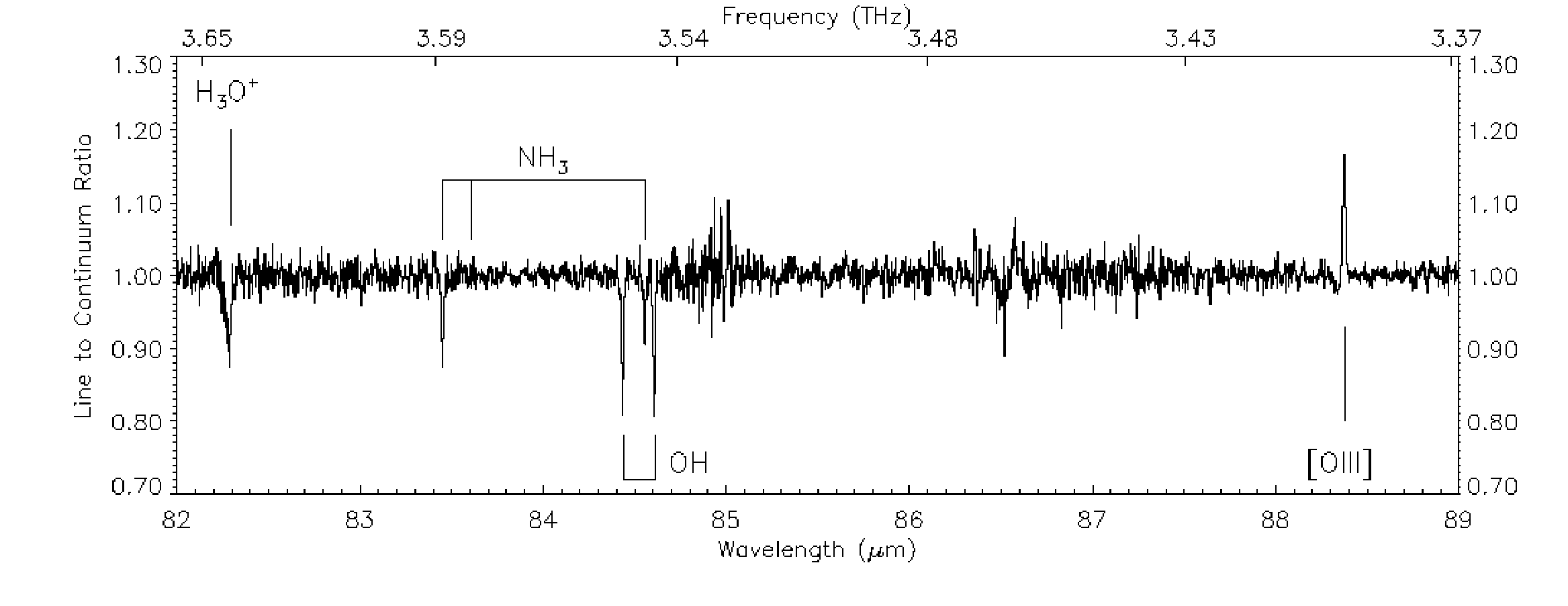}}
\resizebox{14.8cm}{!}{\includegraphics{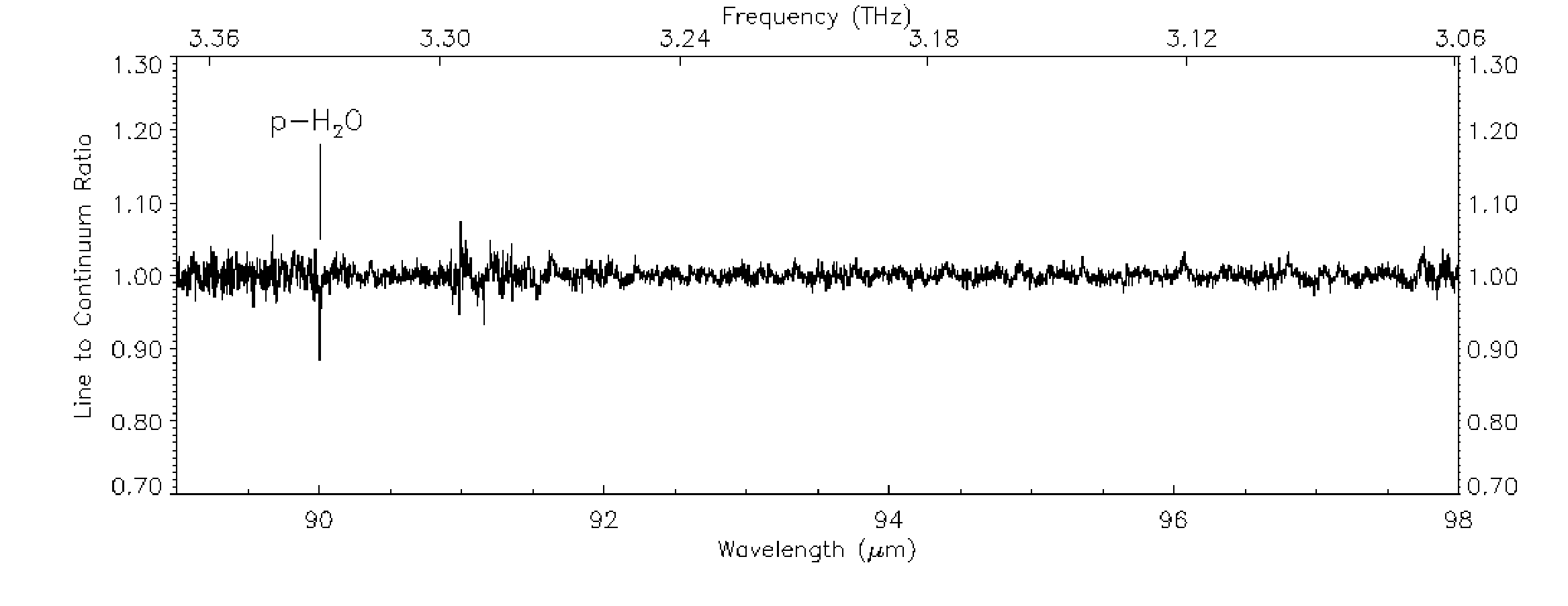}}
\resizebox{14.8cm}{!}{\includegraphics{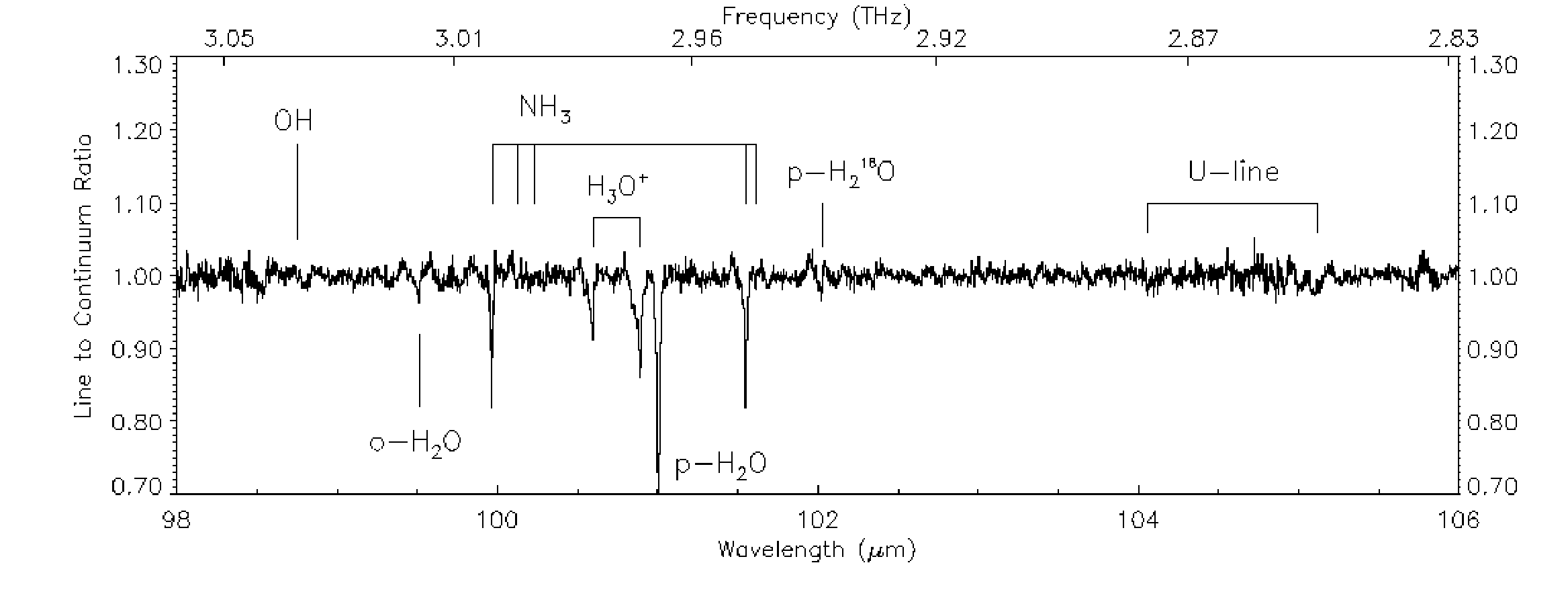}}
\end{figure*}

\begin{figure*}
\resizebox{14.8cm}{!}{\includegraphics{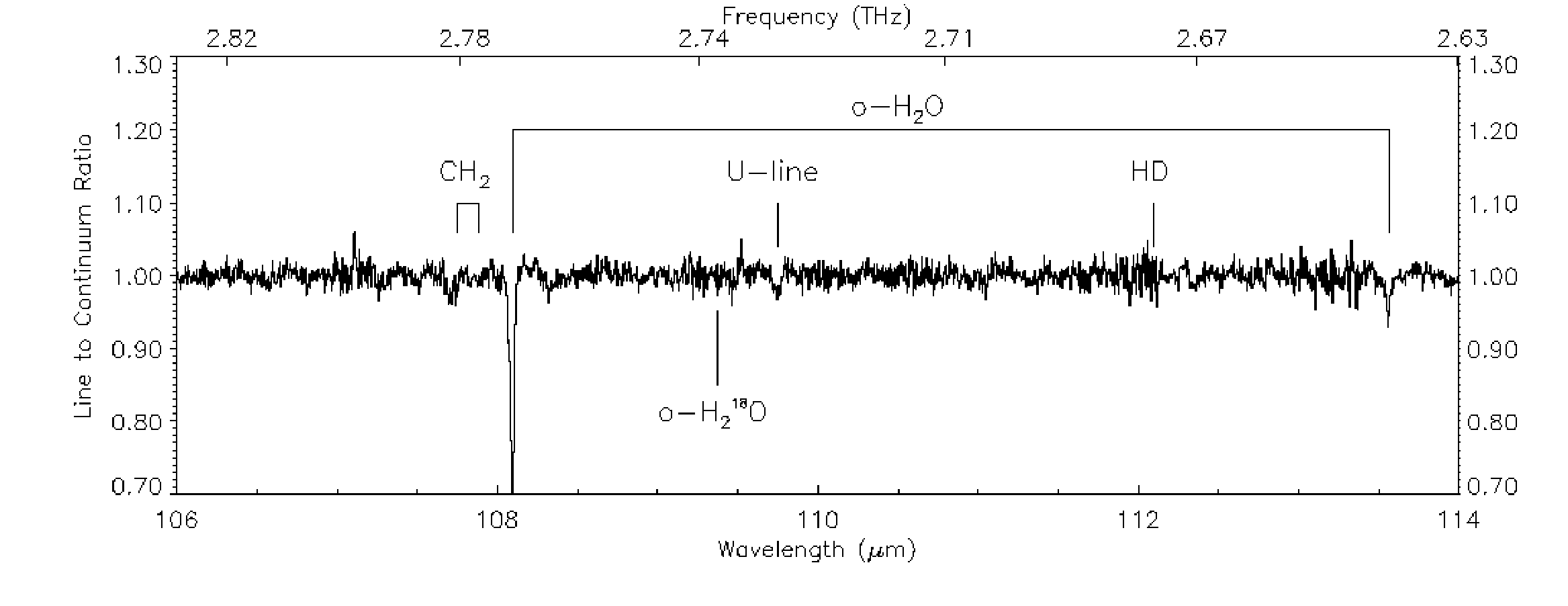}}
\resizebox{14.8cm}{!}{\includegraphics{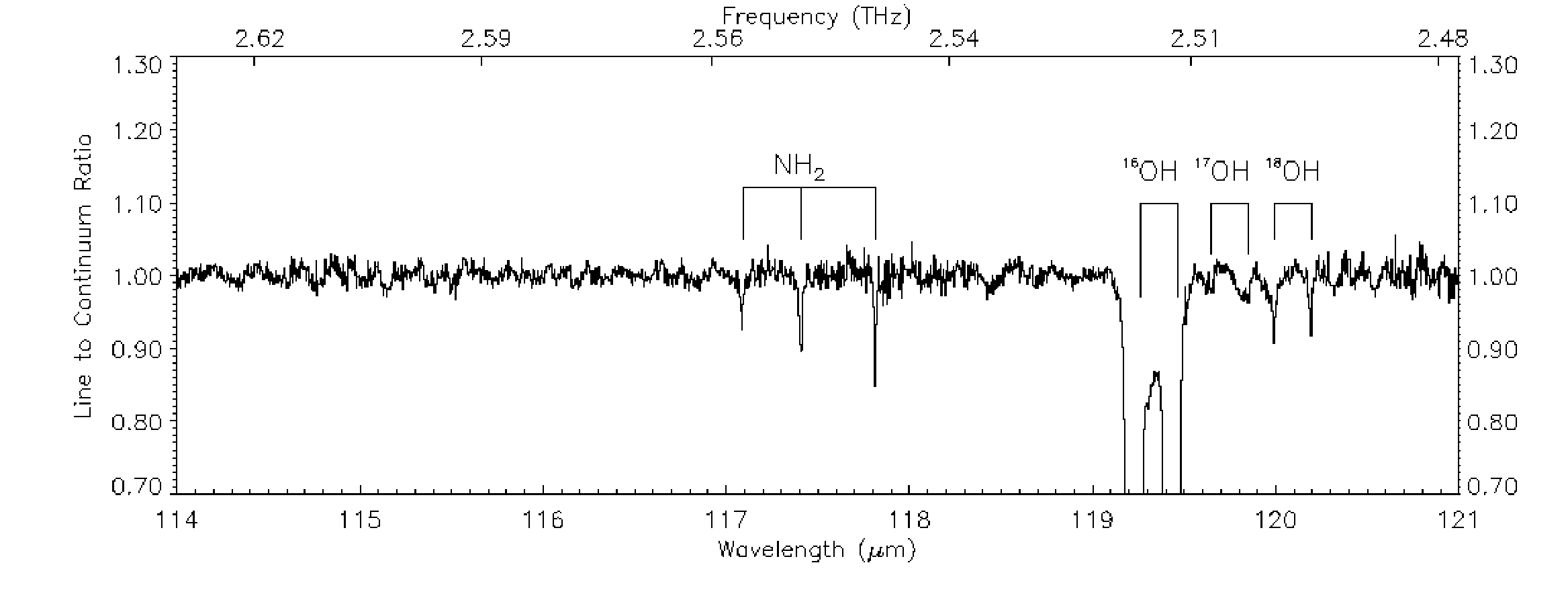}}
\resizebox{14.8cm}{!}{\includegraphics{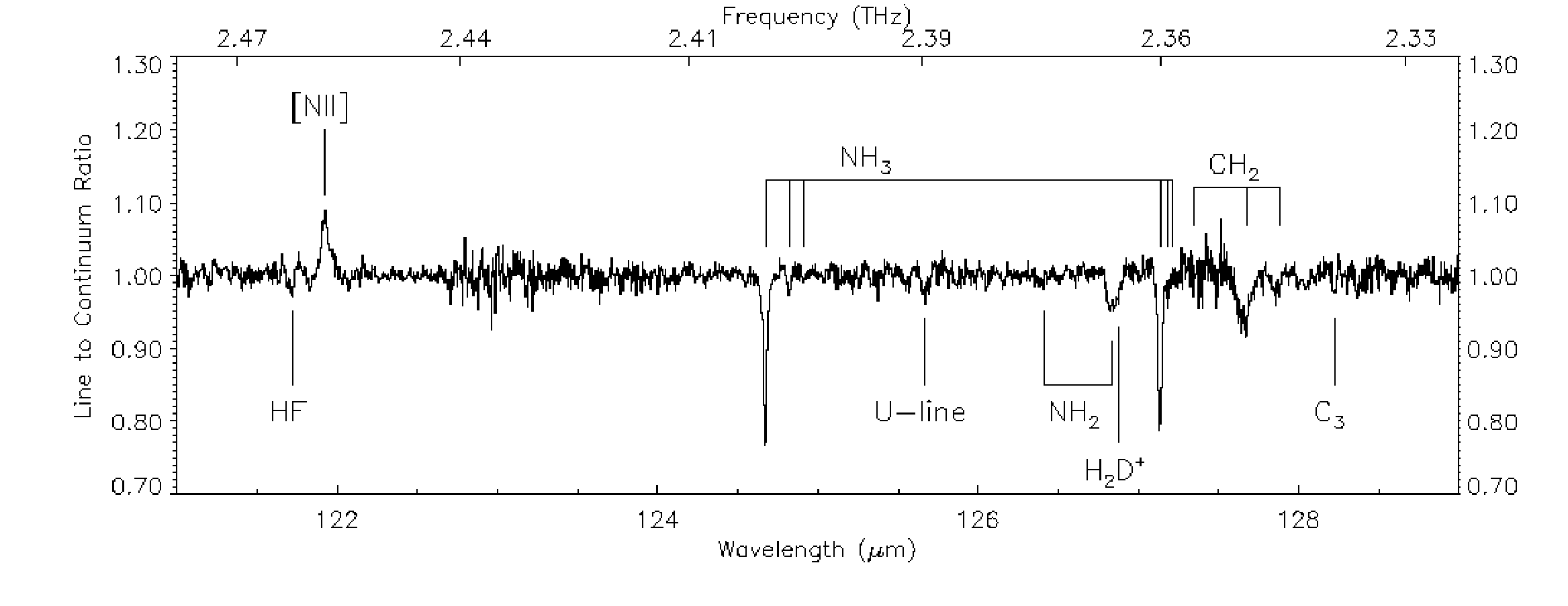}}
\resizebox{14.8cm}{!}{\includegraphics{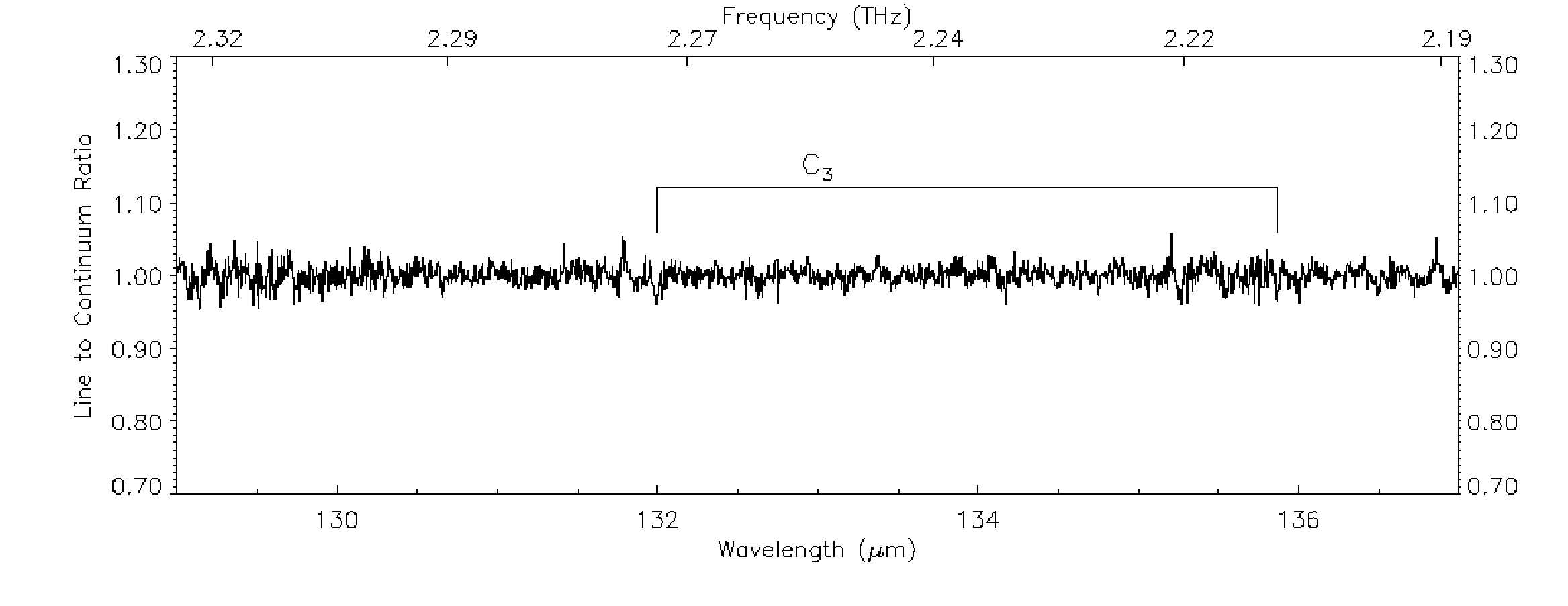}}
\end{figure*}

\begin{figure*}
\resizebox{14.8cm}{!}{\includegraphics{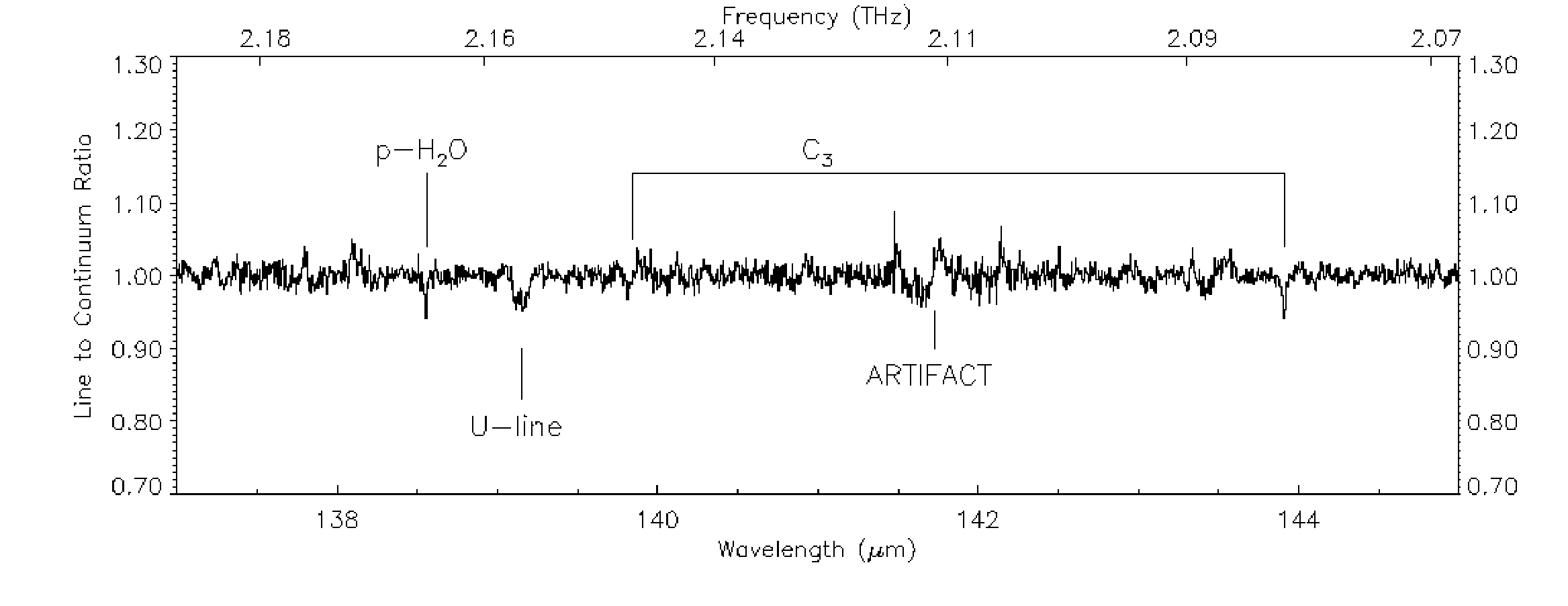}}
\resizebox{14.8cm}{!}{\includegraphics{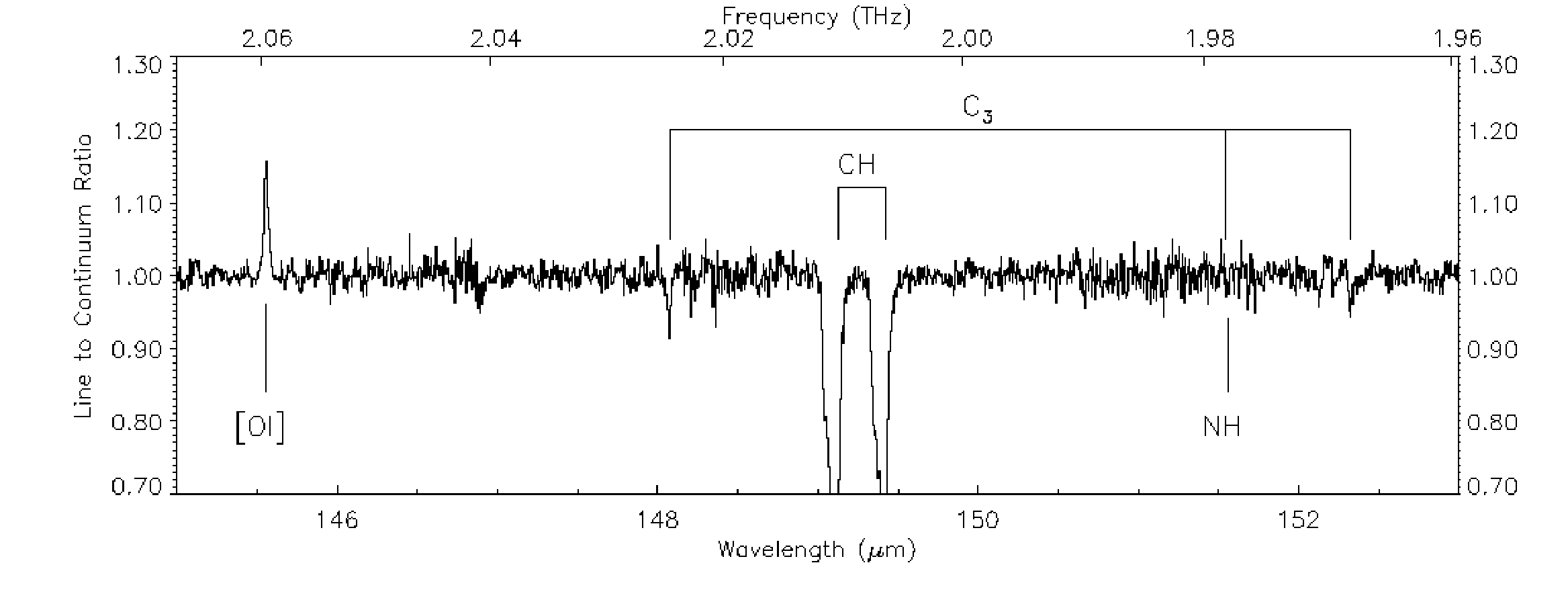}}
\resizebox{14.8cm}{!}{\includegraphics{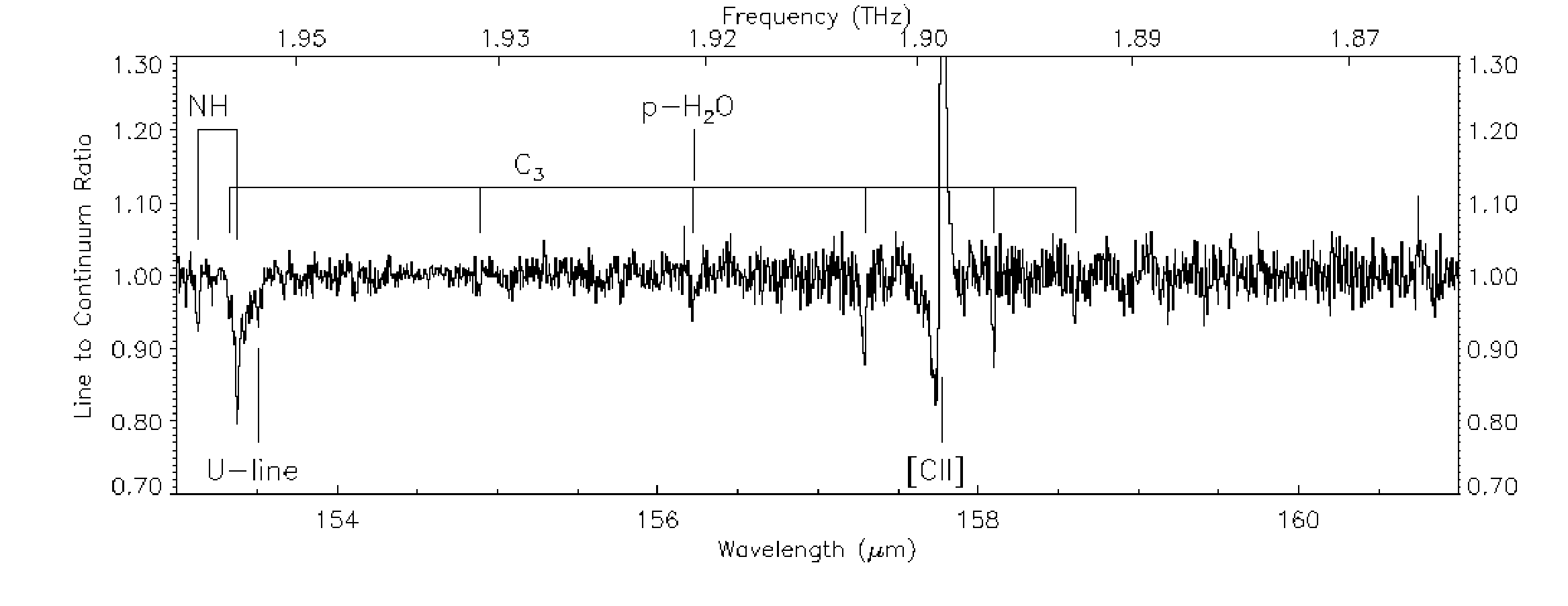}}
\resizebox{14.8cm}{!}{\includegraphics{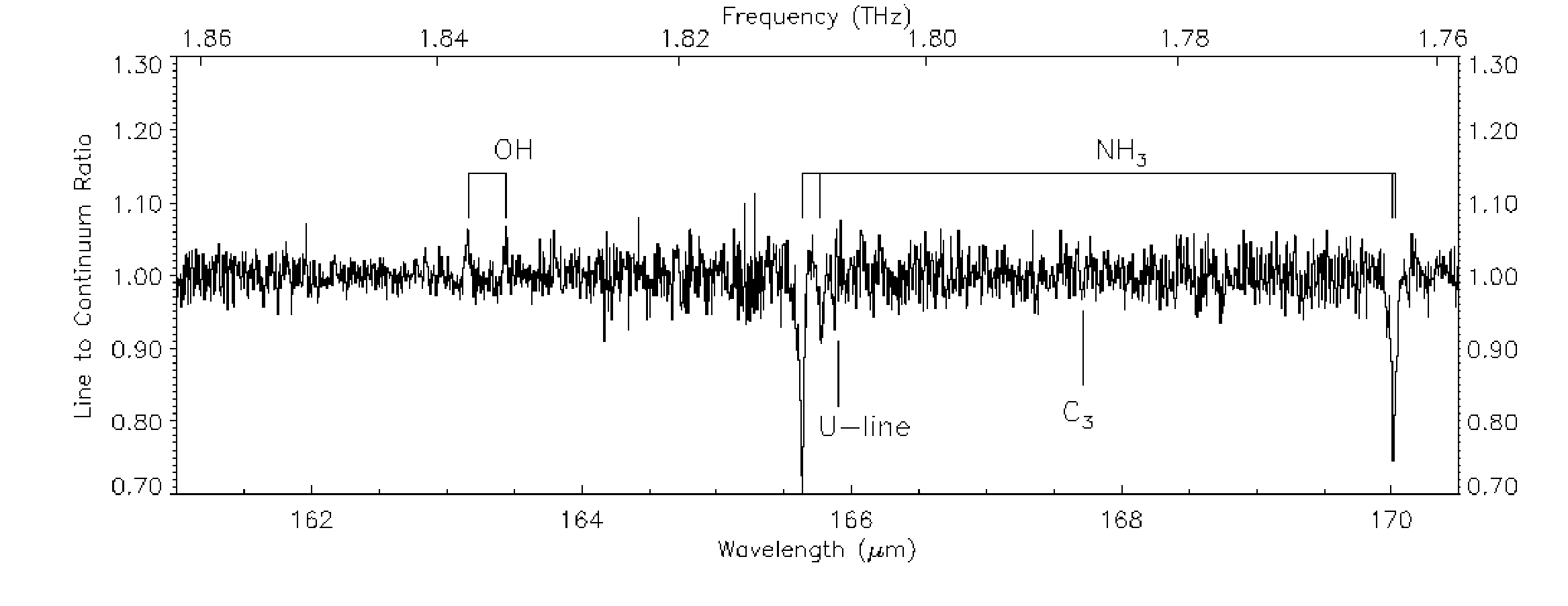}}
\end{figure*}

\begin{figure*}
\resizebox{14.8cm}{!}{\includegraphics{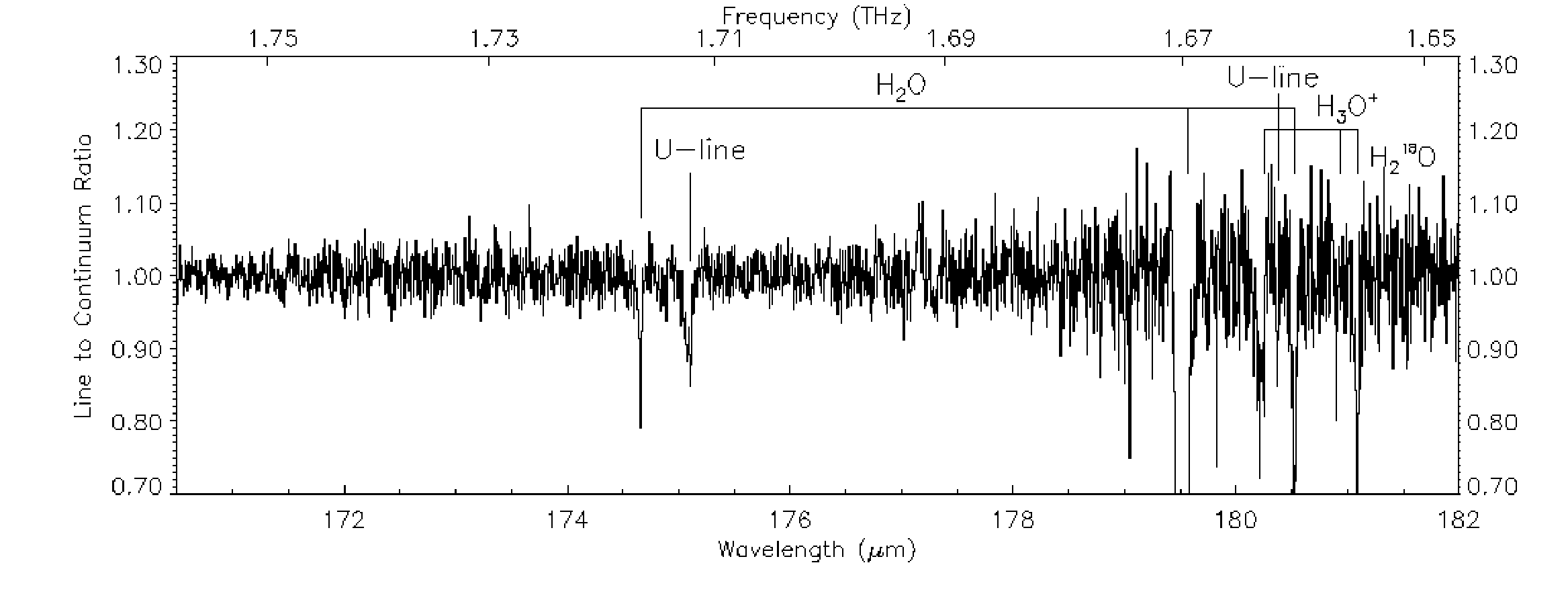}}
\resizebox{14.8cm}{!}{\includegraphics{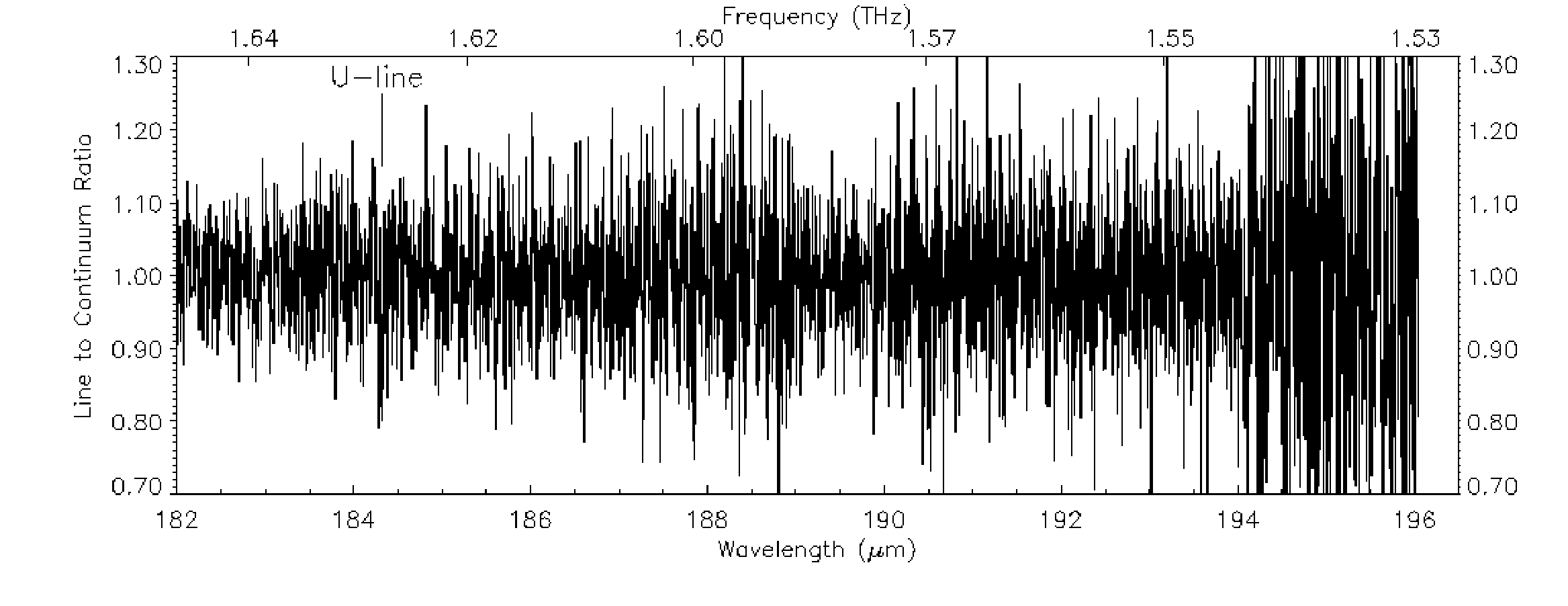}}
\end{figure*}

\section{Observations}

\begin{table*}
\caption{Log of all the Sgr~B2 L03 observations and the corresponding L01 grating observation. The TDT (target dedicated time) number is an unique identifier for each observation. For the L03 data, only the wavelength range recorded on the prime detector is shown. In some observations the specified wavelength coverage was split between two prime detectors (this occurred at the edge of the detector nominal range). The length of the observation in seconds, the observation date, the number of repeated FP mini-scans and the LSR velocity correction applied are shown in the remaining columns.}
\label{alll03}
\begin{tabular}{cccccccc}
\hline
TDT    & Prime Wavelength & FP   & Prime    & Length & Date & Repeated & Velocity Correction \\
Number & Range            & Used & Detector & (s)    &(dd.mm.yy)& Scans    & Applied  \\
       & ($\mu$m)         &      &          &        &      &          & (km~s$^{-1}$) \\
\hline
50400431 & 47.0~--~49.5   & FPS & SW1     & 4646 & 3.4.97 & 3 & $-39.3$\\
50400330 & 49.5~--~52.0   & FPS & SW1/SW2 & 5934 & 3.3.97 & 3 & $-39.6$\\
50400929 & 52.0~--~54.5   & FPS & SW2     & 4309 & 3.4.97 & 3 & $-37.8$\\
50401028 & 54.5~--~57.0   & FPS & SW2     & 4200 & 3.4.97 & 3 & $-37.7$\\
50900327 & 57.0~--~60.0   & FPS & SW2/SW3 & 6492 & 8.4.97 & 3 & $-37.7$\\
50400526 & 60.0~--~63.0   & FPS & SW3     & 4540 & 3.4.97 & 3 & $-38.6$\\
50400725 & 63.0~--~66.0   & FPS & SW3     & 4490 & 3.4.97 & 3 & $-38.2$\\
50400224 & 66.0~--~70.3   & FPS & SW3     & 6184 & 3.4.97 & 3 & $-39.7$\\
\hline
50400823 & 70.3~--~73.0   & FPL & SW4     & 4446  & 3.4.97  & 3 & $-38.0$\\
50400122 & 73.0~--~77.0   & FPL & SW4     & 6364  & 3.4.97  & 3 & $-39.1$\\
50900521 & 77.0~--~81.0   & FPL & SW4/SW5 & 8828  & 8.4.97  & 3 & $-39.0$\\
50900620 & 81.0~--~85.0   & FPL & SW5/LW1 & 3448  & 8.4.97  & 3 & $-38.1$\\
50800819 & 85.0~--~89.0   & FPL & LW1     & 4366  & 7.4.97  & 3 & $-37.6$\\
50800218 & 89.0~--~94.0   & FPL & LW1     & 5232  & 7.4.97  & 3 & $-39.0$\\
50800317 & 94.0~--~99.0   & FPL & LW1     & 5134  & 7.4.97  & 3 & $-38.6$\\
50800416 & 99.0~--~104.0  & FPL & LW1     & 5040  & 7.4.97  & 3 & $-38.3$\\
50800515 & 104.0~--~109.0 & FPL & LW2     & 4708  & 7.4.97  & 3 & $-38.0$\\
50600814 & 109.0~--~115.0 & FPL & LW2     & 5492  & 5.4.97  & 3 & $-37.9$\\
50601013 & 115.0~--~121.0 & FPL & LW2     & 5424  & 5.4.97  & 3 & $-37.4$\\
50601112 & 121.0~--~127.0 & FPL & LW2     & 5244  & 5.4.97  & 3 & $-37.2$\\
50700511 & 127.0~--~133.0 & FPL & LW2/LW3 & 6054  & 6.4.97  & 3 & $-37.7$\\
50700610 & 133.0~--~140.0 & FPL & LW3     & 5812  & 6.4.97  & 3 & $-37.4$\\
47600809 & 140.0~--~147.0 & FPL & LW3     & 5806  & 6.3.97  & 3 & $-38.9$\\
50700208 & 147.0~--~154.0 & FPL & LW3/LW4 & 6418  & 6.4.97  & 3 & $-38.8$\\
50700707 & 154.0~--~161.0 & FPL & LW4     & 5566  & 6.4.97  & 3 & $-37.2$\\
50600506 & 161.0~--~168.0 & FPL & LW4     & 5536  & 5.4.97  & 3 & $-38.6$\\
83800606 & 167.0~--~170.0 & FPL & LW4     & 3342  & 2.3.98  & 4 & $-38.5$\\
83600605 & 170.0~--~174.0 & FPL & LW4     & 4246  & 28.2.98 & 4 & $-38.1$\\
50600405 & 168.0~--~175.0 & FPL & LW4     & 5553  & 5.4.97  & 3 & $-38.9$\\
83600704 & 174.0~--~178.0 & FPL & LW4     & 4096  & 28.2.98 & 4 & $-38.1$\\
84900803 & 178.0~--~182.0 & FPL & LW5     & 5917  & 13.3.98 & 6 & $-39.4$\\
50700404 & 175.0~--~182.0 & FPL & LW4/LW5 & 6146  & 6.4.97  & 3 & $-38.1$\\
84500102 & 182.0~--~189.0 & FPL & LW5     & 10158 & 9.3.98  & 6 & $-39.7$\\
50600603 & 182.0~--~190.0 & FPL & LW5     & 5494  & 5.4.97  & 3 & $-38.3$\\
84700301 & 189.0~--~194.0 & FPL & LW5     & 7060  & 11.3.98 & 6 & $-39.4$\\
50600902 & 189.0~--~196.0 & FPL & LW5     & 5468  & 5.4.97  & 3 & $-37.7$\\
\hline
\multicolumn{8}{l}{L01 observation}\\
28701401 & 43.0~--~196.0  &     &         & 2386  & 30.8.96 &   & \\
\end{tabular}
\end{table*}

\label{lastpage}
\end{document}